\documentclass[
    9pt,
    twocolumn,
    ]{extarticle}

\usepackage[utf8]{inputenc}
\usepackage[
    colorlinks=true,
    linkcolor=black,
    citecolor=black,
    filecolor=black,
    urlcolor=black,
    plainpages=false,
    pdfpagelabels,
    breaklinks=true,
    pdfusetitle,
    ]{hyperref}

\usepackage[
    natbib=true,
    backend=bibtex,
    style=nature,
    citestyle=numeric-comp,
    sorting=none,
    giveninits=true,
    uniquename=init
]{biblatex}

\usepackage{algorithm}
\usepackage{algpseudocode}
\usepackage{amsthm, amsmath, amsfonts, amssymb, amscd}
\usepackage{bbm}
\newcommand{\bm}[1]{\boldsymbol{\mathbf{#1}}}  %
\usepackage{bookmark}
\usepackage{booktabs}
\usepackage[font=footnotesize, labelfont=bf]{caption}
\usepackage{cuted}
\usepackage{enumitem}
\usepackage[capitalize, nameinlink]{cleveref}
    \creflabelformat{equation}{#2\textup{#1}#3}
\newcommand{\nameitref}[1]{\cref{#1}}
\usepackage[sharp]{easylist}
    \ListProperties(
        Hang=true,Space=4pt,Space*=4pt,Hide=1000,
        Margin1=1.5em,Style1*=\textbullet\hskip .5em,
        Margin2=3.7em,Style2*=--\hskip .5em,
        Margin3=5.9em,Style3*=$\ast$\hskip .5em,
        Margin4=7.8em,Style4*=$\cdot$\hskip .5em
        )
\usepackage[a4paper, margin=20pt]{geometry}
\usepackage[docdef=atom]{glossaries-extra}
\usepackage{graphicx}
\usepackage{listings}                                %
\usepackage{mathtools}
\usepackage{cancel}
\usepackage{multicol}
\usepackage{multirow}
\usepackage{parskip}
\usepackage{placeins}                                %
\usepackage{relsize}
\usepackage{setspace}
\usepackage[binary-units]{siunitx}
\usepackage{subcaption}
\usepackage{threeparttable}
\usepackage[dvipsnames]{xcolor}
\usepackage{xparse}
\usepackage{xspace}
\usepackage{tikz}
\usepackage{dsfont}
\usepackage{marvosym}

\makeatletter
\newcounter{algorithmicH}%
\let\oldalgorithmic\algorithmic
\renewcommand{\algorithmic}{%
  \stepcounter{algorithmicH}%
  \oldalgorithmic}%
\renewcommand{\theHALG@line}{ALG@line.\thealgorithmicH.\arabic{ALG@line}}
\makeatother

\usepackage[final]{changes}  %
\definechangesauthor[name=Paul, color=magenta]{PH}
\definechangesauthor[name=Jakob, color=cyan]{JJ}
\definechangesauthor[name=Ismael, color=purple]{IJ}
\definechangesauthor[name=Beni, color=ForestGreen]{BE}
\definechangesauthor[name=Fede, color=red]{FB}
\definechangesauthor[color=orange]{MAP}

\usepackage{lineno}

\bookmarksetup{depth=4}
\definecolor{lightgrey}{gray}{0.9}
\crefname{equation}{Eqn.}{Eqns.}
\Crefname{equation}{Eqn.}{Eqns.}
\setabbreviationstyle[acronym]{long-short}
\glssetcategoryattribute{acronym}{nohyperfirst}{true}

\newacronym{ad}{AD}{automatic differentiation}
    \newcommand{\ad}{\gls{ad}\xspace}
\newacronym{am}{AM}{adjoint method}
    \newcommand{\am}{\gls{am}\xspace}
    \newcommand{\ambptt}{\gls{am}/\gls{bptt}\xspace}
\newacronym{ann}{ANN}{artificial neural network}
    \newcommand{\ann}{\gls{ann}\xspace}
    \newcommand{\anns}{\glspl{ann}\xspace}
\newacronym{asr}{ASR}{automatic speech recognition}
\newacronym{bp}{BP}{backpropagation}
    \newcommand{\bp}{\gls{bp}\xspace}
\newacronym{bptt}{BPTT}{backpropagation through time}
    \newcommand{\bptt}{\gls{bptt}\xspace}
\newacronym{ce}{CE}{cross-entropy}
\newacronym{cifar}{CIFAR}{Canadian Institute For Advanced Research}
\newacronym{cifar10}{CIFAR-10}{CIFAR-10}
    \newcommand{\cifarIO}{\gls{cifar10}\xspace}
\newacronym{cmc}{CMC}{cortical microcircuit}
    \newcommand{\cmc}{\gls{cmc}\xspace}
    \newcommand{\cmcs}{\glspl{cmc}\xspace}
\newacronym{cnn}{CNN}{convolutional neural network}
    
    \newcommand{\cnns}{\glspl{cnn}\xspace}
\newacronym{cvnn}{CVNN}{complex-valued neural network}
    
    \newcommand{\cvnns}{\glspl{cvnn}\xspace}
\newacronym{dl}{DL}{deep learning}
    \newcommand{\dl}{\gls{dl}\xspace}
\newacronym{esn}{ESN}{echo state network}
    \newcommand{\esn}{\gls{esn}\xspace}
    \newcommand{\esns}{\glspl{esn}\xspace}
\newacronym{fa}{FA}{feedback alignment}
    \newcommand{\fa}{\gls{fa}\xspace}
\newacronym{gd}{GD}{gradient descent}
    \newcommand{\gd}{\gls{gd}\xspace}
\newacronym{gle}{GLE}{Generalized Latent Equilibrium}
    \newcommand{\gle}{\gls{gle}\xspace}
\newacronym{gn}{GN}{Gauss-Newton}
    \newcommand{\gn}{\gls{gn}\xspace}
\newacronym{gru}{GRU}{gated recurrent unit}
    \newcommand{\gru}{\gls{gru}\xspace}
    \newcommand{\grus}{\glspl{gru}\xspace}
\newacronym{gsc}{GSC}{Google Speech Commands}
    \newcommand{\gsc}{\gls{gsc}\xspace}
\newacronym{ie}{IE}{instantaneous errors}
\newacronym{le}{LE}{Latent Equilibrium}
    \renewcommand{\le}{\gls{le}\xspace}
\newacronym{li}{LI}{leaky integrator}
\newacronym{lif}{LIF}{leaky integrate-and-fire}
    
\newacronym{lrnn}{LRNN}{locally recurrent neural network}
    
    \newcommand{\lrnns}{\glspl{lrnn}\xspace}
\newacronym{lru}{LRU}{linear recurrent unit}
    
    \newcommand{\lrus}{\glspl{lru}\xspace}
\newacronym{lstm}{LSTM}{long short-term memory}
    \newcommand{\lstm}{\gls{lstm}\xspace}
    \newcommand{\lstms}{\glspl{lstm}\xspace}
\newacronym{lti}{LTI}{linear time-invariant}
    \newcommand{\lti}{\gls{lti}\xspace}
\newacronym{mfcc}{MFCC}{Mel-frequency cepstral coefficient}
\newacronym{mfs}{MFS}{Mel-frequency spectrogram}
\newacronym{ml}{ML}{machine learning}
    \newcommand{\ml}{\gls{ml}\xspace}
\newacronym{mlp}{MLP}{multi-layer perceptron}
    \newcommand{\mlp}{\gls{mlp}\xspace}
    
\newacronym{mnist1d}{MNIST-1D}{MNIST-1D}
    \newcommand{\mnistId}{MNIST-1D\xspace}
\newacronym{mse}{MSE}{mean-squared error}
\newacronym{nla}{NLA}{neuronal least-action}
    \newcommand{\nla}{\gls{nla}\xspace}
\newacronym{nlif}{nLIF}{non-leaky integrate-and-fire}
\newacronym{ode}{ODE}{ordinary differential equation}
\newacronym{pyr}{PYR}{pyramidal}
    \newcommand{\pyr}{\gls{pyr}\xspace}
\newacronym{rflo}{RFLO}{random feedback online learning}
    \newcommand{\rflo}{\gls{rflo}\xspace}
\newacronym{rl}{RL}{reinforcement learning}
\newacronym{rnn}{RNN}{recurrent neural network}

\newacronym{rtrl}{RTRL}{real-time recurrent learning}
    \newcommand{\rtrl}{\gls{rtrl}\xspace}
\newacronym{sgd}{SGD}{stochastic gradient descent}
    \newcommand{\sgd}{stochastic gradient descent\xspace}
\newacronym{snn}{SNN}{spiking neural network}
\newacronym{ssm}{SSM}{state space model}
    
    \newcommand{\ssms}{\glspl{ssm}\xspace}
\newacronym{sst}{SST}{somatostatin-expressing}
    \newcommand{\sst}{\gls{sst}\xspace}
\newacronym{stdp}{STDP}{spike-timing-dependent plasticity}
    \newcommand{\stdp}{\gls{stdp}\xspace}
\newacronym{smape}{sMAPE}{symmetric mean absolute percentage error}
\newacronym{tcn}{TCN}{temporal convolutional network}
    \newcommand{\tcn}{\gls{tcn}\xspace}
    \newcommand{\tcns}{\glspl{tcn}\xspace}
\newacronym{ttfs}{TTFS}{time-to-first-spike}

\theoremstyle{theorem}
\newtheorem{postulate}{Postulate}
\renewbibmacro*{name:andothers}{
    \ifboolexpr{
    test {\ifnumequal{\value{listcount}}{\value{liststop}}}
    and
    test \ifmorenames %
  }
    {\ifnumgreater{\value{liststop}}{1}
       {\finalandcomma}
       {}%
     \andothersdelim\bibstring{andothers}}
 {}
}

\newcommand*\subtxt[1]{_{\textnormal{#1}}}
\DeclareRobustCommand\_{\ifmmode\expandafter\subtxt\else\textunderscore\fi}
\mathchardef\ordinarycolon\mathcode`\:
    \mathcode`\:=\string"8000 %
    \begingroup \catcode`\:=\active{}
    \gdef:{\mathrel{\mathop\ordinarycolon}}
    \endgroup

\newcommand{\Bell}{\bm b_\ell}
\newcommand{\bigO}{\mathcal O}

\newcommand{\bs}[1]{\bm{#1}}
\newcommand{\btau}{\bs{\tau}}
\newcommand{\btaum}{\bs{\tau}^\mathrm{m}}
\newcommand{\btaumell}{\bs{\tau}^\mathrm{m}_\ell}
\newcommand{\btaur}{\bs{\tau}^\mathrm{r}}
\newcommand{\btaurell}{\bs{\tau}^\mathrm{r}_\ell}
\newcommand{\dd}{\text d}

\NewDocumentCommand{\dminusm}{o}{
    \IfNoValueTF{#1}
    {\op{\mathcal D}{-}{\taum}}
    {\op{\mathcal D}{-}{\taum}[#1]}
}
\NewDocumentCommand{\Dminusm}{o}{
    \IfNoValueTF{#1}
    {\op{\bs{\mathcal D}}{-}{\btaum}}
    {\op{\bs{\mathcal D}}{-}{\btaum}[#1]}
}
\NewDocumentCommand{\dminusr}{o}{
    \IfNoValueTF{#1}
    {\op{\mathcal D}{-}{\taur}}
    {\op{\mathcal D}{-}{\taur}[#1]}
}
\newcommand{\dminusrhat}{\op{\widehat{\mathcal D}}{-}{\taur}}
\NewDocumentCommand{\dminusrell}{o}{
    \IfNoValueTF{#1}
    {\op{\mathcal D}{-}{\taurell}}
    {\op{\mathcal D}{-}{\taurell}[#1]}
}
\NewDocumentCommand{\dplusrell}{o}{
    \IfNoValueTF{#1}
    {\op{\mathcal D}{+}{\taurell}}
    {\op{\mathcal D}{+}{\taurell}[#1]}
}
\NewDocumentCommand{\dplusri}{o}{
    \IfNoValueTF{#1}
    {\op{\mathcal D}{+}{\taur_i}}
    {\op{\mathcal D}{+}{\taur_i}[#1]}
}
\NewDocumentCommand{\dplusrj}{o}{
    \IfNoValueTF{#1}
    {\op{\mathcal D}{+}{\taur_j}}
    {\op{\mathcal D}{+}{\taur_j}[#1]}
}
\NewDocumentCommand{\Dminusr}{o}{
    \IfNoValueTF{#1}
    {\op{\bs{\mathcal D}}{-}{\btaur}}
    {\op{\bs{\mathcal D}}{-}{\btaur}[#1]}
}
\NewDocumentCommand{\Dminusrell}{o}{
  \IfNoValueTF{#1}
  {\op{\bs{\mathcal D}}{-}{\btaurell}}
  {\op{\bs{\mathcal D}}{-}{\btaurell}[#1]}
}
\NewDocumentCommand{\DminusrL}{o}{
  \IfNoValueTF{#1}
  {\op{\bs{\mathcal D}}{-}{\btaur_L}}
  {\op{\bs{\mathcal D}}{-}{\btaur_L}[#1]}
}
\NewDocumentCommand{\dplus}{o}{
    \IfNoValueTF{#1}
    {\op{\mathcal D}{+}{\tau}}
    {\op{\mathcal D}{+}{\tau}[#1]}
}
\newcommand{\dplushat}{\op{\widehat{\mathcal D}}{+}{\tau}}
\NewDocumentCommand{\dminus}{o}{
    \IfNoValueTF{#1}
    {\op{\mathcal D}{-}{\tau}}
    {\op{\mathcal D}{-}{\tau}[#1]}
}
\newcommand{\dminushat}{\op{\widehat{\mathcal D}}{-}{\tau}}
\NewDocumentCommand{\dminusmi}{o}{
    \IfNoValueTF{#1}
    {\op{\mathcal D}{-}{\taum_i}}
    {\op{\mathcal D}{-}{\taum_i}[#1]}
}
\NewDocumentCommand{\dplusm}{o}{
    \IfNoValueTF{#1}
    {\op{\mathcal D}{+}{\taum}}
    {\op{\mathcal D}{+}{\taum}[#1]}
}
\newcommand{\dplusmhat}{\op{\widehat{\mathcal D}}{+}{\taum}}
\NewDocumentCommand{\Dplusm}{o}{
    \IfNoValueTF{#1}
    {\op{\bs{\mathcal D}}{+}{\btaum}}
    {\op{\bs{\mathcal D}}{+}{\btaum}[#1]}
}
\NewDocumentCommand{\Dplus}{o}{
    \IfNoValueTF{#1}
    {\op{\bs{\mathcal D}}{+}{\btau}}
    {\op{\bs{\mathcal D}}{+}{\btau}[#1]}
}
\NewDocumentCommand{\dplusr}{o}{
    \IfNoValueTF{#1}
    {\op{\mathcal D}{+}{\taur}}
    {\op{\mathcal D}{+}{\taur}[#1]}
}
\NewDocumentCommand{\dplusmell}{o}{
    \IfNoValueTF{#1}
    {\op{\mathcal D}{+}{\taumell}}
    {\op{\mathcal D}{+}{\taumell}[#1]}
}
\NewDocumentCommand{\Dplusmell}{o}{
    \IfNoValueTF{#1}
    {\op{\bs{\mathcal D}}{+}{\btaum_\ell}}
    {\op{\bs{\mathcal D}}{+}{\btaum_\ell}[#1]}
}
\NewDocumentCommand{\Dplusrell}{o}{
    \IfNoValueTF{#1}
    {\op{\bs{\mathcal D}}{+}{\btaur_\ell}}
    {\op{\bs{\mathcal D}}{+}{\btaur_\ell}[#1]}
}
\NewDocumentCommand{\Dplusrellmo}{o}{
    \IfNoValueTF{#1}
    {\op{\bs{\mathcal D}}{+}{\btaur_{\ell-1}}}
    {\op{\bs{\mathcal D}}{+}{\btaur_{\ell-1}}[#1]}
}
\NewDocumentCommand{\dplusmi}{o}{
    \IfNoValueTF{#1}
    {\op{\mathcal D}{+}{\taum_i}}
    {\op{\mathcal D}{+}{\taum_i}[#1]}
}
\NewDocumentCommand{\Dplusr}{o}{
    \IfNoValueTF{#1}
    {\op{\bs{\mathcal D}}{+}{\btaur}}
    {\op{\bs{\mathcal D}}{+}{\btaur}[#1]}
}
\NewDocumentCommand{\dplusw}{o}{
    \IfNoValueTF{#1}
    {\op{\mathcal D}{+}{\tauw}}
    {\op{\mathcal D}{+}{\tauw}[#1]}
}

\newcommand{\Eell}{\bm e_\ell}

\newcommand{\Einst}{\bm e^\text{inst}}

\newcommand{\Einstell}{\Einst_\ell}

\newcommand{\EtaW}{\bm \eta^W}

\newcommand{\EtaWell}{\EtaW_\ell}

\NewDocumentCommand{\iminus}{o}{
    \IfNoValueTF{#1}
    {\op{\mathcal I}{-}{\tau}}
    {\op{\mathcal I}{-}{\tau}[#1]}
  }
\newcommand{\iminushat}{\op{\widehat{\mathcal I}}{-}{\tau}}
  \NewDocumentCommand{\iminusm}{o}{
    \IfNoValueTF{#1}
    {\op{\mathcal I}{-}{\taum}}
    {\op{\mathcal I}{-}{\taum}[#1]}
  }
  \NewDocumentCommand{\iminusmell}{o}{
    \IfNoValueTF{#1}
    {\op{\mathcal I}{-}{\taumell}}
    {\op{\mathcal I}{-}{\taumell}[#1]}
  }
  \NewDocumentCommand{\iminusmi}{o}{
    \IfNoValueTF{#1}
    {\op{\mathcal I}{-}{\taum_i}}
    {\op{\mathcal I}{-}{\taum_i}[#1]}
  }
  \NewDocumentCommand{\iminusrell}{o}{
    \IfNoValueTF{#1}
    {\op{\mathcal I}{-}{\taurell}}
    {\op{\mathcal I}{-}{\taurell}[#1]}
  }
  \NewDocumentCommand{\Iminusrell}{o}{
    \IfNoValueTF{#1}
    {\op{\bs{\mathcal I}}{-}{\btaur_\ell}}
    {\op{\bs{\mathcal I}}{-}{\btaur_\ell}[#1]}
  }
  \NewDocumentCommand{\iminusri}{o}{
    \IfNoValueTF{#1}
    {\op{\mathcal I}{-}{\taur_i}}
    {\op{\mathcal I}{-}{\taur_i}[#1]}
  }
  \NewDocumentCommand{\Iminusm}{o}{
    \IfNoValueTF{#1}
    {\op{\bs{\mathcal I}}{-}{\btaum}}
    {\op{\bs{\mathcal I}}{-}{\btaum}[#1]}
}
  \NewDocumentCommand{\Iminusmell}{o}{
	\IfNoValueTF{#1}
	{\op{\bs{\mathcal I}}{-}{\btaumell}}
	{\op{\bs{\mathcal I}}{-}{\btaumell}[#1]}
}
\NewDocumentCommand{\iminusr}{o}{
    \IfNoValueTF{#1}
    {\op{\mathcal I}{-}{\taur}}
    {\op{\mathcal I}{-}{\taur}[#1]}
}
\newcommand{\iminusrhat}{\op{\widehat{\mathcal I}}{-}{\taur}}
\NewDocumentCommand{\Iminus}{o}{
    \IfNoValueTF{#1}
    {\op{\bs{\mathcal I}}{-}{\btau}}
    {\op{\bs{\mathcal I}}{-}{\btau}[#1]}
}
\NewDocumentCommand{\Iminusr}{o}{
    \IfNoValueTF{#1}
    {\op{\bs{\mathcal I}}{-}{\btaur}}
    {\op{\bs{\mathcal I}}{-}{\btaur}[#1]}
}
\NewDocumentCommand{\iplusm}{o}{
    \IfNoValueTF{#1}
    {\op{\mathcal I}{+}{\taum}}
    {\op{\mathcal I}{+}{\taum}[#1]}
}
\newcommand{\iplusmhat}{\op{\widehat{\mathcal I}}{+}{\taum}}
\NewDocumentCommand{\Iplusm}{o}{
    \IfNoValueTF{#1}
    {\op{\bs{\mathcal I}}{+}{\btaum}}
    {\op{\bs{\mathcal I}}{+}{\btaum}[#1]}
}
\NewDocumentCommand{\iplusmell}{o}{
    \IfNoValueTF{#1}
    {\op{\mathcal I}{+}{\taumell}}
    {\op{\mathcal I}{+}{\taumell}[#1]}
}
\NewDocumentCommand{\Iplusmell}{o}{
    \IfNoValueTF{#1}
    {\op{\bs{\mathcal I}}{+}{\btaum_\ell}}
    {\op{\bs{\mathcal I}}{+}{\btaum_\ell}[#1]}
}
\NewDocumentCommand{\IplusmL}{o}{
    \IfNoValueTF{#1}
    {\op{\bs{\mathcal I}}{+}{\btaum_L}}
    {\op{\bs{\mathcal I}}{+}{\btaum_L}[#1]}
}
\NewDocumentCommand{\iplusmi}{o}{
    \IfNoValueTF{#1}
    {\op{\mathcal I}{+}{\taum_i}}
    {\op{\mathcal I}{+}{\taum_i}[#1]}
}
\NewDocumentCommand{\iplusr}{o}{
    \IfNoValueTF{#1}
    {\op{\mathcal I}{+}{\taur}}
    {\op{\mathcal I}{+}{\taur}[#1]}
}
\NewDocumentCommand{\iplus}{o}{
    \IfNoValueTF{#1}
    {\op{\mathcal I}{+}{\tau}}
    {\op{\mathcal I}{+}{\tau}[#1]}
}
\newcommand{\iplushat}{\op{\widehat{\mathcal I}}{+}{\tau}}
\NewDocumentCommand{\Iplusr}{o}{
    \IfNoValueTF{#1}
    {\op{\bs{\mathcal I}}{+}{\btaur}}
    {\op{\bs{\mathcal I}}{+}{\btaur}[#1]}
}

\NewDocumentCommand{\op}{mmmo}{
    \IfNoValueTF{#4}
        {{#1}^{#2}_{#3}}
        {{#1}^{#2}_{#3} \left\{ #4 \right\}}
    }

\newcommand{\Rell}{\bm{r}_\ell}
\newcommand{\relu}{\text{ReLU}}
\newcommand{\rin}{r_\text{in}}

\newcommand{\taum}{\tau^\mathrm{m}}
\newcommand{\Taum}{\bm\tau^\mathrm{m}}
\newcommand{\taumell}{\tau^\mathrm{m}_\ell}
\newcommand{\Taumell}{\bm\tau^\mathrm{m}_\ell}
\newcommand{\taur}{\tau^\mathrm{r}}
\newcommand{\Taur}{\bm\tau^\mathrm{r}}
\newcommand{\taurell}{\tau^\mathrm{r}_\ell}
\newcommand{\Taurell}{\bm\tau^\mathrm{r}_\ell}
\newcommand{\tauw}{\tau^w}

\newcommand{\TauW}{\bm\tau^W}

\newcommand{\TauWell}{\TauW_\ell}
\newcommand{\tb}[1]{\textbf{#1}}
\newcommand{\thetadot}{\dot \theta}
\newcommand{\TT}{\mathrm{T}}

\newcommand{\udot}{\dot u}
\newcommand{\Udot}{\dot{\bm u}}

\newcommand{\Uell}{\bm u_\ell}
\newcommand{\Uelldot}{\dot{\bm u}_\ell}

\newcommand{\Vell}{\bm v_\ell}
\newcommand{\Wdot}{\dot W}
\newcommand{\Well}{\bm W_\ell}

\renewcommand{\odot}{\circ}

\title{\Huge{\tb{Backpropagation through space, time and the brain}}}

\newcommand{\affilPhysio}{1}
\newcommand{\sharedFirst}{*}
\newcommand{\mailSymbol}{\Letter}
\author{
    \\[0em]
    \normalsize{
        Benjamin Ellenberger\textsuperscript{\sharedFirst,\,\affilPhysio},
        Paul Haider\textsuperscript{\sharedFirst,\,\affilPhysio,\,\mailSymbol},
    }
    \\
    \normalsize{
        Federico Benitez\textsuperscript{\affilPhysio},
        Jakob Jordan\textsuperscript{\affilPhysio},
        Kevin Max\textsuperscript{\affilPhysio},
        Ismael Jaras\textsuperscript{\affilPhysio},
        Laura Kriener\textsuperscript{\affilPhysio},
    }
    \\
    \normalsize{
        Mihai A. Petrovici\textsuperscript{\affilPhysio,\,\mailSymbol}
    }
    \\[0.4cm]
    {\normalfont\small
      \textsuperscript{\sharedFirst} These authors contributed equally
    }
    \\[0.1cm]
    {\normalfont\small
        \textsuperscript{\affilPhysio}\,Department of Physiology, University of Bern, 3012 Bern, Switzerland.
    }
    \\[0.1cm]
    \normalfont{\small
        \textsuperscript{\mailSymbol} \texttt{paul.haider@unibe.ch}, \texttt{mihai.petrovici@unibe.ch}
    }
}

\begin{document}
\date{}
\maketitle

\begin{abstract}
How physical neuronal networks, bound by spatio-temporal locality constraints, can perform efficient credit assignment, remains an intriguing question.
Both backward- and forward-propagation algorithms rely on assumptions that violate this locality in various ways.
We introduce Generalized Latent Equilibrium (GLE), a framework for fully local spatio-temporal credit assignment in physical, dynamical neuronal networks.
From an energy based on neuron-local mismatches, we derive neuronal dynamics via stationarity and parameter dynamics as gradient descent.
The result is an online approximation of backpropagation through space and time in deep networks of cortical microcircuits with continuously active, local synaptic plasticity.
GLE exploits dendritic morphology to enable complex information storage and processing in single neurons, as well as their ability to react in anticipation of their future input.
This ``prospective coding'' enables the computation of spatio-temporal convolutions in the forward direction and the approximation of adjoint variables in the backward stream.

\end{abstract}

\section{Introduction}
\label{sec:intro}

The world in which we have evolved appears to lie in a Goldilocks zone of complexity: it is rich enough to produce organisms that can learn, yet also regular enough to be learnable by these organisms in the first place.
Still, regular does not mean simple; the need to continuously interact with and learn from a dynamic environment in real time faces these agents -- and their nervous systems -- with a challenging task.

One can view the general problem of dynamical learning as one of \emph{constrained minimization}, i.e., with the goal of minimizing a behavioral cost with respect to some learnable parameters (such as synaptic weights) within these nervous systems, under constraints given by their physical characteristics.
For example, in the case of biological neuronal networks, such dynamical constraints may include leaky integrator membranes and nonlinear output filters.
The standard approach to this problem in \dl uses \sgd, coupled with some type of \ad algorithm that calculates gradients via reverse accumulation~\cite{baydinAutomaticDifferentiationMachine2018, pearlmutterGradientCalculationsDynamic1995}.
These methods are well-known to be highly effective and flexible in their range of applications.

If the task is time-independent, or can be represented as a time-independent problem, the standard error \bp algorithm provides this efficient backward differentiation~\cite{linnainmaaTaylorExpansionAccumulated1976, werbosApplicationsAdvancesNonlinear1982, rumelhartLearningRepresentationsBackpropagating1986}.
We refer to this class of problems as \emph{spatial}, as opposed to more complex \emph{spatio-temporal} problems, such as sequence learning.
For the latter, the solution to constrained minimization can be sought through a variety of methods.
When the dynamics are discrete, the most commonly used method is \bptt~\cite{pinedaGeneralizationBackpropagationRecurrent1987, werbosBackpropagationTimeWhat1990}.
For continuous-time problems, optimization theory provides a family of related methods, the most prominent being the \am~\cite{kelley1962method}, Pontryagin's maximum principle~\cite{Todorov2006,Chachuat2007}, and the Bellman equation~\cite{Sutton2018}.

\begin{figure}[t!]
    \centering
    \includegraphics[width=0.45\textwidth]{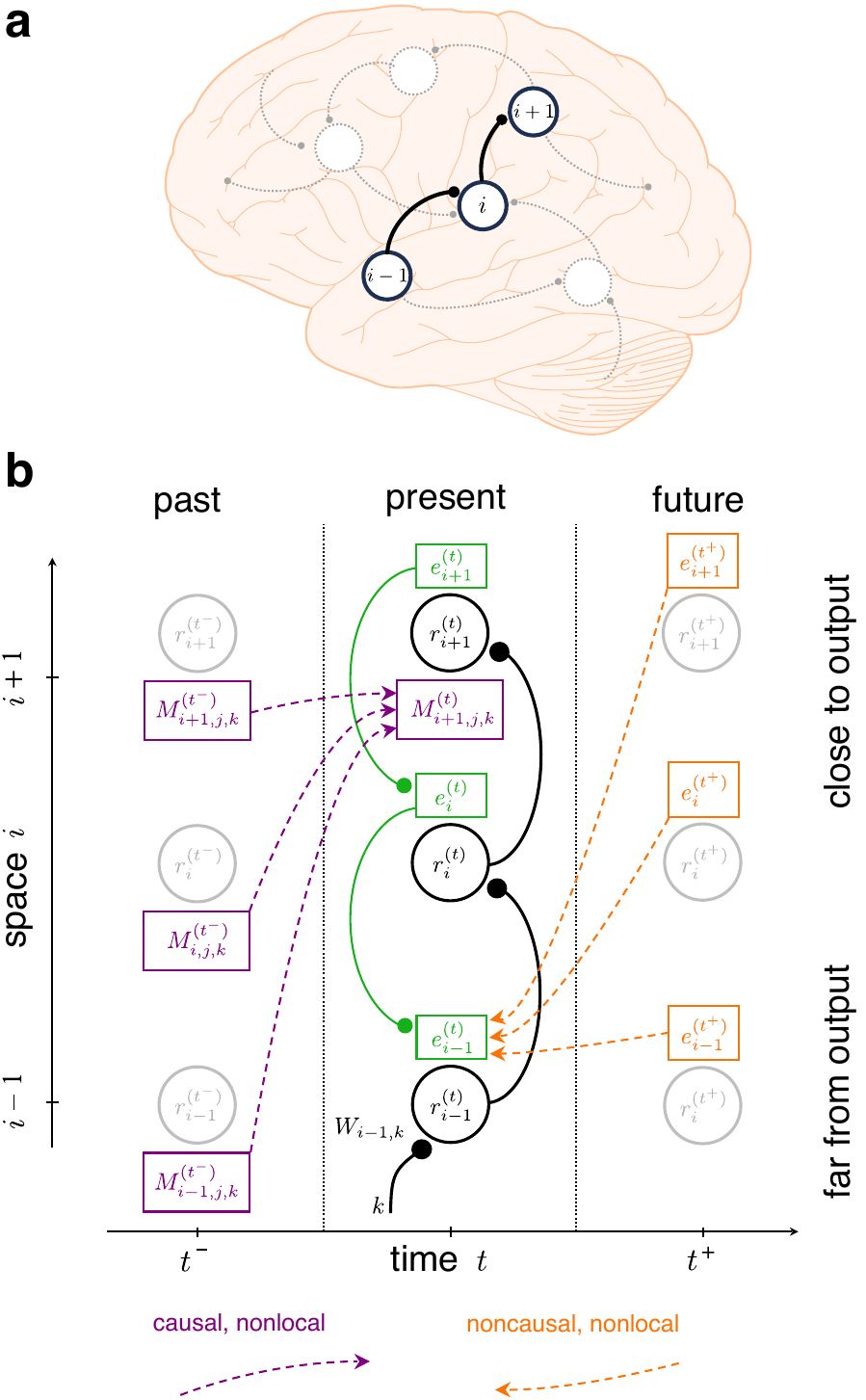}
    \caption{
        \tb{The problem of locality in spatio-temporal credit assignment.}
        \tb{(a)} To illustrate the different learning algorithms, we consider three neurons within a larger recurrent network.
        The neuron indices are indicative of the distance from the output, with neuron $i+1$ being itself an output neuron, and therefore having direct access to an output error $e_{i+1}$.
        \tb{(b)} Information needed by a deep synapse at time $t$ to calculate an update $\dot w_{i-1, k}^{(t)}$.
        Orange: future-facing algorithms such as BPTT require the states $r_{n}^{(t^{+})}$ of all future times $t^+$ and all neurons $n$ in the network and can therefore only be implemented in an offline fashion.
        These states are required to calculate future errors $e_n^{(t^+)}$, which are then propagated back in time into present errors $e_n^{(t)}$ and used for synaptic updates $\dot w_{i-1,k}^{(t)} \propto e_{i-1}^{(t)} r_k^{(t)}$.
        Purple: past-facing algorithms such as RTRL store past effects of all synapses $w_{jk}^{(t^{-})}$ on all past states $r_n^{(t^-)}$ in an influence tensor $M_{n,j,k}^{(t^-)}$.
        This tensor can be updated online and used to perform weight updates $\dot w_{i-1,k}^{(t)} \propto \sum_{n} e_{n}^{(t)} M_{n,i-1,k}^{(t)}$.
        Note that all synapse updates need to have access to distant output errors.
        Furthermore, the update of each element in the influence tensor requires the knowledge of distant elements and is thus itself nonlocal in space.
        Green: GLE operates exclusively on present states $r_n^{(t)}$.
        It uses them to infer errors $e_n^{(t)}$ that approximate the future backpropagated errors of BPTT.
    }
    \label{fig:stca}
\end{figure}

\am and \bptt have proven to be very powerful methods, but it is not clear how they could be implemented in physical neuronal systems that need to function and learn continuously, in real time, and using only information that is locally available at the constituent components~\cite{lillicrapBackpropagationTimeBrain2019}.
Learning through \ambptt cannot be performed in real time, as it is only at the end of a task that errors and parameter updates can be calculated retrospectively (\cref{fig:stca}).
This requires either storing the entire trajectory of the system (i.e., for all of its dynamical variables) until a certain update time, and/or recomputing the necessary variables during backward error propagation.
While straightforward to do in computer simulations (computational and storage inefficiency notwithstanding), a realization in physical neuronal systems would require a lot of additional, complex circuitry for storage, recall and (reverse) replay.
Additionally, it is unclear how useful errors can be represented and transmitted in the first place, i.e., which physical network components calculate errors that correctly account for the ongoing dynamics of the system, and how these then communicate with the components that need to change during learning (e.g., the synapses in the network).
It is for these reasons that \ambptt sits firmly within the domain of \ml, and is largely considered not applicable for physical neuronal systems, both biological and artificial.
(We use the term \emph{neuronal} to differentiate between physical, time-continuous dynamical networks of neurons and their abstract, time-discrete \ann counterparts.)

Instead, theories of biological (or bio-inspired) spatio-temporal learning focus on other methods, either by using reservoirs and foregoing the learning of deep weights altogether, or by using direct and instantaneous output error feedback, transmitted globally to all neurons in the network, as in the case of FORCE~\cite{sussillo2009generating} and FOLLOW~\cite{gilraPredictingNonlinearDynamics2017}.
Alternatively, the influence of synaptic weights on neuronal activities can be carried forward in time and associated with instantaneous errors, as proposed in \rtrl~\cite{williams1989learning}.
While thereby having the advantage of being temporally causal, \rtrl still violates locality, as this influence tensor relates all neurons to all synapses in the network, including those that lie far away from each other (\cref{fig:stca}).
The presence of this tensor also incurs a much higher memory footprint (cubic scaling for \rtrl vs. linear scaling for \bptt), which quickly becomes prohibitive for larger-scale applications in \ml.
Various approximations of \rtrl alleviate some of these issues~\cite{marschallUnifiedFrameworkOnline2020}, but only at the cost of propagation depth.
We return to these methods in the Discussion, as our main focus here is \ambptt.

Indeed, and contrary to prior belief, we suggest that physical neuronal systems can approximate \bptt very efficiently and with excellent functional performance.
More specifically, we propose the overarching principle of \gle to derive a comprehensive set of equations for inference and learning that are local in both time and space (\cref{fig:stca}).
These equations fully describe a dynamical system running in continuous time, without the need for separate phases, and undergoing only local interactions.
Moreover, they describe the dynamics and morphology of structured neurons performing both \emph{retrospective} integration of past inputs and \emph{prospective} estimation of future states, as well as the weight dynamics of error-correcting synapses, thus linking to experimental observations of cortical dynamics and anatomy.
Due to its manifest locality and reliance on rather conventional analog components, our framework also suggests a blueprint for powerful and efficient neuromorphic implementation.

We thus propose a new solution for the spatio-temporal credit assignment problem in physical neuronal systems, with substantial advantages over previously proposed alternatives.
Importantly, our framework does not differentiate between spatial and temporal tasks and can thus be readily used in both domains.
As it represents a generalization of \le~\cite{haiderLatentEquilibriumUnified2021}, a precursor framework for bio-plausible, but purely spatial computation and learning, it implicitly contains (spatial) \bp as a sub-case.
For a more detailed comparison to \gls{le}, see \nameitref{sec:discussion}.

This manuscript is structured as follows:
In \nameitref{sec:framework}, we first propose a set of four postulates from which we derive network structure and dynamics.
This is strongly inspired by approaches in theoretical physics, where a specific energy function provides a unique reference from which everything else about the system can be derived.
We then discuss the link to \ambptt in \nameitref{sec:adjoint}.
Additionally, we show how our framework describes physical networks of neurons, with implications for both cortex and hardware (\nameitref{sec:circuits}).
Subsequently, we discuss various applications, from small-scale setups that allow an intuitive understanding of our network dynamics (\nameitref{sec:chain} and \nameitref{sec:smallnetworks}), to larger-scale networks capable of solving difficult spatio-temporal classification problems (\nameitref{sec:largescale}) and chaotic time series prediction (\nameitref{sec:mackeyglass}).
Finally, we elaborate on the connections and advantages of our framework when compared to other approaches in \nameitref{sec:discussion}.

\section{Results}
\label{sec:results}

\subsection{The GLE framework}
\label{sec:framework}

At the core of our framework is the realization that biological neurons are capable of performing two fundamental temporal operations.
First, as is well-known, neurons perform temporal integration in the form of low-pass filtering.
We describe this as a ``retrospective'' operation and denote it with the operator
\begin{equation}
    \iminusm[x](t) \coloneqq \frac 1 \taum \int_{-\infty}^t x(s) \exp \left( -\frac{t-s}{\taum} \right) \dd s \; ,
    \label{eqn:iminus}
\end{equation}
where $\taum$ represents the membrane time constant and $x$ the synaptic input.

The second temporal operation is much less known but well-established physiologically~\cite{hodgkinQuantitativeDescriptionMembrane1952, pucciniIntegratedMechanismsAnticipation2007, kondgenDynamicalResponseProperties2008, plesserEscapeRateModels2000, pozzorini2013temporal, brandt2024prospective}: neurons are capable of performing temporal differentiation, an inverse low-pass filtering that phase-shifts inputs into the future, which we thus name ``prospective'' and denote with the operator
\begin{equation}
    \dplusr[x](t) \coloneqq \left( 1 + \taur \frac{\dd}{\dd t} \right) x(t) \; .
    \label{eqn:dplus}
\end{equation}
The time constant $\taur$ is associated with the neuronal output rate, which, rather than being simply $\varphi(u)$ (with $u$ the membrane potential and $\varphi$ the neuronal activation function), takes on the prospective form $r = \varphi(\dplusr[u])$.

In brief, this prospectivity can arise from two distinct mechanisms.
On one hand, it follows as a direct consequence of the output nonlinearity in spiking neurons~\cite{plesserEscapeRateModels2000}.
In more complex neurons that are capable of bursting, the input slope also directly affects the spiking output \cite{kepecsBurstingNeuronsSignal2002}.
On the other hand, prospectivity appears when the neuronal membrane (alternatively but equivalently, its leak potential or firing threshold) is negatively coupled to an additional retrospective variable.
Such variables include, for example, the inactivation of sodium channels, or slow adaptation (both spike frequency and subthreshold) currents~\cite{brandt2024prospective,pozzorini2013temporal}, thus giving neurons access to a wide range of prospective horizons.
For more detailed, intuitive explanations of these mechanisms, we refer to the Supplement, \nameitref{si:prospectivity}.
For a more  technical discussion of prospectivity in neurons with adaptation currents, we refer to  \nameitref{sec:adaptation} in the Methods.
We also note that in analog neuromorphic hardware, adaptive neurons are readily available~\cite{aamir2018mixed,rubino2019ultra,shaban2021adaptive,billaudelle2022accurate}, but a direct implementation of an inverse low-pass filter would evidently constitute an even simpler and more efficient solution.

Importantly, the retrospective and prospective operators $\iminusm$ and $\dplusr$ have opposite effects; in particular, for $\taum = \taur$, they are exactly inverse, which forms the basis of \le~\cite{haiderLatentEquilibriumUnified2021}.
In that case, the exact inversion of the low-pass filtering allows the network to react instantaneously to a given input, which can solve the relaxation problem for spatial tasks. However, this exact inversion also precludes the use of neurons for explicit temporal processing in spatio-temporal tasks, which represents the main focus of this work.

We can now define the \gle framework as a set of four postulates, from which the entire network structure and dynamics follow.
The postulates use these operators to describe how forward and backward prospectivity work in biological neuronal networks.
As we show later, the dynamical equations derived from these postulates approximate the equations derived from \ambptt, but without violating causality, with only local dependencies and without the need for learning phases.

\vspace{\parskip}
\begin{postulate}
    The canonical variables describing neuronal network dynamics are $\dplusmi[u_i(t)]$ and $\dplusri[u_i(t)]$, where each neuron is denoted with the subscript $i \in \{ 1,\ldots, n\}$.
\end{postulate}

This determines the relevant dynamical variables for the postulates below.
They represent, respectively, the prospective voltages with respect to the membrane and rate time constants.
Importantly, each neuron $i$ can in principle have its own time constants $\tau^{x}_i$, and the two are independent, so in general $\taum_{i} \neq \taur_{i}$.
This is in line with the biological mechanisms for retro- and prospectivity, which are also unrelated, as discussed above.
\begin{figure}[!t]
    \centering
    \includegraphics[width=0.42\textwidth]{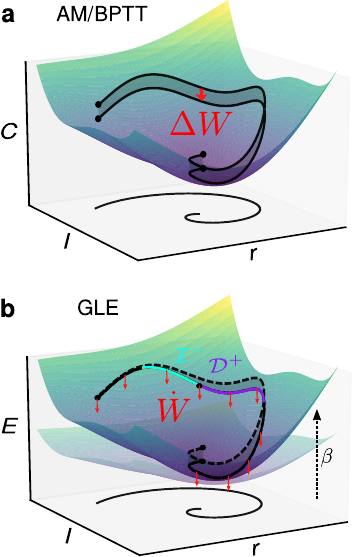}
    \caption{
        \tb{Comparison between \am and \gle.}
        Network dynamics define trajectories (black) in the cost/energy landscape, spanned by external inputs $I$ and neuron outputs $r$.
        Parameter updates (red, here: synaptic weights) reduce the cost/energy along these trajectories.
        \tb{(a)} \am records the trajectory between two points in time and calculates the total update $\Delta W$ that reduces the \textit{integrated} cost along this trajectory.
        \tb{(b)} \gle calculates an approximate cost gradient \textit{at every point in time}, by taking into account past network states (via retrospective coding, $\iminus$) and estimating future errors from the current state (via prospective coding, $\dplus$).
        Learning is thus fully online and can gradually reduce the energy in real-time, with the (real) trajectory slowly dropping away from the (virtual) trajectory of a network that is not learning (dashed line).
    }
    \label{fig:energy}
\end{figure}

\vspace{\parskip}
\begin{postulate}
    A neuronal network is fully described by the energy function
    \begin{align}
        E(t) &= \frac12 \sum_i \lVert e_i(t) \rVert^2  + \beta C(t) \; ,
        \label{eqn:GLEE}
    \end{align}
    where $e_i(t) = \dplusmi[u_i(t)] - \sum_j W_{ij} \varphi ( \dplusrj[ u_j(t)] ) - b_i$ is the mismatch error of neuron $i$.
    $W_{ij}$ and $b_i$ respectively denote the components of the weight matrix and bias vector, $\varphi$ the output nonlinearity and $\beta$ a scaling factor for the cost.
    The cost function $C(t)$ is usually defined as a function of the rate and hence of $\dplusri[u_{i}]$ of some subset of output neurons.
\end{postulate}

This approach is inspired by physics and follows a time-honored tradition in \ml, dating back to Boltzmann machines~\cite{ackleyLearningAlgorithmBoltzmann1985} and Hopfield networks~\cite{hopfieldNeuralNetworksPhysical1982} as well as computational neuroscience model such as, e.g., Equilibrium Propagation~\cite{scellierEquilibriumPropagationBridging2017} and Predictive Coding frameworks~\cite{whittingtonApproximationErrorBackpropagation2017}.
The core idea of these approaches is to define a specific energy function that provides a unique reference from which everything else follows. In physics, this is the Hamiltonian, from which the dynamics of the system can be derived; in our case, it is a measure of the ``internal tension'' of the network, from which we derive the dynamics of the network and its parameters.
Under the weak assumption that the cost function can be factorized, this energy is simply a sum over neuron-local energies $E_i(t) = \frac12 e_i^2(t) + \beta C_i(t)$.
Each of these energies represent a difference between a neuron's own prospective voltage, i.e., what membrane voltage the neuron predicts for its near future ($\dplusmi[u_i(t)]$), and what its functional afferents (and bias) expect it to be ($\sum_j W_{ij} \varphi ( \dplusrj[ u_j(t)] ) + b_i$), with the potential addition of a teacher nudging term for output neurons that is related to the cost that the network seeks to minimize ($\beta C_i(t)$).

\vspace{\parskip}
\begin{postulate}
    Neuron dynamics follow the stationarity principle
    \begin{equation}
        \iminusmi[\frac{\partial E}{\partial \dplusmi[u_i]}] + \iminusri[\frac{\partial E}{\partial \dplusri[u_i]}] = 0 \; .
        \label{eqn:GLEudot}
    \end{equation}
\end{postulate}

As the two rates of change (the partial derivatives) are with respect to prospective variables, with temporal advances determined by $\taum$ and $\taur$, they can be intuitively thought of as representing quantities that refer to different points in the future -- loosely speaking, at $t+\taum$ and $t+\taur$.
To compare the two rates of change on equal footing, they need to be pulled back into the present by their respective inverse operators $\iminusm$ and $\iminusr$.
It is the equilibrium of this mathematical object, otherwise not immediately apparent (hence: ``latent'') from observing the network dynamics themselves (see below), that gives our framework its name.
It is also easy to check that for the special case of $\taum_{i} = \taur_{i} \quad \forall i$, \gle reduces to \le~\cite{haiderLatentEquilibriumUnified2021} -- hence the `generalized' nomenclature.

\vspace{\parskip}
\begin{postulate}
    Parameter dynamics follow \gls{gd} on the energy
    \begin{equation}
        \bs{\thetadot} = - \eta_{\bs{\theta}} \frac{\partial E}{\partial \bs{\theta}}
        \label{eqn:GLEthetadot}
    \end{equation}
    with individual learning rates $\eta_{\bs{\theta}}$.
\end{postulate}

Parameters include $\bs \theta = \{\bs W, \bs b, \btaum, \btaur\}$, with boldface denoting matrices, vectors, and vector-valued functions.
These parameter dynamics are the equivalent of plasticity, both for synapses ($W_{ij}$) and for neurons ($b_i, \taum_i, \taur_i$).
The intuition behind this set of postulates is illustrated in \cref{fig:energy}.
Without an external teacher, the network is unconstrained and simply follows the dynamics dictated by the input; as there are no errors, both cost $C$ and energy $E$ are zero.
As an external teacher appears, errors manifest and the energy landscape becomes positive; its absolute height is scaled by the coupling parameter $\beta$.
While neuron dynamics $\bm \udot$ trace trajectories across this landscape, plasticity $\bm \thetadot$ gradually reduces the energy along these trajectories (cf. \cref{fig:energy}).
Thus, during learning, the energy landscape (more specifically, those parts deemed relevant by the task of the network, lying on the state subspace traced out by the trajectories during training) is gradually lowered, as illustrated by the faded surface.
Ultimately, after learning, the energy will ideally be pulled down to zero, thus implicitly also reducing the cost, because it is a positive, additive component of the energy.
Beyond this implicit effect, we  later show how the network dynamics derived from these postulates also explicitly approximate gradient descent on the cost.

The four postulates above fully encapsulate the \gle framework.
From here, we can now take a closer look at the network dynamics and see how they enable the sought transport of signals to the right place and at the right time.

\subsection{Network dynamics}
\label{sec:dynamics}

With our postulates at hand, we can now infer dynamical and structural properties of neuronal networks that implement \gle.
We first derive the neuronal dynamics by applying the stationarity principle (Postulate 3, \cref{eqn:GLEudot}) to the energy function (Postulate 2, \cref{eqn:GLEE}):
\begin{align}
  \taum_i \udot_i = - u_i &+ \textstyle\sum_j W_{ij} \varphi \left( \dplusrj[u_j] \right) + b_i \nonumber \\
  &+ \underbrace{\dplusmi\Bigl\{\iminusri\Bigl\{\varphi'_{i} \textstyle\sum_{j}W_{ji} e_{j}\Bigr\}\Bigr\}}_{e_{i}}
  \; ,
    \label{eqn:udot}
\end{align}
where \(\varphi'_{i}\) is a shorthand for the derivative of the activation functional evaluated at \(\dplusri[u_{i}]\).
For a detailed derivation, we refer to \nameitref{sec:meth-networkdynamics} in the Methods.
This is very similar to conventional leaky integrator dynamics, except for two important components: first, the use of the prospective operator for the neuronal output, which we already connected to the dynamics of biological neurons above, and second, the additional error term.
With this, we have two complementary representations for the error term $e_i$. First, as mismatch between the prospective voltage and the basal inputs (cf.~\cref{eqn:GLEE}), describing how errors couple two membranes, and second, as a function of other errors $e_j$, as given by the error propagation equation (cf.~\cref{eqn:udot}).
Thus, in the \gls{gle} framework, a single neuron performs four operations in the following order: (weighted) sum of presynaptic inputs, integration (retrospective), differentiation (prospective), and the output nonlinearity.
The timescale associated with retrospectivity is the membrane time constant $\taum$, whereas prospectivity is governed by $\taur$.
This means that even if the membrane time constant is fixed, as may be the case for certain neuron classes or models thereof, single neurons can still tune the time window to which they attend by adapting their prospectivity.
This temporal attention window can lie in the past (retrospective neurons, $\taur < \taum$), in the present (instantaneous neurons, $\taur = \taum$, as described by \le), but also in the future (prospective neurons, $\taur > \taum$).
These neuron classes can, for example, be found in cortex \citep{mainen_model_1995,gerstner_neuronal_2014} and hippocampus \citep{traub_neuronal_1991,linaro2018dynamical};
for a corresponding modeling study, we refer to \cite{brandt2024prospective}.
The prospective capability becomes essential for error propagation, as we discuss below, while the use of different attention windows allows the learning of complex spatio-temporal patterns, as we show in action in later sections.
Note that for $\taur = 0$, we recover classical leaky integrator neurons as a special case of our framework.

\Cref{eqn:udot} also suggests a straightforward interpretation of neuronal morphology and its associated functionality.
In particular, it suggests that separate neuronal compartments store different variables: a somatic compartment for the voltage $u_i$, and two dendritic compartments for integrating $\sum_j W_{ij} r_j$ and $e_i$, respectively.
This separation also gives synapses access to these quantities, as we discuss later on.
Further below, we also show how this basic picture extends to a microcircuit for learning and adaptation in \gle networks.

The error terms in \gle also naturally include prospective and retrospective operators.
As stated in Postulate 2 (\cref{eqn:GLEE}), the total energy of the system is a sum over neuron-local energies.
If we now consider a hierarchical network, these terms can be easily rearranged into the form (see Methods and SI for a detailed derivation)
\begin{equation}
    \bm e_\ell = \underbrace{\Dplusmell\Bigl\{\Iminusrell}_{\text{temporal BP}}\Bigl\{\underbrace{\bm \varphi'_{\ell} \odot \bm W_{\ell+1}^{\TT} \bm e_{\ell+1}}_{\text{spatial BP}}\Bigr\}\Bigr\} \; ,
    \label{eqn:erec}
\end{equation}
where $\ell$ denotes the network layer and $\varphi'$ denotes the derivative of $\varphi$ evaluated at $\Dplusr[\bm u_\ell]$.
In this form, the connection to \gls{bp} algorithms becomes apparent.
For $\taur = \taum$, the operators cancel and \Cref{eqn:erec} reduces to the classical (spatial) error \bp algorithm, as already studied in~\cite{haiderLatentEquilibriumUnified2021}.
When $\taur \neq \taum$ however, the error exhibits a switch between the two time constants when compared to the forward neuron dynamics (\cref{eqn:udot}): whereas forward rates are retrospective with $\taum$ and prospective with $\taur$, backward errors invert this relationship.
In other words, backward errors invert the temporal shifts induced by forward neurons.
As we discuss in the following section, it is precisely this inversion that enables the approximation of \ambptt.

As for the neuron dynamics, parameter dynamics also follow from the postulates above.
For example, synaptic plasticity is obtained by applying the gradient descent principle (Postulate 4, \cref{eqn:GLEthetadot} with respect to synaptic weights $\bs W$) to the energy function (Postulate 2, \cref{eqn:GLEE}):
\begin{equation}
    \Wdot_{ij} = \eta_{W} e_{i} r_{j} = \eta_{W} (\dplusmi[u_i] - v_i) r_{j} \; ,
    \label{eqn:Wdot}
\end{equation}
where $v_i = \sum_j W_{ij} r_j$ is the membrane potential of the dendritic compartment that integrates bottom-up synaptic inputs.
Such three-factor error-correcting rules have often been discussed in the context of biological \dl (see \cite{richards2019deep} for a review).
For a more detailed biological description of our specific type of learning rule, we refer to \cite{urbanczik2014learning}.
Notice that parameter learning is neuron-local, and that we are performing \gd explicitly on the energy $E$, and only implicitly on the cost $C$.
This is a quintessential advantage of the energy-based formalism, as the locality of \gle dynamics is a direct consequence of the locality of the postulated energy function.
This helps provide the physical and biological plausibility that other methods lack.
In \nameitref{sec:circuits}, we discuss how \gle dynamics relate to physical neuronal networks and cortical circuits, but first, we show how these dynamics effectively approximate \ambptt.

\subsection{GLE dynamics implement a real-time approximation of AM/BPTT}
\label{sec:adjoint}

\begin{figure}[bt]
    \centering
    \includegraphics[width=0.46\textwidth]{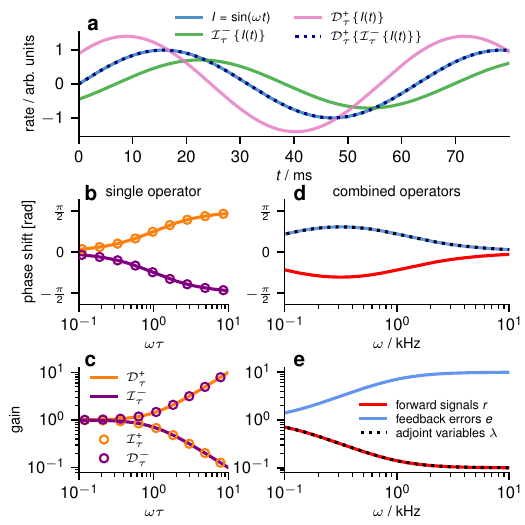}
    \caption{
        \tb{Comparison of \gle and \ambptt in Fourier space.}
        \tb{(a)} Effect of the individual and combined GLE operators $\iminus$ and $\dplus$ with shared time constant $\tau$ on a single frequency component of an input current $I$.
             $\iminus$ generates a negative phase shift (towards later times) and sub-unit gain.
             $\dplus$ is its exact inverse and generates a positive phase shift (towards earlier times) and supra-unit gain.
        \tb{(b)} Phase shift and
        \tb{(c)} gain of all four temporal operators in \gle and \ambptt across a wide range of the frequency spectrum.
             Note how the prospective operators $\dplus$ and $\iplus$ (orange) have the same shift but inverse gain; the same holds for the retrospective operators $\iminus$ and $\dminus$ (purple).
        \tb{(d)} Phase shift and
        \tb{(e)} gain of the combined operators as they appear in the neuron dynamics.
             Here, we choose an example forward neuron (blue) with a retrospective attention window ($\taum = 10 \taur$).
             Both the associated \gle errors $e$ (blue) and the  adjoint variables $\lambda$ (dotted) are prospective and precisely invert this phase shift, albeit with a different gain.
        }
    \label{fig:fourier}
\end{figure}

The learning capabilities of \gle arise from the specific form of the errors encapsulated in the neuron dynamics, which make the similarity to \ambptt apparent, as discussed below.
For a detailed derivation of the following relationships, we refer to Methods and the SI.

Just like in \gle, learning in \ambptt is error-correcting:
\begin{equation}
    \Delta \bs W_\ell^\text{AM} = \int_{0}^{T} \bs \lambda_\ell \bs r_{\ell-1}^\TT \,\dd t\; ,
\end{equation}
where the continuous-time adjoint variables $\bs \lambda$ in \am are equivalent to the time-discrete errors in \bptt.
While typically calculated in reverse time, as for the backpropagated errors in \bptt, for the specific dynamics of cost-decoupled GLE networks ($\beta = 0 \Rightarrow \bs e = \bs 0 \Rightarrow E = 0$) it is possible to write the adjoint dynamics in forward time as follows:
\begin{equation}
    \bs \lambda_\ell = \Iplusmell[\Dminusrell[\bs \varphi'_\ell \odot \bs W_{\ell+1}^\TT \bs \lambda_{\ell+1}]] \; .
    \label{eqn:lambda}
\end{equation}
Here, we use adjoint operators $\dminus[x(t)] = \left(1 - \tau \frac{\dd }{\dd t}\right) x(t)$ and $\iplus[x(t)] = \frac{1}{\tau} \int_{t}^{\infty} x(s) \, e^{\frac{t-s}{\tau}}\,\dd s$ to describe the hierarchical coupling of the adjoint variables.
We note that the adjoint dynamics (\cref{eqn:lambda}) can also be derived in our \gle framework by simply replacing $\iminusmi$ with $\dminusmi$ in Postulate 3.

Note the obvious similarity between \cref{eqn:lambda,eqn:erec}.
The inner term $\bs \varphi'_\ell \odot \bs W_{\ell+1}^\TT \bs \lambda_{\ell+1}$ is identical and describes backpropagation through space.
The outer operators perform the temporal backpropagation, enacting the exact opposite temporal operations compared to the representation neurons (first retrospective with $\taur$ then prospective with $\taum$, cf. \cref{eqn:udot}).

The intuition is as follows.
Training \anns on sequential data usually requires unrolling the network in time and then backpropagating errors via \ambptt, as given by the adjoint equations (\cref{eqn:erec}).
However, these adjoint equations effectively operate in reverse time, i.e., are non-local in time and thus not applicable for real-time learning in physical neuronal systems.
This nonlocality (noncausality) is due to the operator $\iplus$, which effectively calculates an integral over the future.
(The operator $\dminus$ is unproblematic in this regard.)
By replacing $\iplus$ by $\dplus$, \gle solves the noncausality problem.
This of course happens at the expense of precision, because the extrapolation carried out by $\dplus$ cannot look arbitrarily far into the future.
Still, $\dplus$ maintains much of the computation performed by $\iplus$ in Fourier space, as discussed below.
The \gle framework further replaces $\dminus$ by $\iminus$.
This is not an algorithmic necessity, as $\dminus$ is fully local in time, but the resulting dynamics map nicely to the retrospective component of biologically observed neuronal membrane dynamics, which are better described as leaky integrators rather than negative differentiators.

How \gle can perform an online approximation of \ambptt is best seen in frequency space, where we can analyze how the combined temporal operators affect individual Fourier components (see also \cref{fig:fourier}).
For a single such component -- a sine wave input of fixed angular frequency $\omega$ -- each operator causes a temporal (phase) shift: the retrospective operator $\iminus$ causes a shift of the input signal towards later times, while the prospective operator causes an inverse shift towards earlier times.
These phase shifts are exactly equal to those generated by the adjoint operators $\dminus$ and $\iplus$.
Therefore, in terms of temporal shift, the \gle errors are perfect replicas of the adjoint variables derived from exact \gd; this is the most important part of the temporal backpropagation in \am.

In terms of gain, \gle and \ambptt are inverted.
For smaller angular frequencies $\omega \tau \lesssim 1$, this approximation is very good and the gradients are only weakly distorted.
We will later see that in practice, for hierarchical networks with sufficiently diverse time constants, successful learning does not strictly depend on this formal requirement.
What appears more important is that, even for larger $\omega \tau$, \gle errors always conserve the sign of the correct adjoints, so the error signal always remains useful; moreover, higher-frequency oscillations in the errors tend to average out over time, as we demonstrate in simulations below.

\subsection{Cortical / neuromorphic circuits}
\label{sec:circuits}
\begin{figure}[bt]
    \centering
    \includegraphics[width=0.45\textwidth]{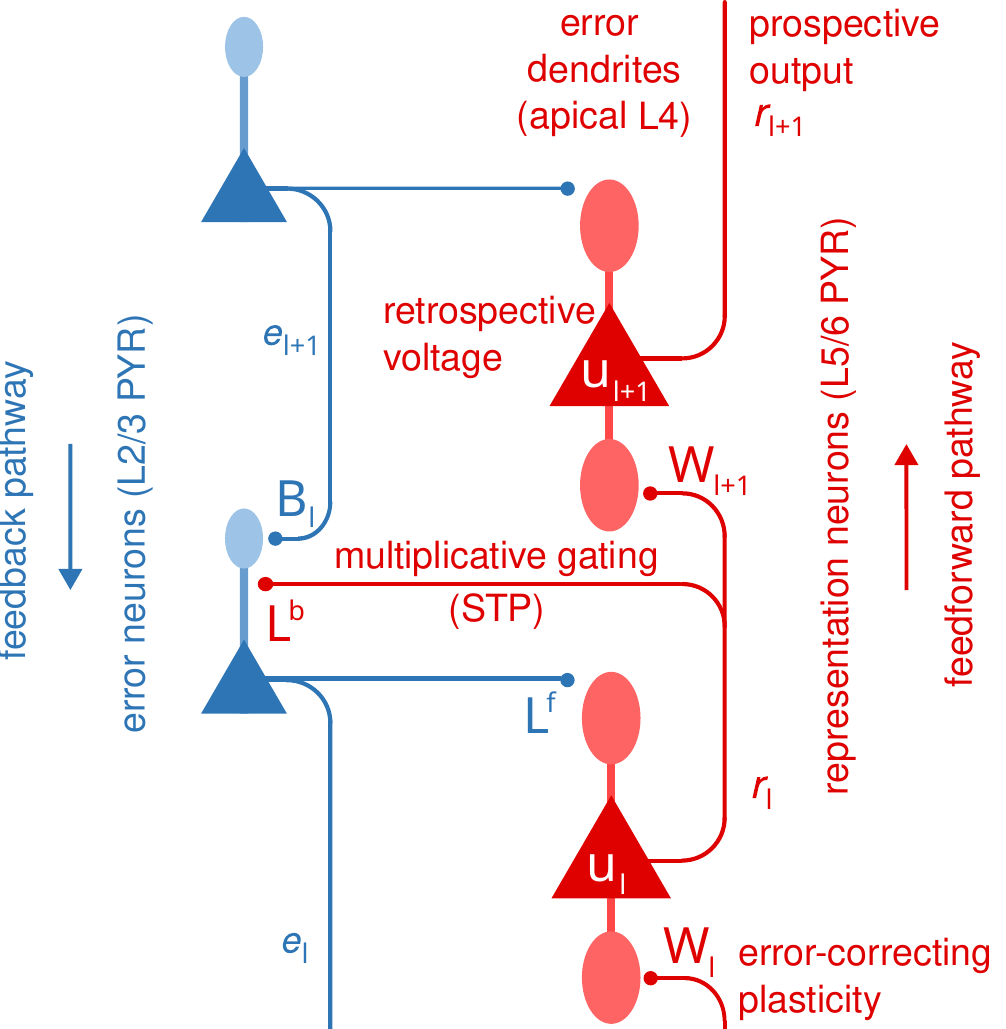}
    \caption{
        \tb{Microcircuit implementation of \gle: key components.}
        Representation neurons form the forward pathway (red), error neurons form the backward pathway (blue).
        Both classes of neurons are \pyr, likely located in different layers of cortex.
        Lateral connections enable information exchange and gating between the two streams.
        The combination of retrospective membrane and prospective output dynamics allow these neurons to tune the temporal shift of the transmitted information.
        Errors are also represented in dendrites, likely located in the apical tuft of signal neurons, enabling local three-factor plasticity to correct the backpropagated errors.
    }\label{fig:mc}
\end{figure}
As shown above, \gle backward (error) dynamics engage the same sequence of operations as those performed by forward (representation) dynamics: first integration $\iminus$, then differentiation $\dplus$.
This suggests that backward errors can be transmitted by the same type of neurons as forward signals~\cite{markov2014anatomy, shai2015physiology}, which is in line with substantial experimental evidence that demonstrates the encoding of errors in L2/3 \pyr neurons~\cite{zmarz2016mismatch, fiser2016experience, attinger2017visuomotor, kellerPredictiveProcessingCanonical2018, ayaz2019layer, gillon2024responses}.
Note that correct local error signals are only possible with neurons that are capable of both retrospective ($\iminus$) and prospective ($\dplus$) coding -- the core element of the \gle framework.

\begin{figure*}[!htb]
    \centering
    \includegraphics[width=0.96\textwidth]{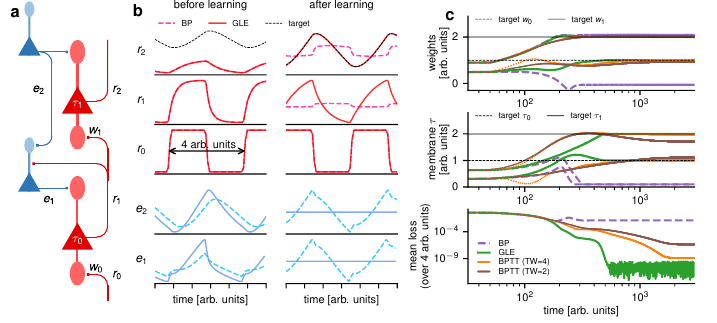}
    \caption{
        \tb{Learning with \gle in a simple chain.}
        \tb{(a)} Network setup.
        A chain of two retrospective representation neurons (red) learns to mimic the output of a teacher network (identical architecture, different parameters).
        In \gle, this chain is mirrored by a chain of corresponding error neurons (blue), following the microcircuit template in \cref{fig:mc}.
        We compare the effects of three learning algorithms: \gle (green), \bp with instantaneous errors (purple) and \bptt (point markers denote the discrete nature of the algorithm; pink, brown and orange denote different truncation windows (TW)).
        \tb{(b)} Output of representation neurons ($r_i$, red) and error neurons ($e_i$, blue) for \gle and instantaneous \bp (BP).
        Left: before learning (i.e., both weights and membrane time constants are far from optimal).
        Right: after learning.
        \tb{(c)} Evolution of weights, time constants and overall loss.
        Fluctuations at the scale of $10^{-10}$ are due to limits in the numerical precision of the simulation.
    }
    \label{fig:chain}
\end{figure*}
This symmetry between representation and error suggests a simple microcircuit motif that repeats in a ladder-like fashion, with L2/3 \pyr error neurons counterposing L5/6 \pyr representation neurons (\cref{fig:mc}).
Information transmitted between the two streams provides these neurons with all the necessary local information to carry out \gle dynamics.
In particular, error neurons can elicit the representation of corresponding errors in dendritic compartments of representation neurons, allowing forward synapses to access and correct these errors through local plasticity.
Recent evidence for error representation in apical dendrites provides experimental support for this component of the model \cite{francioniVectorizedInstructiveSignals2023}.

The correct propagation of errors requires two elements that can be implemented by static lateral synapses.
First, error neuron input needs to be multiplicatively gated by the derivative of the corresponding representation neuron's activation function $\varphi$.
This can either happen through direct lateral interaction, or through divisive (dis)inhibition, potentially carried out by somatostatin (SST) and parvalbumin (PV) interneuron populations~\cite{wilson2012division, seybold2015inhibitory, lee2013disinhibitory, dorsett2021impact}, via synapses that are appropriately positioned at the junction between dendrites and soma.
The required signal $\varphi'$ can be generated and transported in different ways depending on the specific form of the activation function.
For example, if $\varphi = \relu$, lateral weights can simply be set to $\bs L^{b} = \bs 1$.
For sigmoidal activation functions, $\varphi'$ can be very well approximated by synapses with short-term plasticity (e.g.,~\cite{petrovici2016form}, Eqn. 2.80).
The second requirement regards the communication of the error back to the error dendrites of the representation neurons; this is easily achieved by setting $\bs L^{f} = \bs 1$.

Ideally, synapses responsible for error transport in the feedback pathway need to mirror forward synapses: $\bs B
_\ell = \bs W^\TT_{\ell+1}$ (cf. \cref{eqn:erec}).
This issue is known as the weight transport problem and has been already addressed extensively in literature.
While it can, to some extent, be mitigated by \fa~\cite{lillicrapRandomFeedbackWeights2014}, improved solutions to the weight transport problem that are both online and local have also been recently proposed~\cite{kolen1994backpropagation,akrout2019deep,meulemans2022minimizing, maxLearningEfficientBackprojections2023,gierlichWeightTransportSpike2025}.
We return to the issue of weight symmetrization in \nameitref{sec:scaling}.

We now proceed to demonstrate several applications of the \gle framework.
First, we illustrate its operation in small-scale examples, to provide an intuition of how \gle networks can learn to solve non-trivial temporal tasks.
Later on, we discuss more difficult problems that usually require the use of sophisticated \dl methods and compare the performance of \gle with the most common approaches used for these problems in \ml.

\begin{figure*}[!hbt]
  \centering
  \includegraphics[width=0.98\textwidth]{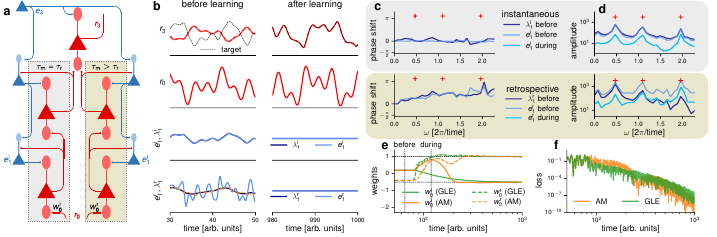}
  \caption{
    \tb{Error propagation and learning with \gle in a small hierarchical network.}
    \tb{(a)} Network setup.
    A network with one output neuron and two hidden layers (red) learns to mimic the output of a teacher network (identical architecture, different input weights to the first hidden layer).
    Each hidden layer contains one instantaneous ($\taum=\taur=1$) and one retrospective ($\taum = 1$, $\taur = 0.1$) neuron.
    In \gle, the corresponding error pathway (blue) follows the microcircuit template in \cref{fig:mc}.
    The input is defined by a superposition of three angular frequency components $\omega \in \{0.49, 1.07, 1.98\}$.
    Here, we compare error propagation, synaptic plasticity and ultimately the convergence of learning under \gle and \am.
    \tb{(b)} Input and output rates ($r$, red), along with bottom layer errors ($e$, light blue) and adjoints ($\lambda$, dark blue) before and during the late stages of learning.
    A running average over $e$ is shown in orange.
    \tb{(c)} Phase shifts (compared to the output error $e_3$) and
    \tb{(d)} amplitudes of bottom layer errors and adjoints across a wide range of their angular frequency spectrum before and during learning. The moments ``before'' and ``during'' learning are marked by vertical dashed lines in panel e.
    Top: $e_1^\text{i}$ and $\lambda_1^\text{i}$ for the instantaneous neuron.
    Bottom: $e_1^\text{r}$ and $\lambda_1^\text{r}$ for the retrospective neuron.
    Note that due to the nonlinearity of neuronal outputs, the network output has a much broader distribution of frequency components compared to the input (with its three components highlighted by the red crosses $\color{red}+$).
    Error amplitudes are shown at two different moments during learning.
    \tb{(e)} Evolution of the bottom weights ($w_0^\text{i}, w_0^\text{r}$) and
    \tb{(f)} of the loss during learning.
    The vertical dashed lines mark the snapshots at which adjoint and error spectra are plotted above.
  }
  \label{fig:small2}
\end{figure*}

\subsection{A minimal GLE example}
\label{sec:chain}

As a first application, we study learning in a minimal teacher-student setup. The network consists of a forward chain of two neurons (depicted in red in \cref{fig:chain}a) provided with a periodic step function input .
The task of the student network is to learn to mimic the output of a teacher network with identical architecture but different parameters, namely different weights and membrane time constants.
Prospective time constants $\taur$ are not learned and set to zero for both student and teacher such that the neurons are simple leaky integrators.
The target membrane time constants $\taum$ are chosen to be on the scale of the dominant inverse frequency of the input signal, such that they cause a significant temporal shift without completely suppressing the signal.
Due to these slow transient membrane dynamics, the task is not solvable using instantaneous backpropagation, but requires true temporal credit assignment instead.

We compare three different solutions to this problem:
(1) a \gle network with time-continuous dynamics, including synaptic and neuronal plasticity;
(2) standard error \bp using instantaneous error signals $e_i = \varphi'_i w_i e_{i+1}$;
(3) truncated \bptt through the discretized neuron dynamics using PyTorch's \texttt{autograd} functionality for different truncation windows.

We first note that the \gle network learns the task successfully and quickly, in contrast to instantaneous \bp(\cref{fig:chain}c).
To understand why, it is  instructive to compare its errors to the instantaneous ones  ($e_{i}$ panels in \cref{fig:chain}b).
The instantaneous errors (\bp, dashed lines) are always in sync with the output error, but their shape and timing become increasingly desynchronized from the neuronal inputs as they propagate toward the beginning of the chain, because they do not take into account the lag induced by the representation neurons.
Thus, the correct temporal coupling between errors and presynaptic rates required by plasticity (cf.~\cref{eqn:Wdot}, see also \cref{eqn:wdotmeth,eqn:bdotmeth,eqn:taumdotmeth,eqn:taurdotmeth} in Methods) is corrupted and learning is impaired.
Note that instantaneous errors are already a strong assumption and themselves require a form of prospectivity~\cite{haiderLatentEquilibriumUnified2021}; without any prospectivity, learning performance would be even more drastically compromised.
In contrast, the \gle errors gradually shift forward in time, matching the phase and shape of the respective neuronal inputs, and thus allowing the stable learning of all network parameters, weights and time constants alike.

Here, we can also see an advantage of \gle over the classical \bptt solution (\cref{fig:chain}c).
Despite only offering an approximation of the exact gradient calculated by \bptt, it allows learning to operate continuously, fully online.
As discussed above, \bptt needs to record a certain period of activity before being able to calculate parameter updates.
If this truncation window is too short, it fails to capture longer transients in the input and learning stalls or diverges (brown and pink, respectively).
Only with a sufficiently long truncation window does \bptt converge to the correct solution (orange), but at the cost of potentially exploding gradients and/or slower convergence due to the resulting requirement of reduced learning rates.
For a demonstration of the noise robustness of this setup, we refer to \nameitref{sec:supp-smallnets} in the Supplement.

\subsection{Small GLE networks}
\label{sec:smallnetworks}
To better visualize how errors are computed and transmitted in more complex \gle networks, we now consider a teacher-student setup with two hidden layers, each with one instantaneous and one retrospective neuron (\cref{fig:small2}a).
Through this combination of fast and slow pathways between network input and output, such a small setup can already perform quite complex transformations on the input signal (\cref{fig:small2}b).
From the perspective of learning an input-output mapping, this can be stated as the output neuron having access to multiple time scales of the input signal, despite the input being provided to the network as a constant stream in real time.
This is essential for solving the complex classification problems that we describe later.

To isolate the effect of error backpropagation into deeper layers, we keep all but the bottom weights fixed and identical between teacher and student.
The goal of the student network is to mimic the output of the teacher network by adapting its own bottommost forward weights.
We then compare error dynamics and learning in the \gle network with exact gradient descent on the cost as computed by \ambptt.

While both methods converge to the correct target (\cref{fig:small2}b and e), they don't necessarily do so at the same pace, since \ambptt cannot perform online updates, as also discussed above.
Also, \gle error propagation is only identical to the coupling of adjoint variables (\ambptt) for instantaneous neurons with $\taum=\taur$ ($e_1^\text{i} = \lambda_1^\text{i}$).
In general, this is not the case ($e_1^\text{r} \neq \lambda_1^\text{r}$), as \gle errors tend to overemphasize higher frequency components in the signal (cf. also \nameitref{sec:adjoint} and \cref{fig:small2}d).
However, this only occurs for slow, retrospective neurons, which only need to learn the slow components of the output signal.
For sufficiently small learning rates, plasticity in their afferent synapses effectively integrates over these oscillations and lets them adapt to the relevant low-frequency components.
Indeed, the close correspondence to \ambptt is reflected in the average \gle errors, which closely track the corresponding adjoints.
While it was not necessary to make use of such additional components in our simulations, high-amplitude high-frequency oscillations in the error signals could be mitigated by several simple mechanisms, including saturating activation functions for the error neurons, input averaging in the error dendrites or synaptic filtering of the plasticity signal.

Following the analysis in the previous section, our simulations now demonstrate how \gle errors encode the necessary information for effective learning.
Most importantly, \gle errors and adjoint variables (\ambptt) have near-identical timing, as shown by the alignment of their phase shifts across the signal frequency spectrum (\cref{fig:small2}c).
Moreover, for the errors of the retrospective neurons, these phase shifts are positive with respect to the output error, thus demonstrating the prospectivity required for the correct temporal alignment of inputs and errors.
The amplitudes of $e$ and $\lambda$ also show distinct peaks at the same angular frequencies, corresponding to the three components of the input signal that need to be mapped to the output (\cref{fig:small2}d).
These signals can thus guide plasticity in the correct direction, gradually learning first the slow and then the fast components of the input-output mapping.
This is also evinced by \cref{fig:small2}e, where the input weights of the retrospective neurons $w_0^\text{r}$ are the first to converge.
The ensuing reduction of the slow error components provides the input weights of the instantaneous neurons $w_0^\text{i}$ with cleaner access to the fast error components -- the only ones that they can actually learn -- which allows them to converge as well.

\subsection{Challenging spatio-temporal classification}
\label{sec:largescale}
\begin{figure}[t!]
    \centering
    \includegraphics[width=0.48\textwidth]{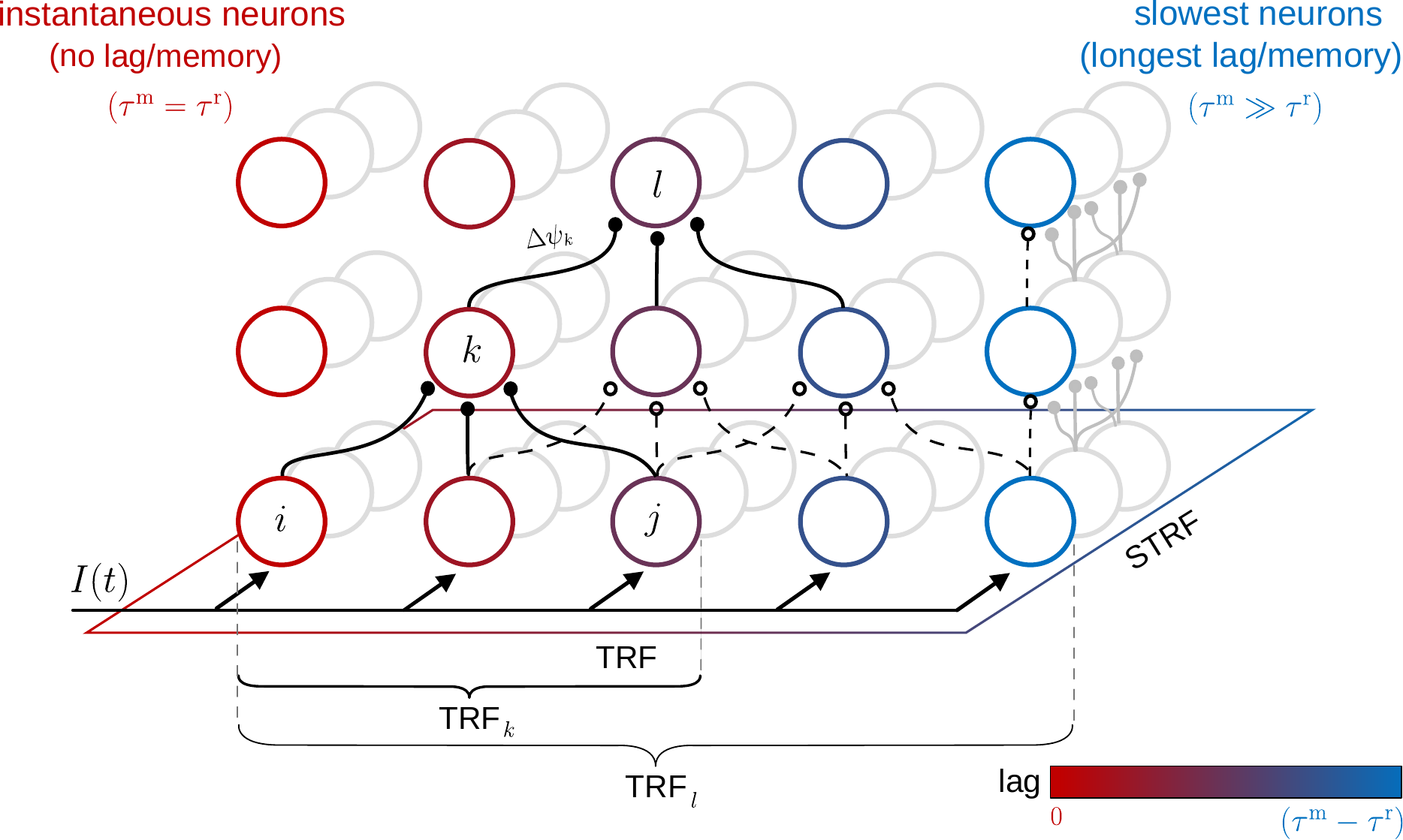}
    \caption{
        \tb{Neuronal diversity fosters complex temporal attention windows in \gle networks.}
        In the simplest case, a single input signal $I(t)$ is fed into a \gle network and all neurons in the bottom layer have access to the same information stream.
        However, the output of each neuron generates a temporal shift, depending on its time constants $\taum$ and $\taur$ (as highlighted by the neuron colors).
        Different chains of such neurons thus provide neurons in higher levels of the hierarchy with a set of attention windows across the past input activity.
        Synaptic and neuronal adaptation shape the nature of these temporal receptive fields (TRFs).
        For multidimensional input, neuron populations (gray) encode the additional spatial dimension and neuronal receptive fields become spatiotemporal (STRFs).
        }
    \label{fig:tempconv}
\end{figure}

We now demonstrate the performance of \gle in larger hierarchical networks, applied to difficult spatio-temporal learning tasks, and compare it to other solutions from contemporary \ml.
An essential ingredient for enabling complex temporal processing capabilities in \gle networks is the presence of neurons with diverse time constants $\taum$ and $\taur$ (\cref{fig:tempconv}).
Each of these neurons can be intuitively viewed as implementing a specific temporal attention window, usually lying in the past, proportionally to $\taum - \taur$.
More specifically, for a given angular frequency component of the input $\omega$, this window is centered around a phase shift of $\arctan(\omega \taur) -\arctan(\omega \taum) \approx \omega (\taur - \taum)$ for $\omega \tau \lesssim 1$ (cf. Methods).
By connecting to multiple presynaptic partners in the previous layer, a neuron thus carries out a form of temporal convolution, similarly to \tcns~\cite{bai2018empirical}.
Importantly however, \tcns rely on some additional mechanism that allows for the mapping of temporal signals to spatial representations, for example by using buffers or delays.
In other words, \tcns do not process their input online, but require it to be rolled out in space and then act like a conventional convolutional network, without any temporal dynamics.
Furthermore, network depth is an essential prerequisite for solving the tasks discussed below.
Deeper networks also allow longer chains of such neurons to be formed, thus providing the output with a diverse set of complex transforms on different time intervals distributed across the past values of the input signal.
\gle effectively enables deep networks to learn a useful set of such transforms.

\subsubsection{MNIST-1D}
\label{sec:mnist1d}

\begin{figure*}[htbp]
    \centering
    \includegraphics[width=0.96\textwidth]{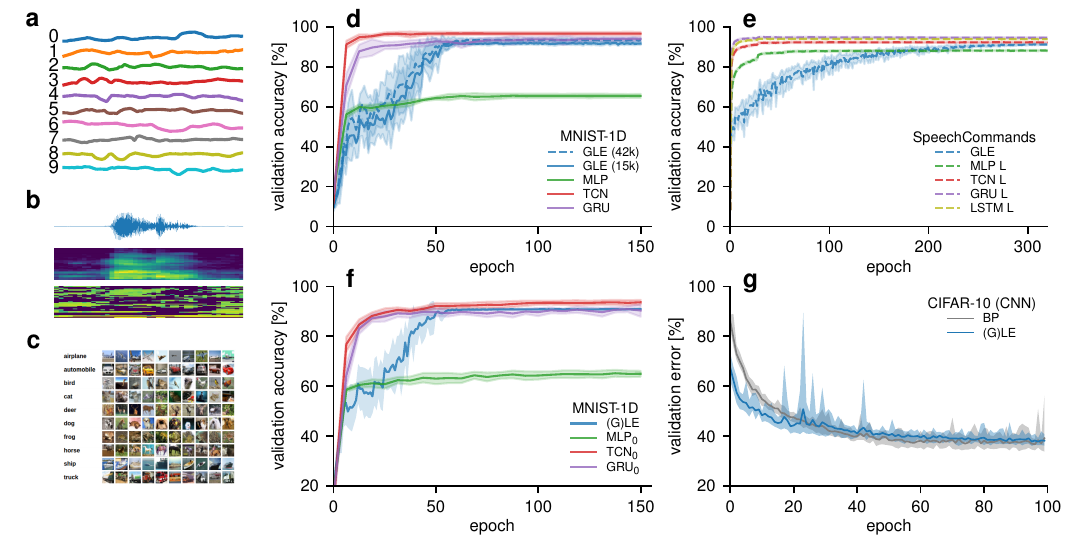}
    \caption{
        \tb{\gle for challenging spatio-temporal classification problems.}
        Averages and standard deviations measured over 10 seeds.
        Top row: samples from the \tb{(a)} \mnistId, \tb{(b)} GSC~-- including raw and preprocessed input -- and \tb{(c)} \cifarIO datasets.
        \tb{(d)} Performance of various architectures on \mnistId.
        Here, we used a higher temporal resolution for the input than in the original reference~\cite{greydanusScalingDeepLearning2020}.
        \tb{(e)} Performance of various architectures on GSC.
        \tb{(f)} Performance of a (G)LE LagNet architecture on \mnistId.
        For reference, we also show the original results from~\cite{greydanusScalingDeepLearning2020} denoted with the index 0.
        \tb{(g)} Performance of a (G)LE convolutional network on \cifarIO~(taken from~\cite{haiderLatentEquilibriumUnified2021}) and comparison with \bp.
    }
    \label{fig:largescale}
\end{figure*}

We first consider the \mnistId~\cite{greydanusScalingDeepLearning2020} benchmark  for temporal sequence classification.
Other than the name itself and the number of classes, \mnistId bears little resemblance to its classical namesake.
Here, each sample is a one-dimensional array of floating-point values, which can be streamed as a temporal sequence into the network (see \cref{fig:largescale}a for examples from each class).
This deceptively simple setup entails two difficult challenges.
First, only a quarter of each sample contains meaningful information; this chunk is positioned randomly within the sample, every time at a different position.
Second, independent noise is added on top of every sample at multiple frequencies, which makes it difficult to remove by simple filtering.
To allow a direct comparison between the different algorithms, we use no preprocessing in our simulations.

We first note that a \mlp fails to appropriately learn to classify this dataset, reaching a validation accuracy of only around 60\%.
This is despite the perceptron having access to the entire sequence from the sample at once, unrolled from time into space.
This highlights the difficulty of the \mnistId task.
More sophisticated \ml architectures yield much better results, with \tcns~\cite{bai2018empirical} and \grus~\cite{cho2014learning} achieving averages of over 90\%.
Notably, both of these models need to be trained offline, that is, they need to process the entire sequence before updating their parameters, with \tcns in particular requiring a mapping of temporal signals to spatial representations beforehand, and \grus requiring offline \bptt training with direct access to the full history of the network.

In contrast, \gle networks are trained online, with a single neuron streaming the input sequence to the network and, and the network updating its parameters in real time.
The network consists of six hidden layers with a mixture of instantaneous and retrospective neurons in each layer, and a final output layer of 10 instantaneous neurons.
We use either 53 or 90 neurons per layer, leading to a total of 15 thousand (as the \mlp) or 42 thousand parameters, respectively.

When faced with short informative signals embedded in a sea of noise, for which the target is always on even in the absence of meaningful information, the online learning advantage of \gle networks represents an additional challenge to learning.
Since only a fraction of each input actually contains a meaningful signal, \gle networks must be capable of remembering these informative combinations of inputs and targets throughout the uninformative portions of their training.
However, because they learn continuously, they also do so during uninformative times, which ultimately slows down learning.
In contrast, conventional \ml models only receive target errors at the end of the sequence or train a readout layer on the unrolled activations of the penultimate layer.
Moreover, GLE networks solve a more complex task than conventional models in that they learn both the temporal convolutions and the ultimate classification task simultaneously.
This manifests as an increase in convergence time, but not in ultimate performance, as \gle maintains an overall good online approximation of the true gradients for updating the network parameters.

Thus, despite facing a significantly more difficult task compared to the methods that have access to the full network activity unrolled in time, \gle achieves highly competitive classification results, with an average validation accuracy of $93.5\pm0.9\%$ for the larger network and $91.7\pm0.8\%$ for the smaller network.
The \gls{ann} baselines achieve an average validation accuracy of $65.5\pm1.0\%$ for the \mlp, $96.7\pm0.9\%$ for the \tcn and $94.0\pm1.1\%$ for the \gru, respectively.
A more detailed comparison of the performance of \gle with the reference methods can be found in \cref{tab:mnist1d-results} of the Supplement.

\subsubsection*{Google Speech Commands}
\label{sec:gsc}

To simultaneously validate the spatial and temporal learning capabilities of \gle, we now apply it to the \gsc dataset~\cite{wardenspeech2018}.
This dataset consists of 105'829 one-second long audio recordings of 35 different speech commands, each spoken by thousands of people.
In the v2.12 version of this dataset, the usual task is to classify 10 different speech commands in addition to a silence and an unknown class, which comprises all remaining commands.
The raw audio signal is transformed into a sequence of 41 \glspl{mfs}; this sequence constitutes the temporal dimension of the dataset and is streamed to the network in real-time.
Each of these spectrograms has 32 frequency bins, which are presented as 32 separate inputs to the network, thus constituting the spatial dimension of the dataset.

\cref{fig:largescale}e compares the performance of \gle to several widely used references: \mlp, \tcn, \gru (as used for \mnistId) and, additionally, \lstm networks~\cite{hochreiter1997long} -- all trained with a variant of \bp.
Similarly as on the \mnistId dataset, the \gle network is trained online and updates its parameters in real time, while the reference networks can only be trained offline; furthermore, the \mlp and \tcn networks do not receive the input as a real-time stream, but rather as a full spectro-temporal ``image'', by mapping the temporal dimension onto an additional spatial one.
Here, the \gle network receives its input through 32 neurons streaming 32 \gls{mfs} bins to the network.
The network consists of three hidden layers with a mixture of instantaneous and retrospective neurons in each layer, and a final output layer of twelve instantaneous neurons.
As with \mnistId, we see how our \gle networks surpass the \mlp baseline and achieve a performance that comes close to the references, with an average test accuracy of $91.44\pm0.23\%$.
The \gls{ann} baselines achieve an average test accuracy of $88.00\pm0.25\%$ for the \mlp, $92.32\pm0.28\%$ for the \tcn, $94.93\pm0.25\%$ for the \gru, and $94.00\pm0.19\%$ for the \lstm, respectively.
We thus conclude that, while offering clear advantages in terms of biological plausibility and online learning capability, \gle remains competitive in terms of raw task performance.
We also note that, in contrast to the reference baselines, \gle achieves these results without additional tricks such as batch or layer normalization, or the inclusion of dropout layers.
A more detailed comparison of the performance of \gle with the reference methods can be found in \cref{tab:gsc-results} of the Supplement.

\subsubsection{GLE for purely spatial problems}
\label{sec:timetospace}

The above results explicitly exploit the temporal aspects of \gle and its capabilities as an online approximation of \ambptt.
However, \gle also contains purely spatial \bp as a subcase, and the presented network architecture can lend itself seamlessly to spatial tasks such as image classification.
In cases like this, where temporal information is irrelevant, one can simply take the \le limit of \gle by setting $\taum = \taur$ for all neurons in the network~\cite{haiderLatentEquilibriumUnified2021}.
In the following, we demonstrate these capabilities in two different scenarios.
Note that \gle learns in a time-continuous manner in all of these cases as well, with the input being presented in real-time.

First, we return to the \mnistId dataset, but adapt the network architecture as follows.
The 1D input first enters a non-plastic preprocessing network module consisting of several parallel chains of retrospective neurons.
The neurons in each chain are identical, but different across chains: the fastest chain is near-instantaneous with $\taum \to 0$, while the slowest chain induces a lag of about 1/4 of the total sample length.
The endpoints of these chains constitute the input for a hierarchical network of instantaneous neurons ($\taum = \taur$).
By differentially lagging the input stream along the input chains, this configuration approximately maps time to space (the output neurons of the chains).
This offers the hierarchical network access to a sliding window across the input -- hence the acronym ``LagNet'' for this architecture -- and changes the nature of the credit assignment problem from temporal to spatial.
While the synaptic weights in the chains are fixed, those in the hierarchical network are trained with \gle, which in this scenario effectively reduces to \le.
As shown in \cref{fig:largescale}f, \gle is capable of training this network to achieve competitive performance with the reference methods discussed above.

As a second application to purely spatial problems, we focus on image classification.
Since \gle does not assume any specific connectivity pattern, we can adapt the network topology to specific use cases.
In \cref{fig:largescale}g we demonstrate this by introducing convolutional architectures (LeNet-5~\cite{lecun1989backpropagation}) and applying them to the CIFAR10~\cite{krizhevsky2009learning} dataset.
With test errors of $(38.0 \pm 1.3)\%$, \gle is again on par with \anns with identical structure at $ (39.4 \pm 5.6)\%$.
We therefore conclude that, as an extension of \le, \gle naturally maintains its predecessor's competitive capabilities for online learning of spatial tasks.

\subsection{Scaling, noise and symmetry}
\label{sec:scaling}

\begin{figure*}[!htb]
  \centering
  \includegraphics[width=0.9\textwidth]{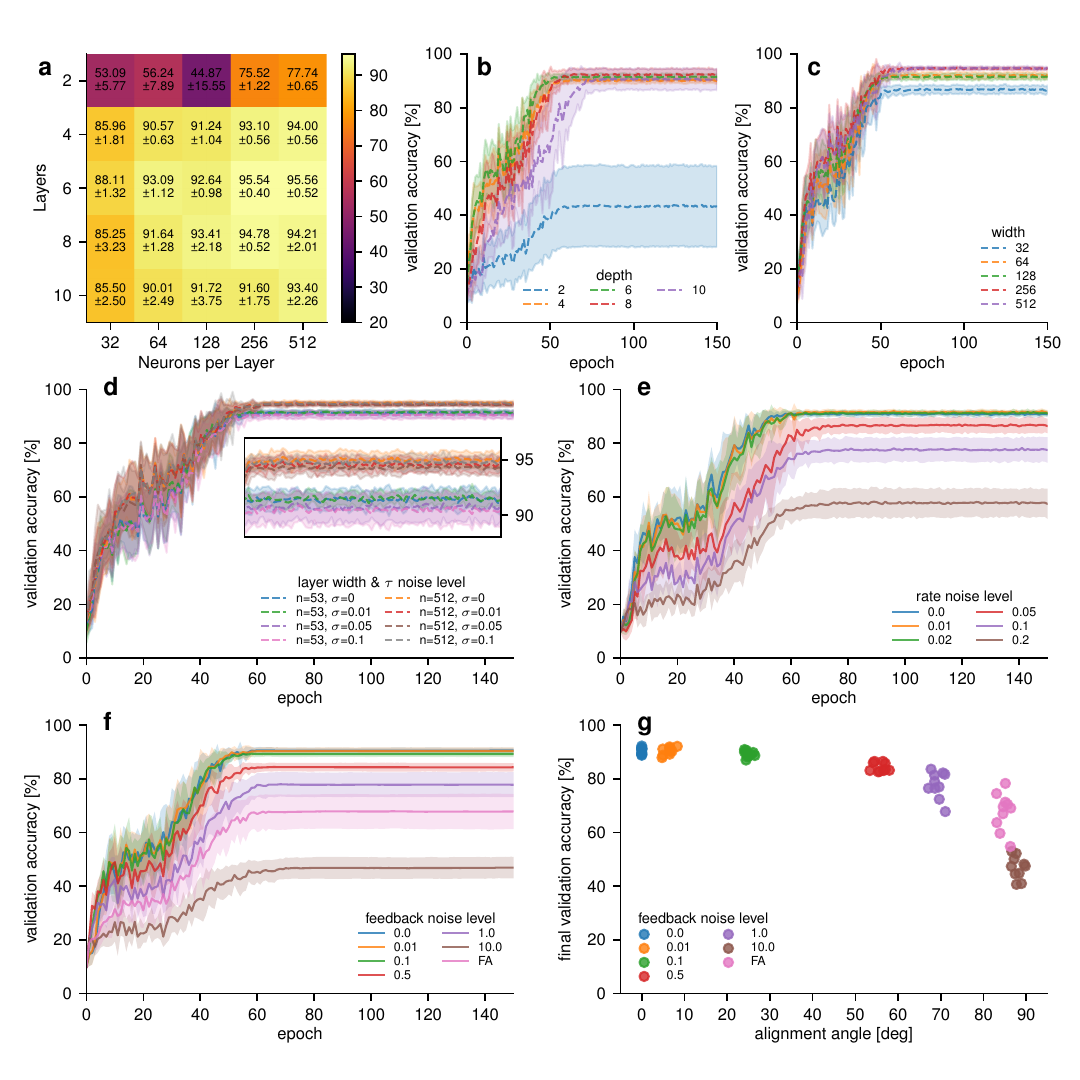}
  \caption{
    \tb{\gle network scaling and robustness to different types of noise.}
    Averages and standard deviations measured over 10 seeds.
    All figures depict validation accuracy on the \mnistId dataset.
    \tb{(a)} Heatmap of scaling in network width and depth.
    \tb{(b)} Learning with 128 neurons per layer across different depths.
    \tb{(c)} Learning in 6-layer networks for different layer widths.
    \tb{(d)} Robustness to spatial noise on $\taum$ and $\taur$ in a six layer network for two different layer widths $n$.
    Zoomed inlay of epoch 60 to 150.
    \tb{(e)} Robustness to correlated noise on all feedforward rates $r_{i}$ for different noise levels corresponding to the fraction of the input signal amplitude.
    \tb{(f)} Robustness to feedback misalignment $\bm{B}_{\ell} = \bm{W}_{\ell+1}^\TT + \mathcal{N}(0,\sigma^2)$ for different noise levels $\sigma$.
    \tb{(g)} Final validation accuracy for different noise levels $\sigma$ plotted against the corresponding alignment angle $\angle(\bm{B}_{\ell}, \bm{W}_{\ell+1}^{\mathrm{T}})$ for the individual layers $\ell \in [1,\ldots,L-1]$.
    Note that \fa performs better than the largest noise levels, because the constant randomization of feedback weights prevents \fa to take effect.
  }
  \label{fig:noise_scaling}
\end{figure*}

To evaluate the effect of network size on classification performance, we trained \gle networks of different widths and depths on the \mnistId dataset (\cref{fig:noise_scaling}a-c).
Because this dataset requires significant depth for neurons to be able to tune their temporal attention windows, good performance can only be reached for depths of 4 layers and above.
The network width plays only a secondary role, with relatively small performance gains for wider layers.
The small decline in performance for larger networks is likely caused by overfitting due to overparametrization.

The robustness of \gle to input level noise has already been implicitly demonstrated, as both \mnistId and \gsc datasets include either explicitly applied noise or implicit measurement noise and speaker variance.
However, whether biological or biologically inspired, analog systems are always subject to additional forms of noise.
Spatial noise refers to neuronal variability, caused by either natural biological growth processes or fixed-pattern noise in semiconductor photolitography.
Temporal noise refers to variability in all transmitted signals, usually due to quantum or thermal effects, and typically modeled as a wide-spectrum random process.
In the following, we discuss the robustness of \gle with respect to both of these effects during learning of the \mnistId task.

First, we established baseline performances for two different network sizes without added noise.
To model spatial variability, we then introduced Gaussian noise with increasing levels of variance to the time constants, as all other parameters were optimized through learning \cref{fig:noise_scaling}d.
The effect on training accuracy was insignificant, despite the highest simulated noise level of 0.1 being relatively large compared to the time constants ranging from 0.2 to 1.2 (cf. \cref{tab:largescale-params} in the Methods).
The effects of temporal variability were investigated by adding correlated noise to all feedforward rates \(r_i\) (\cref{fig:noise_scaling}e).
This type of noise can be much more detrimental than spatial variability, as it gradually and irrevocably destroys information while it passes through the network.
We noticed a gradual performance decline for noise levels of above 2\%, with classification performance dropping by about 10\% at a noise level of 10\%.
Note that, across layers, this sums up to a total noise-to-signal ratio of about 60\%, which would represent a major impediment for any network model.

While \fa \cite{lillicrapRandomFeedbackWeights2014} is known to have issues with scaling in depth (as we shall also see below), several learning algorithms for active weight alignment have been proposed to explain how biological and bio-inspired systems could solve this problem locally.
However, these cannot always guarantee a perfect alignment of forward and backward weights $\bs B_\ell = \bs W^\TT_{\ell+1}$.
We thus study the performance of \gle networks under varying levels of weight asymmetry.
To allow a general assessment independent of particular alignment algorithms, we train \gle networks with backward weights that are noisy versions of the forward weights (\cref{fig:noise_scaling}f).
We resample this noise after each epoch, thus making the problem harder by ensuring that the forward weights cannot mitigate this asymmetry by aligning with the feedback weights during forward learning (as in \fa).
\Cref{fig:noise_scaling}g shows how the performance of \gle degrades gracefully with increasing weight misalignment.
In particular, for alignment angles $\angle(\bs{B}_{\ell}, \bs{W}_{\ell+1}^\TT) < \ang{30}$, network performance is barely affected.
This lies well within the alignment capabilities of multiple alignment algorithms, as we show in \nameitref{si:weight_transport} of the SI.

Altogether, these results illustrate the robustness of the \gle framework to realistic types of noise, thereby substantiating its plausibility as a model of biological networks, as well as its applicability to suitable analog neuromorphic devices.

\subsubsection{Chaotic time series prediction}
\label{sec:mackeyglass}

\begin{figure*}[!htb]
  \centering
  \includegraphics[width=0.96\textwidth]{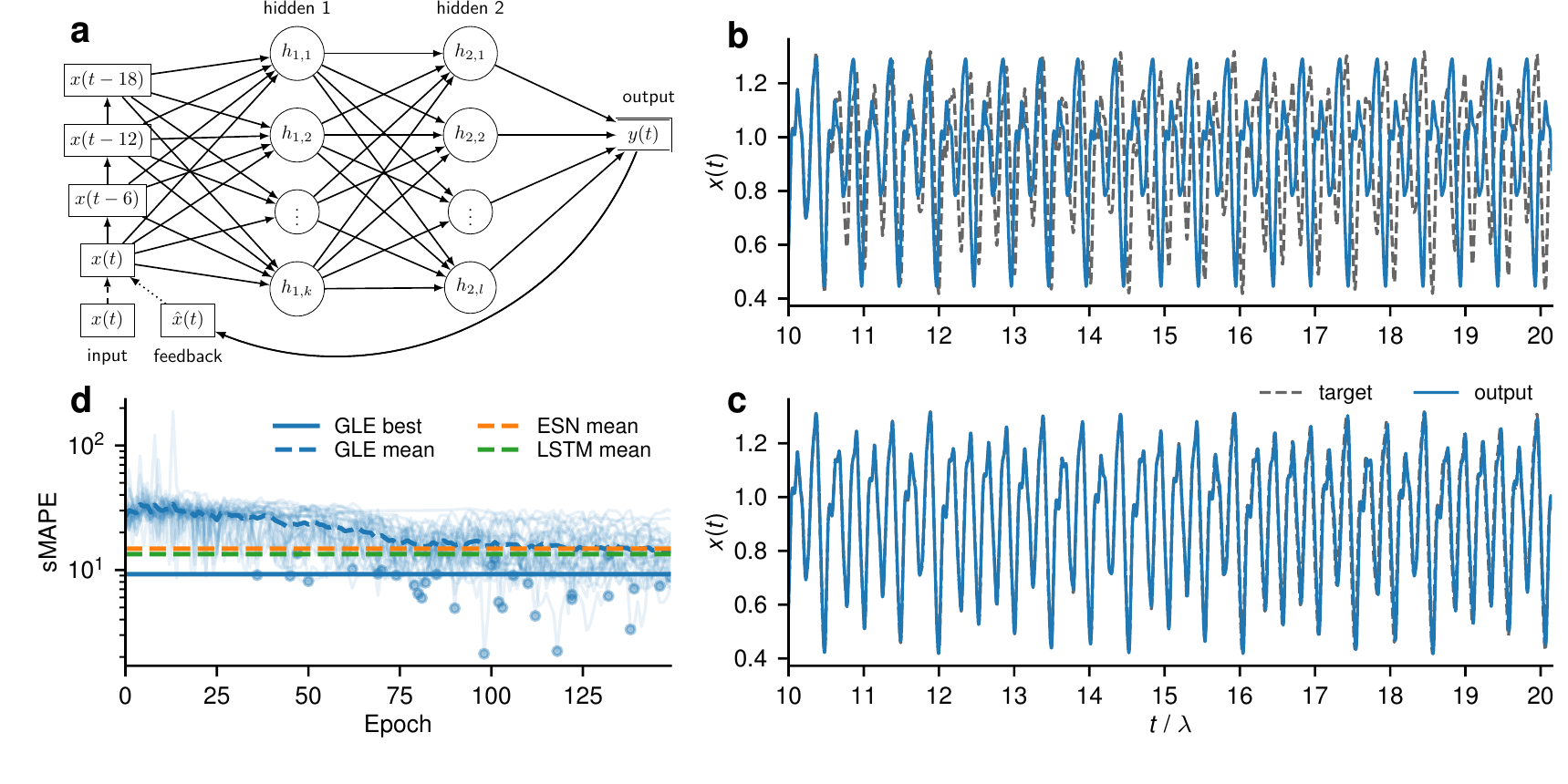}
  \caption{
    \tb{GLE for chaotic time series prediction.}
    \tb{(a)} \gle network architecture with 93 hidden neurons per layer.
    Input neurons are delayed w.r.t. each other by $\Delta = 6$ (same units as $\lambda=197$).
    \tb{(b,c)} Target sequence (dashed black) vs. network output (solid blue) after 40 (b) and 138 (c) epochs of training.
    \tb{(d)} sMAPE loss over the course of training for 30 different sequences compared to \esn and \lstm results as described in~\cite{yikNeurobenchFrameworkBenchmarking2025}.
    The best performance for each sequence is marked as a blue dot.
  }
  \label{fig:mackeyglass}
\end{figure*}

As a final application, we now turn to a sequence prediction task, where an autoregressively recurrent \gle network (with its output feeding back into its input) is trained to predict the continuation of a time series based on its previous values.
To this end, we use the Mackey-Glass dataset~\cite{mackeyOscillationChaosPhysiological1977}, a well-known chaotic time series described by a delayed differential equation that exhibits complex dynamics.
\Cref{fig:mackeyglass}a shows the network setup, which consists of four input neurons, two hidden layers with a mixture of instantaneous and retrospective neurons, and one output neuron.
Each Mackey-Glass sequence has a length of 20 Lyapunov times $\lambda$, with $\lambda=197$ for our chosen parametrization of the Mackey-Glass equation.
The first half of the sequence is used for training, while the second half is used for autoregressive prediction.
The resulting requirement of predicting the continuation of the time series across 10 Lyapunov times (for which initial inaccuracies are expected to diverge by at least a factor of $e^{10}$) is what makes this problem so difficult.
During training, the network's target $y^*(t)$ is given by the next value in the time series $x(t+1)$ based on the ground truth $x(t)$.
After training, the external input $x(t)$ is removed after the first 10 $\lambda$ and the network needs to predict the second half of the sequence (which it has never seen during training) using its own output as an input.
To provide an intuition for how the network learns, we consider its prediction at an earlier and at a later stage of training.
\Cref{fig:mackeyglass}b shows the target sequence and the network's output after 40 epochs of training.
The network has already learned a periodic pattern that is superficially similar to parts of the sequence, but the output diverges from the target sequence after only half a Lyapunov time.
From here, it takes another 100 epochs (\cref{fig:mackeyglass}c) for the network's output to closely match the target sequence over the full period of 10 $\lambda$.
\Cref{fig:mackeyglass}d shows the \gls{smape}, a commonly used metric for evaluating the performance of time series prediction models, over the course of training and averaged over 30 different sequences and initializations.
The final average \gls{smape} of 14.25\% for our \gle network is on par with the 14.79\% for \esns and 13.37\% for \lstms reported in \cite{yikNeurobenchFrameworkBenchmarking2025}.
If we average the lowest \gls{smape} over the course of training, we find that our \gle networks can achieve an even lower \gls{smape} of 9.25\%.

\section{Discussion}
\label{sec:discussion}

We have presented \gle, a novel framework for spatio-temporal computation and learning in physical neuronal networks.
Inspired by well-established approaches in theoretical physics, \gle derives all laws of motion from first principles: a global network energy, a conservation law, and a dissipation law.
Unlike more traditional approaches, which aim to minimize a cost defined only on a subset of output neurons, our approach is built around an energy function which connects all relevant variables and parameters of all neurons in the network.
This permits a unified view on the studied problem and creates a tight link between the dynamics of computation and learning in the neuronal system.
The extensive nature of the energy function (i.e.,~its additivity over subsystems) also provides an important underpinning for the locality of the derived dynamics.

In combination, these dynamics ultimately yield a local, online, real-time approximation of \ambptt -- to our knowledge, the first of its kind.
Moreover, they suggest a specific implementation in physical neuronal circuits, thus providing a possible template for spatio-temporal credit assignment in the brain, as well as blueprints for dedicated hardware implementations.
This shows that, in contrast to conventional wisdom, physical neuronal networks can implement future-facing algorithms for temporal credit assignment.
More recently, tentative calls in this direction have indeed been formulated \cite{lillicrapBackpropagationTimeBrain2019}, to which \gle provides an answer.

In the following, we highlight some interesting links to other models in \ml, discuss several biological implications of our model, and suggest avenues for improvement and extension of our framework.

\subsection{Connection to related approaches}

\paragraph{Latent Equilibrium (LE)}
\label{sec:le}

    As the spiritual successor of \le\cite{haiderLatentEquilibriumUnified2021}, \gle inherits its energy-based approach, as well as the derivation of dynamics from energy conservation and minimization.
    \gle also builds on the insight from \le that prospective coding can undo the low-pass filtering of neuronal membranes.
    The core addition of \gle lies in the separation of prospective and retrospective coding, leading to an energy function that depends on two types of canonical variables instead of just one, and to a conservation law that accounts for their respective time scales.
    The resulting neuron dynamics can thus have complex dependencies on past and (estimated) future states, whereas information processing in \le is always instantaneous.
    Functionally, this allows neurons to access their own past and future states (through appropriate prospective and retrospective operators), thereby allowing the network to minimize an integrated cost over time, whereas \le only minimizes an instantaneous cost.
    
    Because \le can be seen as a limit case of \gle, the new framework offers a more comprehensive insight into the effectiveness of its predecessor, and also answers some of the questions left open by the older formalism.
    Indeed, the \gle analysis demonstrates that \le errors are an exact implementation of the adjoint dynamics of the forward system for equal time constants, further confirming the solid grounding of the method, and helping explain its proven effectiveness.
    Conversely, \gle networks can learn to evolve towards \le via the adaptation of time constants if required by the task to which the framework is applied.
    While \le also suggests a possible mechanism for learning the coincidence of time scales, \gle provides a more versatile and rigorous learning rule that follows directly from the first principles on which the framework is based.
    More generally speaking, by having access to local plasticity for all parameters in the network, \gle networks can either select (through synaptic plasticity) or adapt (through neuronal plasticity) neurons and their time constants in order to achieve their target objective.
    
    Thus, \gle not only extends \le to a much more comprehensive class of problems (spatio-temporal instead of purely spatial), but also provides it with a better theoretical grounding, and with increased biological plausibility.
    Recently, \cite{fayyaziProspectiveMessagingLearning2024} also proposed an extension of \le by incorporating hard delays in the communication channel between neurons.
    In order to learn in this setting, they propose to either learn a linear estimate of the future errors (similar to our prospective errors) or to learn an estimate of future errors by a separate network.
    In contrast to this, \gle implements a biologically plausible mechanism for retrospection, and a self-contained error prediction which we show gives a good approximation to the exact \am solution.

\paragraph{\Gls{nla}}
    
    Similarly inspired by physics, but following a different line of thought, the \nla principle~\cite{sennNeuronalLeastactionPrinciple2023} also uses prospective dynamics as a core component.
    It uses future discounted membrane potentials $\tilde u = \iplusm[u]$ as canonical variables for a Lagrangian $L$ and derives neuronal dynamics as associated Euler-Lagrange equations.
    A simpler but equivalent formulation places \nla firmly within the family of energy-based models such as \le and \gle, where $L$ is replaced by an equivalent energy function $E$ that sums over neuron-local errors and from which neuronal dynamics can be derived by applying the conservation law $\dplusm[\partial E / \partial u] = 0$.
    
    The most important difference to \gle is that \nla cannot perform temporal credit assignment.
    Indeed, other than imposing a low-pass filter on its inputs, an \nla network effectively reacts instantaneously to external stimuli and can neither carry out nor learn temporal sequence processing.
    This is an inherent feature of the \nla framework, as the retrospective low-pass filter induced by each neuronal membrane exactly undoes the prospective firing of its afferents.
    
    This property also directly implies that, except for the initial low-pass filter on the input, all neurons in the network need to share a single time constant for both prospective and retrospective dynamics.
    In contrast, both \le and \gle successively lift this strong entanglement by modifying the energy function and conservation law.
    \le correlates retro- and prospectivity within single neurons and allows their matching to be learned, thus obviating the need for globally shared time constants, while \gle decouples these two mechanisms, thus enabling temporal processing and learning, as discussed above.

\paragraph{RTRL and its approximations}

    \rtrl~\cite{williamsGradientbasedLearningAlgorithms1995} is a past-facing algorithm that implements online learning by recursively updating a tensor $M_{ijk}$ that takes into account the influence of every synapse $w_{jk}$ on every neuronal output $r_i$ in the network.
    As evident from the dimensionality of this object, this requires storing $\bigO (N^3)$ floating-point numbers in memory (where $N$ is the number of neurons in the network).
    Because this is much less efficient than future-facing algorithms, \rtrl is rarely used in practice, and the manifest non-locality of the influence tensor also calls into question its biological plausibility.
    Nonetheless, several approximations of \rtrl have been recently proposed, with the aim of addressing these issues~\cite{marschallUnifiedFrameworkOnline2020}.
    
    A particularly relevant algorithm of this kind is \rflo~\cite{murrayLocalOnlineLearning2019}, in which a synaptic eligibility trace is used as a local approximation of the influence tensor, at the cost of ignoring dependencies between distant neurons and synapses.
    With a reduced memory scaling of $\bigO (N^2)$, this puts \rflo at significant advantage over \rtrl, while closing the distance to \bptt (with $\bigO(NT)$, where $T$ is the length of the learning window).
    In its goal of reducing the exact, but nonlocal computation of gradients to an approximate, but local solution, \rflo shares the same spirit as \gle.
    In the following, we highlight several important differences which we consider to give \gle both a conceptual and a practical advantage.
    
    First, the neuron membranes and synaptic eligibilities in \rflo are required to share time constants, in order for the filtering of the past activity to be consistent between the two.
    In cortex, this would imply the very particular neurophysiological coincidence of synaptic eligibility trace biochemistry closely matching the leak dynamics of efferent neuronal membranes.
    In contrast, the symmetry requirements of \gle are between neurons of the same kind (\pyr cells), which share their fundamental physiology, and whose time constants can be learned locally within the framework of \gle.
    
    Second, the dimensionality reduction of the influence tensor proposed by \rflo also puts it at a functional disadvantage, because the remaining eligibility matrix only takes into account first-order synaptic interactions between directly connected neurons.
    This is in contrast to \gle, which can propagate approximate errors throughout the entire network.
    This flexibility also allows \gle to cover applications over multiple time scales, from purely spatial classification to slow temporal signal processing.
    Moreover, \gle accomplishes this within a biologically plausible, mechanistic model of error transmission, while admitting a clear interpretation in terms of cortical microcircuits.
    Finally, \gle's additional storage requirements only scale linearly with $N$ (one error per neuron at any point in time), which is even more efficient than \bptt.

    While both \rflo and \gle are inspired by and dedicated to physical neuronal systems, both biological and artificial, it might still be interesting to also consider their computational complexity for digital simulation, especially given the multitude of digital neuromorphic systems \cite{schuman2017survey} and \ann accelerators \cite{deng2020model} capable of harnessing their algorithmic capabilities.
    For a single update of their auxiliary learning variables (eligibility traces / errors), both \rflo and \gle incur a computational cost of $\bigO(N^2)$.
    Thus, for an input of duration $T$, all three algorithms -- \rflo, \gle and \bptt (without truncation) -- are on par, with a full run having a computational complexity of $\bigO(N^2T)$.

\paragraph{\Glspl{ssm}}

    \gle networks are linked to \lrnns (see, e.g.,~\cite{campolucciCausalBackPropagation1996}).
    In recent years there has been renewed interest in similar models in the form of \lrus~\cite{orvietoResurrectingRecurrentNeural2023, orvietoUniversalityLinearRecurrences2024}, which combine the fast inference characteristics of \lrnns with the ease of training and stability stemming from the linearity of their recurrence.
    These architectures are capable of surpassing the performance of Transformers in language tasks involving long sequences of tokens~\cite{guMambaLinearTimeSequence2023, deGriffinMixingGated2024}.
    
    More specifically, \gle networks are closely linked to \lrus with diagonal linear layers~\cite{gupta2022diagonal} and \ssms~\cite{gu2021combining, gu2021efficiently}, where a linear recurrence only acts locally at the level of each neuron, as realized by the leaky integration underlying the retrospective mechanism.
    Additionally, the inclusion of prospectivity enables the direct passthrough of input information across layers as in \ssms (see discussion in Methods).
    \gle could thus enable online training of these models, similarly to how \cite{zucchetOnlineLearningLongrange2023} demonstrate local learning with \rtrl, but with the added benefits discussed above.
    Since such neuron dynamics are a de-facto standard for neuromorphic architectures~\cite{schuman2017survey}, \gle could thus open an interesting new application area for these systems, especially in light of their competitive energy footprint \cite{wunderlich2019demonstrating, goltzFastEnergyefficientNeuromorphic2021}.

    As \gle is strongly motivated by biology, it currently only considers real-valued parameters for all connections, including the self-recurrent ones, in contrast to the complex-valued parameters in \lrus.
    However, an extension of \gle to complex activities appears straightforward, by directly incorporating complex time constants into prospective and retrospective operators, or, equivalently, by extending them to second order in time.
    With this modification, \gle would also naturally extend to the domain of \cvnns~\cite{lee2022complex}.

\subsection{Neurophysiology}

Neuronal prospectivity is a core component of the \gle framework.
Prospective coding in biological neurons is supported by considerable experimental and theoretical evidence~\cite{mainen_model_1995,  gerstner_neuronal_2014, traub_neuronal_1991, plesserEscapeRateModels2000, brette_simulation_2007, pozzorini2013temporal, brandt2024prospective} on both short and long time scales.
We note that our Fourier analysis (\nameitref{sec:adjoint} above and Methods \nameitref{sec:adaptation}) also offers a rigorous account of prospectivity in neuron models with multiple, negatively coupled variables such as the Hodgkin-Huxley mechanism or adaptation currents.

In the specific GLE implementations used to solve the more challenging spatiotemporal classification tasks, we explicitly employ neurons with both short and long (effective) integration time constants.
This is necessary since the tasks  require access to (almost) instantaneous stimulus information, as well as to a temporal context in which to interpret it.
While we do not explicitly model spikes, these two ``classes'' of neurons can be interpreted through the concepts of integration and coincidence detection \citep{abeles1982role,konig1996integrator,ratte2013impact}.
Our results thus suggest how the cortex can benefit from neurons in both operating modes, as they can provide orthogonal pieces of information to guide behavior.

\gle further predicts functional aspects of \pyr neuron morphology, as well as \cmcs for signal and error propagation.
In these \cmcs, \pyr cells are responsible for the transmission of both representation and error signals, as supported by ample experimental data \cite{zmarz2016mismatch, fiser2016experience, attinger2017visuomotor, kellerPredictiveProcessingCanonical2018, ayaz2019layer, gillon2024responses}.
Moreover, the morphological separation of the cell body into multiple distinct units, including soma, basal and apical trees, corresponds to a functional separation that allows the simultaneous representation of different pieces of information -- bottom-up input, top-down errors and the resulting integrated signal -- within the same cell \cite{francioniVectorizedInstructiveSignals2023}.
This is also what gives synapses local access to this information, allowing the implementation of the proposed three-factor, error-correcting plasticity rule \cite{urbanczik2014learning}.
Furthermore, by representing forward activites and backward errors in separate pathways, these \cmcs are capable of robust learning.

While building on many insights from previous proposals for \cmcs, most notably~\cite{sacramentoDendriticCorticalMicrocircuits2018, haiderLatentEquilibriumUnified2021}, we argue that our proposed model features significant improvements.
Our model does not require nerve cells belonging to two different classes (\sst interneurons and \pyr cells) to closely track each others' activity, which is more easy to reconcile with the known electrophysiology of cortical neurons.
This also makes training more robust and obviates the need to copy neuronal activities when training the network for more complex tasks.
Additionally, the original \cmc model for error backpropagation~\cite{sacramentoDendriticCorticalMicrocircuits2018} suffers from a relaxation problem, as already addressed in~\cite{haiderLatentEquilibriumUnified2021}.
As it subsumes the capabilities of \le, \gle is inherently able to alleviate this problem.

Most importantly, these other \cmc models are, by construction, only capable of solving purely spatial classification problems.
We have shown that \gle \cmcs can perform spatial and temporal tasks across a range of scales, by adapting the network parameters to the characteristic temporal and spatial timescales of the problem at hand.
All of the above observations notwithstanding, it is worth noting that \gle can also apply to these alternative \cmc models by implementing prospective coding in the apical dendrites of the \pyr representation neurons rather than in \pyr error neurons.

\subsection{Open questions and future work}

While the efficacy of \gle relies on correctly phase-shifting the backpropagated errors for all frequency components of the signal, their frequency-dependent gain diverges from the one required for exact gradient descent.
This is not surprising, as a perfect match of gains would require the kind of perfect knowledge of the future that is available to \ambptt.
A reasonable approximation of \am can be said to hold for $\omega \tau \lesssim 1$, but outside of this regime there is no guarantee for convergence to a good solution.
Our simulations clearly demonstrate that such solutions exist and can be achieved, but it would be preferable to have a more robust mechanism for controlling gain discrepancies at high frequencies.
Some straightforward and sufficient mitigation strategies might include a simple saturation of error neuron activations or the inclusion of a small synaptic time constants as proposed in \cite{haiderLatentEquilibriumUnified2021}.
However, primarily for neuromorphic realizations of \gle, we expect \lti system theory to provide more elegant solutions, such as Bessel or all-pass (active) filters, with favorable phase-response properties and circuit-level implementations.

On a more technical note, gain amplification in prospective neurons requires particular attention in discrete-time forward-Euler simulations, where fast transients can cause a breakdown of the stability assumption for finite, fixed-size time steps on which this integration method relies.
In physical, time-continuous neuronal systems, this effect is naturally mitigated by the finite time constants of all physical components, including, for example, the synapses themselves, as mentioned above.

Ideally, to ensure a close correspondence of synaptic updates between \gle and \am, errors should not disrupt representations.
This can be easily achieved in the limit of weak teacher coupling $\beta \ll 1$ and/or weak somato-dendritic coupling $\gamma$.
Indeed, for larger networks and more complex tasks requiring fine tuning of neuronal activities and weights, we observed an increasing sensitivity of our networks to these parameters.
To avoid these effects, we simply operated our large-scale simulations in the $\beta \to 1, \gamma \to 0$ regime.
As this sensitivity also manifests as a consequence of the high-frequency gain amplification, we expect it to be correspondingly mitigated by the above-mentioned solutions.

To guarantee a perfect match between error and representation signals, pairs of \pyr neurons in the two pathways ideally require an exact inversion of time constants $\taum \leftrightarrow \taur$.
This relationship also needs to be maintained when time constants are learned.
Similarly to the weight transport problem, we expect a certain degree of robustness to some amount of symmetry breaking \cite{lillicrapRandomFeedbackWeights2014}.
However, it would be preferable to have an additional local adaptation mechanism to ensure scalability, while maintaining full compatibility with the locality constraints of physical neuronal systems.
To this end, we expect that local solutions based on decay during adaptation \cite{kolen1994backpropagation}, mirroring \cite{akrout2019deep}, and especially correlations \cite{maxLearningEfficientBackprojections2023} are likely to apply here as well, even more so as the matching needs to develop between active, reciprocally connected and physically proximal neurons as opposed to passive, uncoupled synapses.

\gle can be naturally extended to (more) complex neuron dynamics, as already discussed in the context of \lrus and \cvnns.
Even more importantly, a \gle mechanism for (sparse) spiking dynamics as opposed to (population) rates is of eminent interest.
The recently described family of solutions for \bp through spike times, including surrogate methods \cite{cramer2022surrogate}, exact solutions \cite{goltzFastEnergyefficientNeuromorphic2021} and, notably, adjoint dynamics \cite{wunderlich2021event} suggest several starting points for deriving corresponding \gle operators.

In the presented simulations, we only applied \gle to hierarchical networks, in order to highlight its versatility in switching between spatial, temporal and spatiotemporal classification tasks without changing the underlying architecture.
We expect these capabilities to extend naturally to problems of sequence generation and motor control.
Moreover, the theoretical framework of \gle is architecture-agnostic, and the inclusion of lateral recurrence represents an obvious next step.
We expect the approximation of \ambptt to be adaptable to this scenario as well, but a dedicated proof and demonstration is left for future work.

\section{Conclusion}

With Generalized Latent Equilibrium, we have proposed a new and flexible framework for inference and learning of complex spatio-temporal tasks in physical neuronal systems.
In contrast to classical \ambptt, but still rivaling its performance, \gle enables efficient learning through fully local, phase-free, on-line learning in real time.
Thus, \gle networks can achieve results competitive with well-known, powerful \ml architectures such as \grus, \tcns and \cnns.

Our framework carries implications both for neuroscience and for the design of neuromorphic hardware.
For the brain, \gle provides a rigorous theory and experimental correlates for spatio-temporal inference and learning, by leveraging an interplay between retrospective and prospective coding at the neuronal level.
For artificial implementations, its underlying mechanics and demonstrated capabilities may constitute powerful assets in the context of autonomous learning on low-power neuromorphic devices.

\section{Methods}

\subsection{Derivation of the network dynamics}
\label{sec:meth-networkdynamics}
The neuron dynamics in \cref{eqn:udot} are derived from the stationarity condition stated in \cref{eqn:GLEudot}. The partial derivatives of the energy function $E$ with respect to the prospective voltages $\dplusmi[u_i]$ and $\dplusri[u_i]$ are given by
\begin{equation}
  \frac{\partial E}{\partial \dplusmi[u_i]} = e_{i}
\end{equation}
and
\begin{equation}
  \frac{\partial E}{\partial \dplusri[u_i]} = - \varphi'(\dplusri[u_i]) \sum_{j} W_{ji} e_j  + \beta \frac{\partial C}{\partial \dplusri[u_i]} \; .
\end{equation}
Substituting these into the stationarity condition from \cref{eqn:GLEudot}
\begin{align}
  0 &= \iminusmi[\frac{\partial E}{\partial \dplusmi[u_i]}] + \iminusri[\frac{\partial E}{\partial \dplusri[u_i]}] \\
    &= \iminusmi[e_{i}] + \iminusri[- \varphi'(\dplusri[u_i]) \sum_{j} W_{ji} e_j + \beta \frac{\partial C}{\partial \dplusri[u_i]}]
\end{align}
yields the neuronwise error dynamics
\begin{align}
  e_{i} = \dplusmi[\iminusri[e^{\mathrm{inst}}_{i}]]
  \label{eqn:error_dynamics_neuronwise}
\end{align}
with the instantaneous error
\begin{align}
  e^{\mathrm{inst}}_{i} = \varphi'(\dplusri[u_i]) \sum_{j} W_{ji} e_j - \beta \frac{\partial C}{\partial \dplusri[u_i]} \; ,
\end{align}
where we used the identity $\mathcal{D}^{+}_{\tau}\left\{ \mathcal{I}^{-}_{\tau} \left\{x\right\}\right\} (t) = x(t)$.
Substituting the definition of the mismatch error $e_{i}$ (\cref{eqn:GLEE}) produces the neuronal dynamics from \cref{eqn:udot}.

Instead of writing these equations for individual neurons $i$, we can also write them in vector form for layers $\ell$ as this is more convenient for the following derivations and, in addition, makes the similarity to \gls{bp} more apparent.
Then, $\forall \ell < L$ the vector of \textit{instantaneous errors} $e_{\ell}^{\mathrm{inst}}$ is the backpropagated error signal
\begin{equation}
  \bs{e}_{\ell}^{\mathrm{inst}} = \bs{\varphi}'_{\ell} \odot \bs{W}_{\ell+1}^{\TT} \bs{e}_{\ell+1} \qquad \forall \ell < L \; ,
  \label{eq:layerwise_instantaneous_error}
\end{equation}
with $\bs{\varphi}'_{\ell} = \bs{\varphi}'(\Dplusrell[\bs{u}_\ell])$ and under the assumption that only neurons in the topmost layer $L$ contribute directly to the cost function $C$. \\
The instantaneous target error for an \gls{mse} or \gls{ce} loss in the last layer is given by
\begin{equation}
  \bs{e}^{\mathrm{inst}}_{L} = - \beta \frac{\partial C}{\partial \bs{\mathcal{D}}^{+}_{\bs{\tau}^{\mathrm{r}}_{L}}\left\{\bs{u}_L\right\}} = \bs{\varphi}'_{L} \odot \beta(\bs{r}^{\mathrm{trg}} - \bs{r}_{L}) \; .
  \label{eq:last_layerwise_instantaneous_error}
\end{equation}

\subsection{Detailed parameter dynamics}
\label{sec:meth-parameterdynamics}
Using the layerwise instantaneous errors from \cref{eq:layerwise_instantaneous_error} and the layerwise target errors from \cref{eq:last_layerwise_instantaneous_error} we can express the neuronwise error dynamics in \cref{eqn:error_dynamics_neuronwise} in vector form as
\begin{align}
  \bs{e}_{\ell} &= \Dplusmell[\Iminusrell[\bs{e}_{\ell}^{\mathrm{inst}}]] \label{eqn:error_dynamics_layerwise} \\
  &=
  \begin{cases}
    \Dplusmell[\Iminusrell[\bs{\varphi}'_{\ell} \odot \bs{W}_{\ell+1}^{\TT} \bs{e}_{\ell+1} ]] & \text{for $\ell < L$} \\
    \Dplusmell[\Iminusrell[\bs{\varphi}'_{L} \odot \beta(\bs{r}^{\mathrm{trg}} - \bs{r}_{L})]] & \text{for $\ell = L$} \; .
    \end{cases} \nonumber
\end{align}
Parameter dynamics are derived from gradient descent on the energy function $E$ (\cref{eqn:GLEthetadot}):
\begin{equation}
  \bs{\dot{\theta}}_{\ell} \propto - \nabla_{\bs{\theta}_{\ell}} E = - \left(\frac{\partial E}{\partial \bs{\theta}_{\ell}}\right)^{\TT} = - \sum_{k\in{\{0,\ldots,L\}}} \bs{e}_{k} \left(\frac{\partial \bs{e}_{k}}{\partial \bs{\theta}_{\ell}}\right)^{\TT} \; .
\end{equation}
Depending on the specific parameter $\bs{\theta}_{\ell}$, we can work out its direct influence on the error via the partial derivative $\frac{\partial \bs{e}_{\ell}}{\partial \bs{\theta}_{\ell}}$ and obtain
\begin{align}
  \bs{\dot{W}}_{\ell} &\propto - \nabla_{\bs{W}_{\ell}} E = - \bs{e}_{\ell} \left(\frac{\partial \bs{e}_{\ell}}{\partial \bs{W}_{\ell}} \right)^{\TT} = \bs{e}_{\ell}  \bs{r}^{\TT}_{\ell-1}\; , \label{eqn:wdotmeth} \\
  \bs{\dot{b}}_{\ell} &\propto - \nabla_{\bs{b}_{\ell}} E = - \bs{e}_{\ell} \left(\frac{\partial \bs{e}_{\ell}}{\partial \bs{b}_{\ell}} \right)^{\TT} = \bs{e}_{\ell} \; , \label{eqn:bdotmeth} \\
  \bs{\dot{\tau}}^{\mathrm{m}}_{\ell} &\propto - \nabla_{\btaum_{\ell}} E = - \bs{e}_{\ell} \left(\frac{\partial \bs{e}_{\ell}}{\partial \btaum_{\ell}} \right)^{\TT} = - \bs{e}_{\ell} \odot \bs{\dot{u}}_{\ell} \; , \label{eqn:taumdotmeth} \\
  \bs{\dot{\tau}}^{\mathrm{r}}_{\ell} &\propto - \nabla_{\btaur_{\ell}} E = - \bs{e}_{\ell+1} \left(\frac{\partial \bs{e}_{\ell+1}}{\partial \btaur_{\ell}} \right)^{\TT} = \bs{e}^{\mathrm{inst}}_{\ell} \odot \bs{\dot{u}}_{\ell} \; \label{eqn:taurdotmeth} .
\end{align}

\subsection{GLE approximates AM/BPTT in real time}
\label{subsec:GLE_approximates_AM}

In \nameitref{sec:adjoint} we discuss the link between \gle errors $\bm e$ (\cref{eqn:erec}) and adjoint variables $\bm \lambda$ (\cref{eqn:lambda}):
\begin{align}
    \begin{split}
        \bs \lambda_\ell(t) &= \Iplusmell[\Dminusrell[\bs \varphi'_\ell  \odot \bs W_{\ell+1}^\TT \bs \lambda_{\ell+1}]](t) \\
                            &\approx \Dplusmell[\Iminusrell[\bs \varphi'_\ell  \odot \bs W_{\ell+1}^\TT \bs e_{\ell+1}]](t) = \bs e_\ell(t) \; .
    \end{split}
\end{align}
For a detailed derivation of the adjoint equations, we refer to the Supplement.
In the following, we provide a detailed analysis of this relation in Fourier (angular frequency) space.

For a linear system with input $x$ and output $y$, their relation in Fourier space is defined by the transfer function $H(\omega) = y(\omega)/x(\omega)$.
In our case, these transfer functions correspond to the Fourier transforms of the prospective and retrospective operators, which we denote as $\widehat{\mathcal I}$ and $\widehat{\mathcal D}$.
We first note the following identities:
\begin{align}
    &\left(1 + \tau \frac{\dd}{\dd t}\right) e^{i \omega t + \psi} = (1 + i \omega \tau) e^{i \omega t + \psi} \; , \\
    &\frac1\tau \int_{-\infty}^t e^{-\frac{t-t'}\tau} e^{i \omega t' + \psi} \dd t' = \frac1{1 - i \omega \tau} e^{i \omega t + \psi} \; .
\end{align}
Since any signal can be written as a linear composition of such frequency components $e^{i \omega t + \psi}$, the above relation directly translate to the Fourier transforms our operators.
For a single error neuron with prospective and retrospective time constants $\taum$ and $\taur$, we can thus calculate
\begin{align}
    \begin{split}
        \iminusrhat(\omega) &= \frac{1}{1+i \omega \taur} \; , \quad \dplusmhat(\omega) = 1+i \omega \taum \; , \\
        \dminusrhat(\omega) &= 1-i \omega \taur \;\;\, , \quad \iplusmhat(\omega) = \frac{1}{1-i \omega \taum} \; .
    \end{split}
    \label{eqn:operators-omega}
\end{align}
From here, the phase shifts $\Delta \psi$ and gains $G$ of our operators become apparent.

\begin{figure}[tbp]
    \centering
    \includegraphics[width=0.48\textwidth]{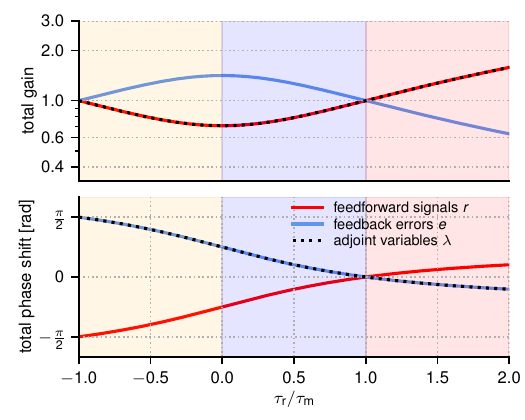}
    \caption{
        \tb{Total gain and phase shift of the respective compositions of operators in \gle and \am.}
        Expressed as a function of $\taur/\taum$ and calculated for $\omega \taum = 1$.
        Colored regions indicate different regimes of interest: negative $\taur$ (orange), effectively retrospective ($\taur < \taum$, blue), effectively prospective ($\taur > \taum$, red).
    }
    \label{fig:fourier-total}
\end{figure}

The phase shifts of $\iminusrhat$ and $\dminusrhat$ are exactly equal; the same is true for the pair $\dplusmhat$ and $\iplusmhat$ (see also \cref{fig:fourier,fig:fourier-total}):
\begin{align}
    \begin{split}
        \Delta \psi (\iminusrhat) &= \Delta \psi (\dminusrhat) = - \arctan(\omega \taur) \; , \\
        \Delta \psi (\dplusmhat)  &= \Delta \psi (\iplusmhat) = \arctan(\omega \taum) \; .
    \end{split}
\end{align}
Thus, in the combinations in which they appear in the \gle errors and adjoint equations, they induce identical phase shifts:
\begin{align}
    \begin{split}
        \Delta \psi ( \dplusmhat \iminusrhat ) &= \arctan(\omega \taum) - \arctan(\omega \taur) \\
                                           &= \Delta \psi (\iplusmhat \dminusrhat ) \; .
    \end{split}
\end{align}

The gains of the operators are given by
\begin{align}
    \begin{split}
        G(\dplushat) = G(\dminushat) = \sqrt{1+(\omega \tau)^2} \; , \\
        G(\iplushat) = G(\iminushat) = \frac{1}{\sqrt{1+(\omega \tau)^2}} \; .
    \end{split}
\end{align}
This means that the \gle errors have an inverse frequency-dependent gain compared to the adjoint variables:
\begin{equation}
    G(\dplusmhat \iminusrhat) = \frac1{G(\iplusmhat \dminusrhat)} = \frac{\sqrt{1 + (\omega \taum)^2}}{\sqrt{1 + (\omega \taur)^2}} \; .
\end{equation}
However, the ratio of \gle and \am gains is bounded by the ratio of prospective and retrospective time constants, so that the discrepancy induced by a \gle error neuron can never exceed~$\left(\taum/\taur\right)^2$.

These results extend to vectors $\bs \lambda_\ell$ and $\bs e_\ell$, where neuronal time constants $\{\taum,\taur\}$ are replaced by the corresponding entries in $\{\btaumell, \btaurell\}$.
As the operators are linear, the above considerations apply straightforwardly to inputs with arbitrary frequency spectra, for which the operators can simply be written as convolutions over frequency space with the expressions in \cref{eqn:operators-omega}.
Overall this shows how, by inducing phase shifts that are identical to \am for every individual frequency component of the signal, \gle errors produce parameter updates that are always in phase with the correct gradients.
Even though their respective frequency-specific gains are inverted with respect to \am, their mismatch is bounded and their sign is conserved.
Therefore, despite distortions at higher frequencies, \gle parameter updates remain well-aligned with their true gradients.
As in \gle the propagation of (feedback) errors and (feedforward) signals is governed by the same operators, both can be easily implemented by leaky integrator neurons with prospective output dynamics.

\subsection{Prospectivity through adaptation}
\label{sec:adaptation}

In general, when neurons have additional variables that couple negatively into the membrane potential, such as certain voltage-gated ionic currents in the Hodgkin-Huxley model, or adaptation currents such as in Izhikevich \cite{izhikevich2003simple} or AdEx \cite{brette2005adaptive} models, prospectivity on various time scales can naturally emerge.
The intuition behind the phenomenon is as follows: an additional variable that performs a low-pass filter over either the neuronal input or the membrane potential produces a negative phase shift and attenuates high-frequency components; if subtracted from the membrane, it has the opposite effect, namely inducing a positive phase shift and increasing the gain of higher-frequency components, thus acting similarly to the prospective operator $\dplus$.
Here, we demonstrate prospectivity in leaky integrator neurons with two different kinds of adaptive currents.

\subsubsection{Voltage-dependent adaptive current}

Consider the 2-variable neuron model
\begin{align}
    \begin{split}
        \taum \dot{u} &= -u + IR - w R \; , \\
        \tauw \dot{w} &= - w + \gamma_{u} u \; ,
    \end{split}
    \label{eqn:adaptive_membrane}
\end{align}
where $u$ is the membrane potential, $w$ the adaptive current with time constant $\tauw$, and $\gamma_{u}$ a coupling factor.
Without loss of generality, we assume $R = 1$ for the membrane resistivity.

\begin{figure}[t!]
    \centering
    \includegraphics[width=0.48\textwidth]{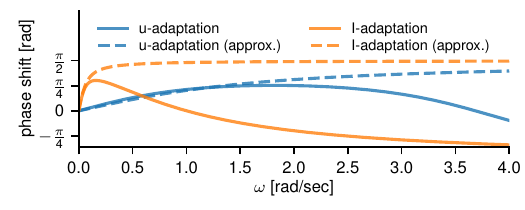}
    \caption{
        \tb{Adaptation enables prospectivity.}
        The first-order analytical approximations are shown with dashed lines.
        Simulation parameters: $\taum = 1$, $\tauw = 0.9$, $\gamma_{u} = 10 \taum/ \tauw$ and $\gamma_{I} = (\taum + 0.9\tauw)/(\taum + \tauw)$.
    }
    \label{fig:adapt}
\end{figure}

As in \nameitref{subsec:GLE_approximates_AM}, we now seek the transfer function $H(\omega) = u(\omega) / I(\omega)$.
To this end, we can simply rewrite the above equations using our differential operators:
\begin{align}
    \begin{split}
        \dplusm[u] &= I - w \; , \\
        \dplusw[w] &= \gamma_{u} u \; .
    \end{split}
\end{align}
We can now apply the Fourier transform to both equations (using \cref{eqn:operators-omega} for the operators) and solve for $u(\omega)$, which yields the transfer function
\begin{equation}
    H(\omega) = \frac{1 + i \omega \tauw}{(1 + i \omega \tauw)(1 + i \omega \taum) + \gamma_{u}} \; .
\end{equation}
The neuronal phase shift is given by $\Delta \psi = \arctan\left(\frac{\text{Im}(H)}{\text{Re}(H)}\right)$.
\Cref{fig:adapt} shows a positive shift (phase advance) over a broad frequency spectrum.

We follow up with an analytical argument for low frequencies.
For this model, $H$ can be approximated by a first-order expansion in $\{ \omega \taum, \omega \tauw \}$:
\begin{equation}
    H(\omega) \approx \frac{1}{1 + \gamma_{u}} + i \omega \frac{\gamma_{u} \tauw  -\taum}{(1 + \gamma_{u})^{2}} \; ,
\end{equation}
which yields a phase shift of
\begin{equation}
    \Delta \psi(\omega) = \arctan \left( \omega \frac{\gamma_{u} \tauw - \taum}{1 + \gamma_{u}}  \right) \; .
\end{equation}
This makes it apparent that for $\gamma_{u} > \frac{\taum}{\tauw}$ the phase shift is positive, so the neuron is prospective.

\subsubsection{Input-dependent adaptive current}
With a small change, the above 2-variable model can be made directly adaptive to the input current:
\begin{align}
    \begin{split}
        \taum \dot{u} &= -u + IR - wR \; , \\
        \tauw \dot{w} &= - w + \gamma_{I} IR \; .
    \end{split}
    \label{eqn:adaptive_input}
\end{align}
The transfer function is now given by
\begin{align*}
    H(\omega) = \frac{1 - \gamma_{I} + i \omega \tauw}{(1 + i \omega \tauw)( 1 + i \omega \taum)} \; ,
\end{align*}
which also yields a phase advance for lower input frequencies \cref{fig:adapt}.

As above, we can do a first-order approximation of $H$ in $\{ \omega \taum, \omega \tauw \}$:
\begin{align*}
    H(\omega) \approx (1 - \gamma_{I}) + i \omega (\gamma_{I} \tauw + (\gamma_{I} - 1)\taum) \; ,
\end{align*}
which yields a phase-shift of
\begin{equation}
    \Delta \psi(\omega) = \arctan \left( \omega \frac{\gamma_{I} \tauw +  (\gamma_{I} - 1)\taum}{1 - \gamma_{I}} \right) \; .
\end{equation}
Thus, for $ \frac{\taum}{\taum + \tauw} < \gamma_{I} < 1 $ the membrane is prospective with respect to its input.

\subsection{Relation to SSMs}
To illustrate the connection between \ssms and \gle, we consider a single layer of the \gle model (and drop the index $\ell$ and explicit time dependence $(t)$ for brevity).
First, we define a shorthand $\bm y$ for prospective states $\Dplusr[\Iminusm[\bm W \bm r]]$ (i.e., the neuronal output before being passed through the nonlinearity) and, as before, we use $\bm u = \Iminusm[\bm W \bm r]$ to denote membrane potentials.
For simplicity, we have ignored the biases.
We can now rewrite $\bm y$ as
\begin{align}
    \begin{split}
        \bm y &= \Dplusr[\Iminusm[\bm W \bm r]] = \Dplusr[\bm u] = \bm u + \bm \taur \odot \frac{\dd}{\dd t} \bm u \\
        &= (\bm 1 - \bm \alpha) \odot \bm u + \bm \alpha \odot \bm W \bm r \; ,
    \end{split}
\end{align}
where we plugged in the integral form of the membrane potential $\bm u$ (cf.~\cref{eqn:iminus}) in the last step, $\bm 1$ denotes the identity vector and $\bm \alpha \coloneq \bm \taur/\bm \taum$.
We use ``$\odot$'' and ``$/$'' to denote element-wise multiplication and division, respectively.
Then, the dynamics of membranes $\bm u$ and prospective states $\bm y$ are given by a system of two coupled equations
\begin{align}
    \begin{split}
        \Udot &= - \frac{\bm 1}{\Taum} \odot \bm u + \frac{\bs 1}{\Taum} \odot \bm W\bm r \; , \\
        \bm y &= (\bm 1 - \bm \alpha) \odot \bm u + \bm \alpha \odot \bm W \bm r \; .
    \end{split}
    \label{eqn:gle-uy}
\end{align}
Similarly, in \ssms, inputs $\bm r$, latent states $\bm u$ and observations $\bm y$ are related by~\cite{gu2021combining}
\begin{align}
    \begin{split}
        \Udot &= \bm A \bm u + \bm B \bm r \; ,\\
        \bm y &= \bm C \bm u + \bm D \bm r \; ,
    \end{split}
    \label{eqn:ssms-uy}
\end{align}
where $\bm A$, $\bm B$, $\bm C$ and $\bm D$ are matrices of appropriate dimensions.
Comparing equations \cref{eqn:gle-uy,eqn:ssms-uy}, it is apparent that membrane potentials in \gle correspond to latent states in \ssms by setting $\bm A = -\operatorname{diag}(\bm 1 / \Taum)$ and $\bm B = \operatorname{diag}(\bm 1 /\Taum) \bm W $.
Furthermore, prospective outputs in \gle correspond to observations in \ssms by setting $\bm C = \operatorname{diag}(\bm 1 - \bm \alpha)$ and $\bm D = \operatorname{diag}(\bm \alpha) \bm W$.
Thus, \gle forward dynamics can implement a specific (diagonal) form of \ssms.

\subsection{Simulation details}

\subsubsection{Numerical integration}

We use forward Euler integration to solve the system of implicit coupled differential equations that determine the network dynamics, which simultaneously includes neuron (membrane potentials for the different compartments) and parameter (synaptic weights, neuronal biases and time constants) dynamics.

Recall the neuronal dynamics of our networks, characterized by feedforward signals at time $t$ (we drop the $(t)$ here for brevity)
\begin{align}
    \Rell &= \bm \varphi(\Dplusrell[\Uell]) \\
          &= \bm \varphi\left(\Dplusrell[\Iminusmell[\Well \bm r_{\ell-1} + \Bell + \Eell]]\right)
\end{align}
with feedback errors
\begin{align}
    \Eell = \Dplusmell[\Vell] = \Dplusmell[\Iminusrell[\Einstell]] \; .
\end{align}
We first approximate the temporal derivatives of the membrane potentials $\Uell(t)$ by a finite difference
\begin{align}
    \Uelldot(t) &= \left[ \frac{1}{\Taumell} \odot \left(-\Uell + \Well \bm r_{\ell-1} + \Bell + \Eell\right) \right](t) \\
                &\approx \frac{\Uell(t + \dd t) - \Uell(t)}{\dd t} \; ,
\end{align}
where the fraction is understood to be component-wise.
This equation can then be solved for the membrane potentials at the next time step:
\begin{align}
    \Uell(t + \dd t) &= \left[ \Uell + \frac{\dd t}{\Taumell} \odot \left(-\Uell + \Well \bm r_{\ell-1} + \Bell + \Eell \right) \right] (t)
\end{align}
Similarly, we can approximate the derivative of the error neuron potentials $\Vell(t)$
and solve for the potentials at the next time step:
\begin{align}
    \Vell(t + \dd t) &= \left[ \Vell + \frac{\dd t}{\Taurell} \odot \left(-\Vell + \Einstell\right) \right] (t)
\end{align}
The learning dynamics (\cref{eqn:wdotmeth,eqn:bdotmeth,eqn:taumdotmeth,eqn:taurdotmeth}) are discretized accordingly.
For example, the update of the synaptic weights $\Well$ is given by
\begin{equation}
    \Well(t + \dd t) = \biggl[ \Well + \underbrace{\frac{\dd t}{\TauWell}}_{\EtaWell} \odot \,\Eell \bm r^{\TT}_{\ell-1} \biggl](t) \; .
\end{equation}
This system of equations is summarized in \cref{alg:GLE}.

\begin{algorithm}
  \small
  \caption{Forward Euler simulation of GLE network}
  \begin{algorithmic}
    \State initialize network parameters $\bm \theta = \{\bm W, \bm b, \Taum, \Taur\}$
    \State initialize network states at $t=0$: $\bm u(0)$, $\bm v(0)$, $\bm r(0)$, $\bm e(0)$
    \For{time step $t$ in $[0, T]$ with step size $\dd t$}
      \For{layer $\ell$ from $1$ to $L$}
        \If{$\ell = L$}
        \State calculate instantaneous target error:
        \State $\bs{e}^{\mathrm{inst}}_{L}(t) \gets \beta\bs{\varphi}'_{L}(t) \odot (\bs{r}^{\mathrm{trg}}(t) - \bs{r}_{L}(t))$
        \Else
        \State propagate feedback error signals:
        \State $\bs{e}^{\mathrm{inst}}_{\ell}(t) \gets \bs{\varphi}'_{\ell}(t) \odot \bs{W}_{\ell+1}^{\TT}(t) \bs{e}_{\ell+1}(t)$
        \EndIf
        \State sum input currents:
        \State $\quad \bs I_\ell(t) = \bs{W}_{\ell}(t) \bs{r}_{\ell-1}(t) + \bs{b}_{\ell}(t) + \gamma\bs{e}_{\ell}(t)$
        \State approximate membrane potential derivatives:
        \State $\quad\Delta \bs{u}_{\ell}(t)  \gets {(\btaum_{\ell}(t))}^{-1} \odot \left(-\bs{u}_{\ell}(t) + \bs I_{\ell}(t)\right)$
        \State update membrane potentials:
        \State $\quad\bs{u}_{\ell}(t + \dd t) \gets \bs{u}_{\ell}(t) + \dd t \Delta \bs{u}_{\ell}(t)$
        \State approximate error potential derivatives:
        \State $\quad\Delta \bs{v}_{\ell}(t) \gets {(\btaur_{\ell}(t))}^{-1} \odot \left(-\bs{v}_{\ell}(t) + \bs{e}^{\mathrm{inst}}_{\ell}(t)\right)$
        \State update error potentials:
        \State $\quad\bs{v}_{\ell}(t + \dd t) \gets \bs{v}_{\ell}(t) + \dd t \Delta \bs{v}_{\ell}(t)$
        \State update prospective error potentials:
        \State $\quad\bs{e}_{\ell}(t + \dd t) \gets \bs{v}_{\ell}(t) + \btaum_{\ell}(t) \odot \Delta \bs{v}_{\ell}(t)$
        \State calculate prospective outputs:
        \State $\quad\bs{r}_{\ell}(t +\dd t) \gets \bs{\varphi}\left(\bs{u}_{\ell}(t) + \btaur_{\ell}(t) \odot \Delta \bs{u}_{\ell}(t)\right)$
        \State update synaptic weights:
        \State $\quad\bs{W}_{\ell}(t +\dd t) \gets \bs{W}_{\ell}(t) + \eta_{W} \bs{e}_{\ell}(t) \bs{r}^{\TT}_{\ell-1}(t)$
        \State update biases:
        \State $\quad\bs{b}_{\ell}(t + \dd t) \gets \bs{b}_{\ell}(t) + \eta_{b} \bs{e}_{\ell}(t)$
        \State update membrane time constants:
        \State $\quad\bs{\tau}_\ell^{\mathrm{m}}(t + \dd t) \gets \bs{\tau}_\ell^{\mathrm{m}}(t) - \eta_{\tau} \bs{e}_{\ell}(t) \odot \Delta \bs{u}_{\ell}(t)$
        \State update prospective time constants:
        \State $\quad\bs{\tau}_\ell^{\mathrm{r}}(t +\dd t) \gets \bs{\tau}_\ell^{\mathrm{r}}(t) + \eta_{\tau} \bs{e}^{\mathrm{inst}}_{\ell}(t) \odot \Delta \bs{u}_{\ell}(t)$
      \EndFor
    \EndFor
  \end{algorithmic}
  \label{alg:GLE}
\end{algorithm}

Note that in addition to the nudging strength $\beta$ that scales the target error, we introduce another parameter $\gamma$ to scale the coupling between the apical dendritic and somatic compartments, thus controlling the feedback of the error signals $\bm e$ into the somatic potentials $\bm u$.

\subsubsection{General simulation details for the \gle networks}

\begin{table*}[!htbp]
    \caption{
        Neuron, network and training parameters used to produce the results shown in \cref{fig:largescale}.
    }
    \label{tab:largescale-params}
    \begin{center}
    \begin{threeparttable}
        \footnotesize
        \begin{tabular}{clcccc}
            &&&&& \\[-0.7em]
            & & \textbf{MNIST1D} & \textbf{MNIST1D} & \textbf{MNIST1D} & \textbf{GSC} \\
            \textbf{Symbol} & \textbf{Parameter name} & \textbf{GLE (15k)} & \textbf{GLE (42k)} & \textbf{LagNet} + \textbf{LE} & \textbf{GLE} \\
            \toprule
            &&&&& \\[-0.7em]
            \multicolumn{5}{l}{\textbf{Dynamic parameters} in arbitrary units} \\ [0.2em]
            $\mathrm{d}t$ & simulation time step & $0.2$ & $0.2$ & $0.2$ & $0.05$ \\
            $T$ & sample duration & $72 = 360\,\mathrm{d}t$ & $72 = 360\,\mathrm{d}t$ & $72 = 360\,\mathrm{d}t$ & $41 \text{mfs/mfcc} = 820\,\mathrm{d}t$ \\
            $\taum$ & membrane time constants    & $[1.2, 1.2, 0.6]$\tnote{1} & [$1.2, 1.2, 0.6]$\tnote{1} & $1.0$ & $[2.4, 0.6, 1.2, 1.8, 2.4]$\tnote{1} \\
            $\taur$ & prospective time constants & $[1.2, 0.2, 0.2]$\tnote{1} & [$1.2, 0.2, 0.2]$\tnote{1} & $1.0$ & $[2.4, 0.1, 0.1, 0.1, 0.1]$\tnote{1} \\
            \midrule
            &&&&& \\[-0.7em]
            \multicolumn{5}{l}{\textbf{Network parameters}} \\ [0.2em]
            $\varphi_{\ell}$ & activation                & $\tanh$ & $\tanh$ & $\tanh$ & $\tanh$ \\
            $\varphi_{L}$ & output activation         & $\mathrm{softmax}$ & $\mathrm{softmax}$ & $\mathrm{softmax}$ & $\mathrm{softmax}$ \\
            $\mathcal{C}$ & cost        & cross-entropy & cross-entropy & cross-entropy & cross-entropy \\
            & architecture              & GLE FC & GLE FC & LagNet\tnote{2} + LE-MLP & GLE FC \\
            & number of hidden layers   & $6$ & $6$ & $2$ & $3$ \\
            & input size                & $1$ & $1$ & $10$ & $32$ \\
            & hidden layer size         & $17 + 18 + 18$ & $30 + 30 + 30$ & $60 + 60$ & $150 + 150 + 150 + 150 + 150$ \\
            & output layer size         & $10$ & $10$ & $10$  & $12$ \\
            $\beta$ & nudging strength      & $1.0$ & $1.0$ & $1.0$ & $1.0$\\
            $\gamma$ & apical-somatic coupling & $0.0$ & $0.0$ & $0.0$ & $0.0$\\
            $\eta_{W,b}$ & learning rate   & $0.01$ & $0.005$ & $0.005$ & $0.005$ \\
            $W_{\mathrm{init}}, b_{\mathrm{init}}$ & initial weights \& biases & \multicolumn{4}{c}{$\mathcal{U}(-\sqrt{k}, \sqrt{k})$ with $k=1/\text{fan-in}$\tnote{3}} \\
            & optimizer                 & \multicolumn{4}{c}{Adam~\cite{kingma2014adam} with PyTorch's default parameters} \\
            & lr scheduler              & \multicolumn{3}{c}{ReduceOnPlateau} & ReduceOnPlateau \\
            & lr scheduler params        & \multicolumn{3}{c}{$\gamma=0.5$, patience=2} & $\gamma=0.9$, patience=3 \\
            \midrule
            &&&&& \\[-0.7em]
            \multicolumn{5}{l}{\textbf{Training parameters}} \\[0.2em]
            & batch size            & 100 & 100 & 100 & 256 \\
            &\# training epochs     & 150 & 150 & 150 & 200 \\
            &\# train samples       & 4000 & 4000 & 4000 & 36923 \\
            &\# validation samples  & 1000 & 1000 & 1000 & 4445 \\
            &\# seeds               & 10 & 10 & 10 & 10 \\
            \bottomrule
        \end{tabular}
        \begin{tablenotes}
            \footnotesize
            \item[1] Time constants of the three neuron populations in each hidden layer.
            \item[2] The LagNet has a single input that projects onto four hidden layers with ten neurons each.
                     The layers are connected with identity weight matrices and not learned.
                     Each hidden layer uses linearly distributed membrane time constants between $\taum = [\mathrm{d}t = 0.2, \ldots, 1.8]$ and equal $\taur = \mathrm{d}t = 0.2$.
                     This allows for a mapping of the temporal sequence to spatially distributed representations.
                     The output of the last hidden layer is fed into the plastic (G)LE network.
        \end{tablenotes}
    \end{threeparttable}
    \end{center}
\end{table*}

Our networks are implemented in Python using the PyTorch library.
All vector quantities with a layer index $\ell$ are implemented as PyTorch tensors.
Neuronal activation functions are hyperbolic tangents.

For the cost function $C$, we either use a \gls{mse}
\begin{equation}
    C = \frac{1}{2} \sum_{i\in L} {(r^{\mathrm{trg}}_i - r_i)}^2
\end{equation}
with outputs
\begin{equation}
    r_{i} = \varphi \left(\mathcal{D}^{+}_{\taur_{i}} \left\{u_i\right\}\right)
\end{equation}
or a \gls{ce} loss
\begin{equation}
    C = - \sum_{i\in L} r^{\mathrm{trg}}_i \log(r_i)
\end{equation}
with softmax outputs
\begin{equation}
    r_i = \exp\left(\mathcal{D}^{+}_{\taur_{i}}\{u_i\}\right) / \sum_{j} \exp\left(\mathcal{D}^{+}_{\taur_{j}}\{u_j\}\right)
\end{equation}
depending on the task.
In both cases, the $\bs{r}^{\text{trg}}$ is either a target rate or a one-hot vector encoding the target class.

In the two teacher-student setups, the target output rate is produced by a teacher network with the same architecture as the student, but with fixed weights and time constants.
Student networks trained with different algorithms are initialized with identical but randomized parameters.
The outputs of teacher and student are compared via the \gls{mse} loss.

Time units are fixed but arbitrary, so we can treat time as unit-free in the following.

\subsubsection{GLE chain (\cref{fig:chain})}

The input signal is a smoothed square wave of amplitude $1$ and period $T = 4$, with a simulation time step of $\dd t = 0.01$.
Batches of 100 samples are generated by shifting the input randomly in time by values between $0$ and $T/2$.
Teacher weights and time constants are $w_0 = 1$, $w_1 = 2$, $\taum_0 = 1$, $\taum_2 = 2$, $\taur_0 = \taur_1 = 0.1$.
The two student networks trained with \gls{gle} and instantaneous \gls{bp} use a learning rate of $0.0001$ and a nudging strength of $\beta = 0.01$.
The three student networks trained with truncated \gls{bptt} and truncation windows of $1$, $2$ and $4$ use learning rates of $0.01$, $0.02$, and $0.04$, respectively.
All networks use the Adam optimizer \cite{kingma2014adam} with default parameters for momentum and decay rates for faster convergence.

\begin{figure*}[htb]
    \centering
    \includegraphics[width=0.98\textwidth]{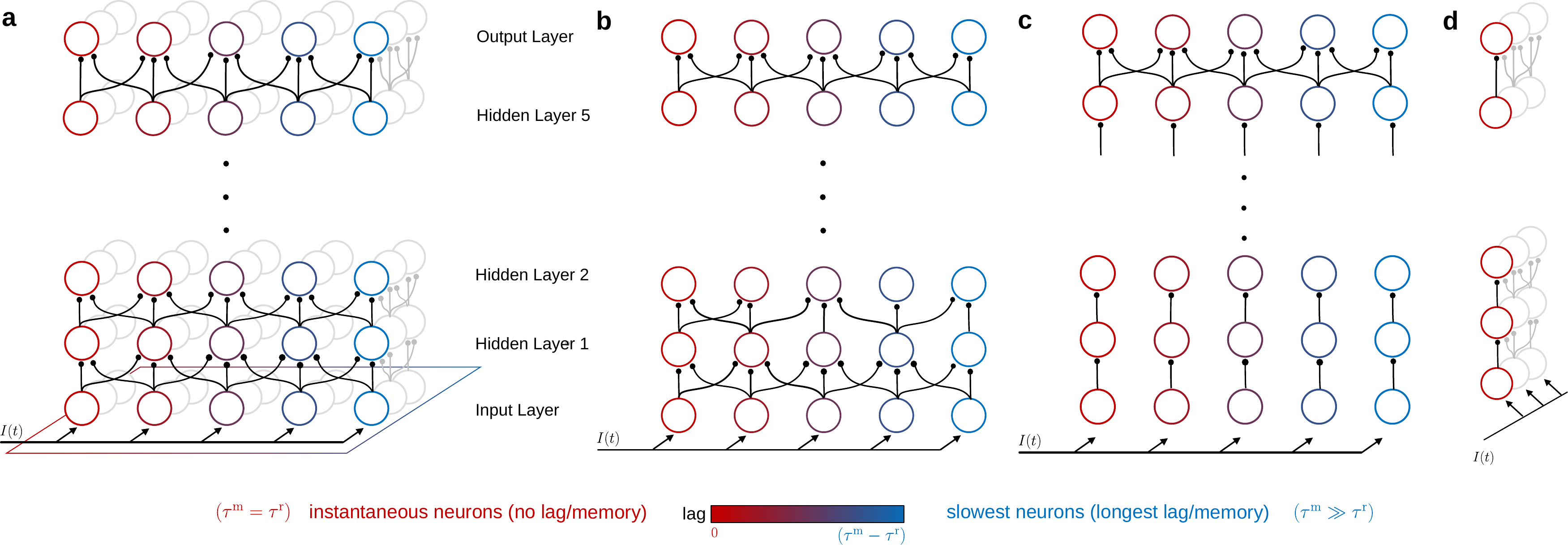}
    \caption{
      \textbf{Architectures of the simulated networks.}
      \textbf{(a)} The \gle network used to produce the results in \cref{fig:largescale}e where multiple input channels project to successive hidden layers with different time constants.  All other networks can be viewed as using a subset of this architecture.
      \textbf{(b)} The \gle network trained on the MNIST-1D dataset uses a single (scalar) input channel.
      This architecture was used to produce the results in \cref{fig:largescale}d and \cref{fig:noise_scaling}a-c.
      \textbf{(c)} "LagNet" architecture used to produce the results in \cref{fig:noise_scaling}f.
      It also receives a single input channel, but the weights of the four bottom layers are fixed to identity matrices.
      This induces ten parallel channels that process the input with different time constants. The \gls{mlp} on top of this LagNet uses instantaneous neurons and is trained with GLE (which in the case of equal time constants \(\taum = \taur\) reduces to \gls{le} as described in~\cite{haiderLatentEquilibriumUnified2021}).
      \textbf{(d)} The \gle network used to tackle spatial problems as per \cref{fig:noise_scaling}g. All neurons are instantaneous, and the network is equivalent to a \le network.
    }
    \label{fig:architectures}
\end{figure*}

\subsubsection{Small networks (\cref{fig:small2})}
The input signal is a composition of three sine signals with frequencies $\omega^{1} = 0.49 $, $\omega^{2} = 1.07 $, and $\omega^{3} = 1.98 $.
To isolate the effect of backpropagated errors over multiple layers, only the bottom weights are initialized randomly and then learned.
The loss depicted in \cref{fig:small2}f is filtered with a rectangular window of width 1.
The \am (see also \cref{alg:AM} in the Supplement) uses a truncation window of duration $3$.
Learning rates are $\eta_{\text{GLE}} = 2 \times 10^{-2}$ and $\eta_{\text{AM}} = 7 \times 10^{-2}$.
We use a teacher nudging strength of $\beta = 1$, and a somato-dendritic coupling of $\gamma = 10^{-3}$.
The frequency resolution in \cref{fig:small2}c,d is given by $\frac{2 \pi}{n \dd t}$, where $n$ is the window length over which the frequency spectrum is calculated.
To calculate the frequency spectrum of the signals before and during the early learning phase, the network is simulated for $n=8000$ steps with fixed weights (yielding a frequency resolution of $\approx 0.078$).

\subsubsection{MNIST-1D (\cref{fig:largescale}d/f and \cref{fig:noise_scaling}a-c/e/f)}

The \gle networks in \cref{fig:largescale}d use a layered architecture with a single input neuron in the first layer, six fully connected hidden layers and an output layer with ten neurons.
Each hidden layer has three different, equally-sized populations of neurons, each with different time constants $\taum$ and $\taur$, corresponding to a fast, medium, and slow population.
The time constants of the slow population \(\tau_{\mathrm{slow}}\) were chosen such that \(2.5 \sum_{\ell \in [1, L]} \tau_{\mathrm{slow}} = L \tau_{\mathrm{slow}} \approx \,\text{pattern duration} = 90\,\dd t\).
This is inspired by~\cite{weidelWaveSenseEfficientTemporal2021} where the factor \(2.5\) was determined empirically.
The (G)LE networks in \cref{fig:largescale}f use a custom, non-plastic LagNet of purely retrospective neurons ($\taur=0$) with different membrane time constants $\taum$ that allow for an approximate, partial mapping of the temporal sequence to the neuronal space.

In the network scaling experiments for \cref{fig:noise_scaling}a-c, the number of layers was chosen from \(\{2, 4, 6, 8, 10\}\), and the total number of neurons per hidden layer from \(\{32, 64, 128, 256, 512\}\).

In the experiments assessing robustness to parameter noise in \cref{fig:noise_scaling}d, a six layer network was used and each neuron's time constants, $\taum$ and $\taur$, were sampled from a normal distribution centered around the original time constants of the respective neuron population, with standard deviations chosen from \(\{0, 0.01, 0.05, 0.1\}\).

In \cref{fig:noise_scaling}e, we added additional correlated noise to the rates of the feedforward neurons including the input and output layers.
The noise was generated from a normal distribution with mean $0$ and standard deviation $2\pi\{0.01, 0.02, 0.05, 0.1, 0.2\}$ for each neuron and time step in the original sequence.
It was correlated over the sequence length of 72 time steps using a Gaussian kernel with a standard deviation of $2$ time steps.
Then, it was interpolated to the new sequence length of 360 time steps and finally added to the individual neuron rates at each time step.
The rescaling factor of $2\pi$ ensures that the standard deviation of the correlated noise corresponds to the reported standard deviations.

For \cref{fig:noise_scaling}f,g we added Gaussian noise to the feedback weight matrices $\bm B_{\ell} = \bm{W}_{\ell+1}^{\TT} + \mathcal{N}(0, \sigma^2)$, sampled from a normal distribution with mean $0$ and standard deviation $\sigma = \{0.01, 0.05, 0.1, 0.2\}$.
This noise is regenerated before every training epoch, except for the \gls{fa} experiment where it is fixed for the entire training.
The alignments reported in \cref{fig:noise_scaling}g are calculated as the angle between the original and the noisy feedback weight matrix, $\angle(\bm A, \bm B) = \arccos\left(\frac{\langle \bm A, \bm B \rangle}{\|\bm A\|_2 \|\bm B\|_2}\right)$, where $\langle \cdot, \cdot \rangle$ denotes the inner product.

Detailed parameters for the different \gls{gle} networks are given in \cref{tab:largescale-params}.

The original publication of the dataset~\cite{greydanusScalingDeepLearning2020} uses a downsampled version of the sequences to 40 time steps (baseline results with index 0 in \cref{fig:largescale}f).
For a more continuous input, we interpolate the original sequences of length 72 to 360 time steps.
To ensure a fair comparison, we also retrain the baseline architectures on these higher-resolution sequences (\cref{fig:largescale}d).
An overview over the different reference models, their size and final performance on the MNIST-1D dataset is shown in \cref{tab:mnist1d-results} in the Supplement.

\subsubsection{Google Speech Commands}

The \gls{gle} networks use a layered architecture with 32 input neurons in the first layer, 3 fully-connected hidden layers and an output layer with twelve neurons.
Each hidden layer has five different populations of neurons, each with different time constants $\taum$ and $\taur$.
Detailed parameters for the different \gls{gle} networks are given in the last column of \cref{tab:largescale-params}.

To provide a fair comparison, the dataset (v2.12) and data augmentation process from~\cite{zhanghello2018} is used to train and test all models.
For a more continuous input, the original sequences are interpolated, increasing the number of time steps by a factor of 20.
The train/validation/test set split ratio is 80:10:10.
The training data is augmented with background noise and random time shifts of up to \SI{100}{ms}.
The length and stride of the Fourier window, the number of frequency bins and the type of Mel frequency representation is optimized for each method independently; all parameters are given in \cref{tab:gsc-results} in the Supplement.

\subsubsection{GLE for purely spatial patterns}
In \cref{fig:largescale}g we compare the validation accuracy of (G)LE (in the instantaneous case where $\taum_{i} = \taur_{i}$ for all neurons $i$) with standard \gls{bp} on a purely spatial task, namely the CIFAR-10 dataset~\cite{krizhevsky2009learning}.
In both cases, the architecture follows LeNet-5 \citep{lecun1998gradient}.
These results are taken from the original publication~\citep{haiderLatentEquilibriumUnified2021}, to which we refer for further details.

\subsubsection{Chaotic time series prediction}
For the Mackey-Glass time series prediction task, we follow the benchmark setup of \cite{yikNeurobenchFrameworkBenchmarking2025} to allow a fair comparison with their results.
The Mackey-Glass time series is generated by the following delay differential equation:
\begin{equation}
  \dot{x}(t) = \beta \frac{x(t-\tau)}{1 + x(t-\tau)^n} - \gamma x(t) \; ,
\end{equation}
with delay constant \(\tau\) and parameters \(\beta\), \(\gamma\), and \(n\).
Depending on the parameters, the Mackey-Glass system can exhibit periodic and chaotic behavior.
As in~\cite{yikNeurobenchFrameworkBenchmarking2025}, we use \(\beta = 0.2\), \(\gamma = 0.1\), \(n = 10\), and \(\tau = 17\) leading to chaotic behavior.
The Lyapunov time $\lambda$ for this system, a measure of the time scale over which two different initial conditions are expected to diverge by at least a factor of $e$, is \(\lambda = 197\).
The NeuroBench framework~\cite{yikNeurobenchFrameworkBenchmarking2025} provides a pregenerated Mackey-Glass time series of a length of 50 $\lambda$, which is used as the training set with a discretization \(75\,\text{steps/$\lambda$}\) corresponding to a time step of approximately \(2.63\).
For the simulation of our \gle networks, we require a much finer discretization of the time series, so we use a time step of \(0.2\) corresponding to \(985\,\text{steps/$\lambda$}\) using the \texttt{jitCDDE} library~\cite{ansmannEfficientlyEasilyIntegrating2018}.
Furthermore, instead of the Euler integration scheme for the simulation of our \gle networks, we use the fourth-order Runge-Kutta method which, together with the finer discretization, allows for improved numerical stability of our simulation.

The original time series of length 50 $\lambda$ is then split into 30 sequences of length 20 $\lambda$, shifted by $0.5\lambda$ time each.
The first 10 $\lambda$ of each sequence are used for training, where the network is trained to predict the next step $x(t+1)$ from input $x(t)$ (and possibly delayed inputs $x(t-\Delta)$).
After the first 10 $\lambda$, the network switches to the autoregressive mode, where the network's prediction of $x(t+1)$, i.e., the output $y(t)$, is fed back as input for the next time step and without providing the true value $x(t+1)$ as a target.
For each of the 30 sequences, a separate network with different random initialization is trained over 150 epochs and then evaluated on the last 10 $\lambda$ of the sequence using the \gls{smape} metric and finally averaged over all 30 sequences.
Our \gle networks use four input neurons, each delayed by $\Delta = 6$ and two hidden layers with 93 neurons each, that is, in total 186 hidden neurons, identical in size to the \esn baseline network.
For details on the \esn and \lstm baselines as well as the benchmark setup, we refer to the original publication~\cite{yikNeurobenchFrameworkBenchmarking2025}.

\section{Data Availability}

All datasets used in this manuscript are publicly available and do not require restricted access.
The \mnistId data used in this study are available in the public GitHub repository \url{https://github.com/greydanus/mnist1d}, see also \cite{greydanusScalingDeepLearning2020}.
The \gsc dataset is publicly accessible from the TensorFlow Speech Commands repository, \url{https://www.tensorflow.org/datasets/catalog/speech_commands}, see also \cite{wardenspeech2018}.
The CIFAR-10 dataset is available from the Kaggle \cifarIO database and \cifarIO repository, \url{https://www.kaggle.com/c/cifar-10} and \url{https://www.cs.toronto.edu/~kriz/cifar.html}. See also \cite{krizhevsky2009learning}.
The Neurobench Mackey-Glass task data generated and used in this study are available in the Neurobench repository \url{https://github.com/neurobench/neurobench}, see also \cite{yikNeurobenchFrameworkBenchmarking2025}.
No data are protected or unavailable due to data privacy laws.
Full details and references to the source data and any processed data files are provided in the Methods and the Supplementary Information.

\section{Code Availability}

Code to reproduce the results presented in this manuscript is available at \url{https://github.com/unibe-cns/gle-code}.

\FloatBarrier
\printbibliography{}
\addcontentsline{toc}{section}{References}

\section*{Acknowledgements}
We thank Simon Brandt for sharing his expertise on prospectivity in mechanistic neuron models, Timo Gierlich for the simulations of spike-based alignment learning, Maxim Kondratenko for proofreading and Walter Senn for many stimulating discussions about various aspects of dendritic computation, energy-based models, error-correcting plasticity and bio-plausible credit assignment.
We further thank Wulfram Gerstner for some nice ideas regarding the assessment of gradient propagation depths.
We gratefully acknowledge funding from
the European Union under grant agreements \#945539 (Human Brain Project, SGA3; PH, LK, MAP) and
\#101147319 (EBRAINS 2.0; PH, JJ, FB, MAP),
the Volkswagen Foundation under the call ``NEXT -- Neuromorphic Computing'' (KM) and
ESKAS for a Swiss Government Excellence Scholarship (IJ).
We would like to express particular gratitude for the ongoing support from the Manfred Stärk Foundation (MAP).
Our work has greatly benefitted from access to the Fenix Infrastructure resources, which are partially funded by the European Union’s Horizon 2020 research and innovation programme through the ICEI project under the grant agreement No.~800858 (MAP).
Furthermore, we thank Marcel Affolter, Reinhard Dietrich and the Insel Data Science Center for the usage and outstanding support of their Research HPC Cluster (BE).

\section*{Author contributions statement}
BE and PH share first authorship of the manuscript.
MAP, PH and BE conceived the core ideas and designed the project.
BE and MAP initiated the project and generated the first results.
PH developed and maintained most of the code.
PH, BE, IJ, JJ and LK performed the simulations.
PH developed the additional autoregressive experiment.
PH, KM, FB, IJ and MAP developed the proof of correspondence between \gle and \am.
All authors contributed to the development of the theory and to the writing and editing of the manuscript.
PH, FB, and MAP were responsible for the manuscript resubmissions.

\section*{Competing Interests Statement}
The authors declare no competing interests.

\newpage

\appendix
\onecolumn
\AddToHook{cmd/section/before}{\FloatBarrier}
\AddToHook{cmd/subsection/before}{\FloatBarrier}
\AddToHook{cmd/subsubsection/before}{\FloatBarrier}
\setcounter{page}{1}  %
\counterwithin*{figure}{section} %
\counterwithin*{table}{section}
\counterwithin*{equation}{section}
\renewcommand{\theequation}{\thesection.\arabic{equation}} 
\renewcommand{\thefigure}{\thesection.\arabic{figure}}
\renewcommand{\thetable}{\thesection.\arabic{table}}
\renewcommand{\figurename}{Supplementary Figure}
\renewcommand{\tablename}{Supplementary Table}
\glsresetall{}

\begin{center}
    \Huge\tb{Backpropagation through space, time and the brain}
\end{center}

\section*{Supplementary Information}

\begin{refsection}
    \section{Biological mechanisms of prospectivity}
\label{si:prospectivity}

While the retrospective (low-pass filter) operator is a common assumption about neuronal dynamics, the prospective properties of neurons are much less frequently used in computational neuroscience.
Nevertheless, they rest on solid experimental and theoretical evidence \cite{hodgkinQuantitativeDescriptionMembrane1952, pucciniIntegratedMechanismsAnticipation2007, kondgenDynamicalResponseProperties2008, plesserEscapeRateModels2000, pozzorini2013temporal, brandt2024prospective, kepecsBurstingNeuronsSignal2002}.
In the following, we provide an intuitive explanation for some of the underlying mechanisms, which sacrifices some mathematical precision in the service of brevity.

One well-established mechanism of prospectivity stems from the spiking behavior of neurons, more specifically from the reset upon threshold crossing (see, e.g., \cite{plesserEscapeRateModels2000}, but also the textbook \cite{gerstner_neuronal_2014}, Sec. 9.4.1 and Fig. 9.10).
The instantaneous firing probability of a neuron is equivalent to the population firing rate of an unconnected ensemble of such neurons.
Consider therefore a large population of identical neurons with constant stimulus that generates some membrane potential $u$.
Under independent balanced Poisson background (equal amounts excitatory and inhibitory input), the membrane potentials in this population will follow a Gaussian distribution centered around $u$.
If one introduces a threshold, those neurons that are suprathreshold spike.
Their proportion is roughly equal to the suprathreshold area of the Gaussian, which depends on $u$, and thus the firing rate is a function of $u$.
Consider now the case in which the stimulus gradually changes such that $u$ increases with time, with some speed $\udot$.
This will cause the Gaussian to gradually shift upwards.
In addition to the area that was initially above the threshold, new area enters the suprathreshold state and thus more neurons fire as they are swept across the threshold.
This happens at a rate that depends on the speed with which the distribution moves upwards, which is determined by $\udot$.
Altogether, the instantaneous firing rate is thus a function $\varphi(u, \udot)$.
The parametrization $\varphi(u + \tau \udot)$ chosen for \gle is an approximation of the more complex expression derived in \cite{plesserEscapeRateModels2000}.

A second mechanism of prospectivity lies in the well-known Hodgkin-Huxley model of action potential generation itself \cite{hodgkinQuantitativeDescriptionMembrane1952}.
The Hodgkin-Huxley model describes the evolution of the membrane potential $u$, along with three gating variables $m$, $h$ and $n$ for voltage-gated Na$^+$ and K$^+$ channels.
Because all of these equations are of type $\tau \dot x = x_0 - x$, without external input, the system decays towards the resting state $x_0$ (in case of the membrane variable, this is the leak potential).
If one injects some input, this resting state changes.
However, if the input is increased very slowly, on a time scale much longer than all time constants $\tau$ that control the reaction speed of the system (i.e., adiabatically), all variables will have sufficient time to adapt and the system will simply follow the (slowly changing) resting state.
This can continue for arbitrarily high input without the neuron ever spiking.
For an input that changes more quickly, the neuron obviously fires.
Whether the neuron fires or not thus depends on the rate with which the input, and therefore its membrane potential, changes -- that is, $\udot$.
For a more detailed description and also simulations depicting this effect, we refer to the textbook \cite{petrovici2016form} (Sec. 2.1.2. and Fig. 2.6).

Third, neuronal adaptation can also lead to prospectivity (\cite{pozzorini2013temporal, brandt2024prospective}, but see also the detailed discussion in \nameitref{sec:adaptation}).
Here, adaptation can be interpreted very generally as the existence of a second variable that is 1) described by some form of low-pass filter over the membrane and 2) couples negatively into the membrane.
This is the case for the adaptation variable in standard adaptive neuron models such as AdEx \cite{brette2005adaptive} or Izhikevich \cite{izhikevich2003simple}, but it also applies to sodium currents in the Hodgkin-Huxley model \cite{brandt2024prospective}.
A low-pass filter preserves lower frequency components while suppressing higher ones; thus, subtracting a low-pass filtered version of a signal from the signal itself suppresses lower frequencies and preserves higher ones.
The negatively coupled adaptation variable therefore acts like an inverse of a low-pass filter, which is indeed what we model with our prospective operator $\dplus$.

\section{Bio-plausibility and the weight transport problem}
\label{si:weight_transport}
Our microcircuit implementation of \gle relies on the backward weights achieving a good approximation of the weight transpose $\bm B \to \bm W^\TT$, i.e., a solution to the classic weight transport problem.
\begin{figure*}
    \centering
    \includegraphics[width=0.9\textwidth]{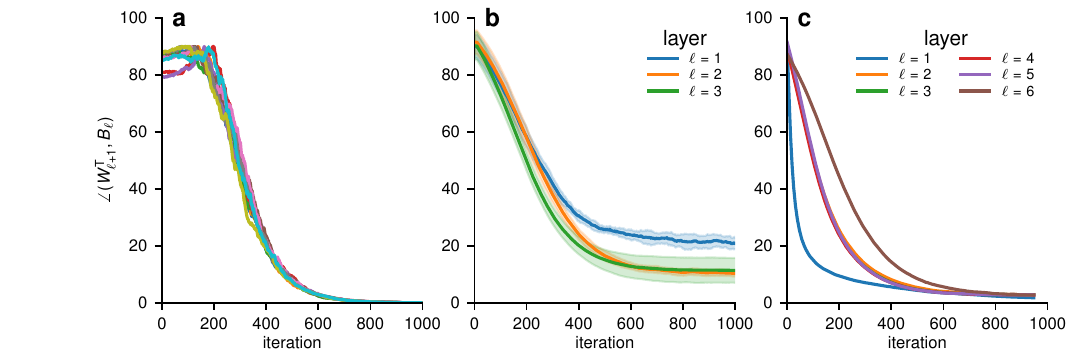}
    \caption{
      Alignment of forward and backward weights using different algorithms.
      \tb{(a)} The Kolen-Pollack algorithm~\cite{kolen1994backpropagation} applied to ten different random initializations of $\bm{B}_{\ell}$ and $\bm{W}_{\ell+1}^{\TT}$.
      \tb{(b)} Phaseless alignment learning applied to a deep microcircuit network with three hidden layers. Adapted from Fig.~3b in~\cite{maxLearningEfficientBackprojections2023}.
      \tb{(c)} Spike-based alignment learning~\cite{gierlichWeightTransportSpike2025} applied to the hidden layers of the \gle network as in \cref{fig:largescale}d.
      All of these algorithms perform weight alignment that is sufficient for optimal performance in the weight noise study from \cref{fig:noise_scaling}.
    }
    \label{fig:alignment_SI}
\end{figure*}

While \fa \cite{lillicrapRandomFeedbackWeights2014} can mitigate this problem to some extent, its capabilities are mostly confined to shallow networks.
Generally speaking, the gradients transported with \fa quickly deteriorate with increasing network depth, so the method does not scale to deeper networks.

Interestingly, this is not necessarily a problem for some recurrent networks, where propagation across a single "layer" of presynaptic partners can be sufficient for good performance and has been used explicitly for biologically plausible models of recurrent learning in both rate-based and spiking networks \cite{murrayLocalOnlineLearning2019, bellec2020solution}.
This is because in these setups, each neuron in the recurrent population is directly connected to the output, from which it receives a target error.
However, a key asset of \gle is its capability of propagating useful errors to much greater depth, so it is all the more important for it to overcome the limitations of \fa.
In the following we thus briefly summarize some very effective strategies for achieving $\bm B \to \bm W^\TT$ in both rate-based and spiking networks.
More specifically, we address the references \cite{meulemans2022minimizing, maxLearningEfficientBackprojections2023, gierlichWeightTransportSpike2025, kolen1994backpropagation, akrout2019deep} from the main text, which explicitly acknowledge the problems with \fa and provide improved, online and local solutions to the weight transport problem.

In \cite{kolen1994backpropagation}, weight updates are applied to $\bm W$ and $\bm B$ simultaneously, accompanied by a small weight decay term.
This ensures that over time, the initial weight values are forgotten and $W_{ij}$ and $B_{ji}$ converge arbitrarily close to each other (\cref{fig:alignment_SI}a).
We note that this algorithm effectively solves the weight transport problem by transporting weight updates, which is often an issue in itself; however, this is possible for \gle networks, as both forward synapses $W_{ij}$ and backward synapses $B_{ji}$ have neuron-local access to the activity $e_i$ of error and $r_j$ of representation neurons needed for these updates (cf. \cref{fig:mc} and \cref{eqn:Wdot}).

In \cite{akrout2019deep}, the weight transpose is learned by a simple Hebbian mechanism of type $\dot B_{ji} \sim r_j r_i$.
Coupled again with weight decay to counteract the diverging nature of purely Hebbian rules, this leads to $\bm B \to \bm W^\TT$ on average.

In \cite{meulemans2022minimizing} and \cite{maxLearningEfficientBackprojections2023}, independent high-frequency noise $\bm{\xi}$ is introduced on top of the informative signal in all neurons.
Such noise is ubiquitous in biological nervous systems and can be easily extracted by high-pass-filtering mechanisms in individual synapses.
These studies demonstrate how local plasticity in $B_{ji}$ can align the local noise $\xi_i$ to the recurrent noise going through the B-W-loop $B_{ji} \varphi'_i W_{ij} \varphi'_j$, ultimately yielding an approximation of either $\bm W^\TT$ for \bp (\cref{fig:alignment_SI}b) or $\bm W^+$ (the Moore-Penrose pseudoinverse) for \gn optimization.

Finally, in \cite{gierlichWeightTransportSpike2025}, near-exact weight transport is demonstrated in spiking networks as well.
Here, the partially stochastic firing of neurons is leveraged by synapses to locally estimate the spike timing cross-correlation of their adjacent neurons.
Since this cross-correlation contains all necessary information about the reciprocal synapses, both $W_{ij}$ and $B_{ji}$ can use it to converge to each other, using a simple anti-Hebbian \stdp rule (\cref{fig:alignment_SI}c).

\section{Error propagation depth analysis in the MNIST-1D experiment}

\begin{figure}[h!]
\centering
\includegraphics[width=0.9\textwidth]{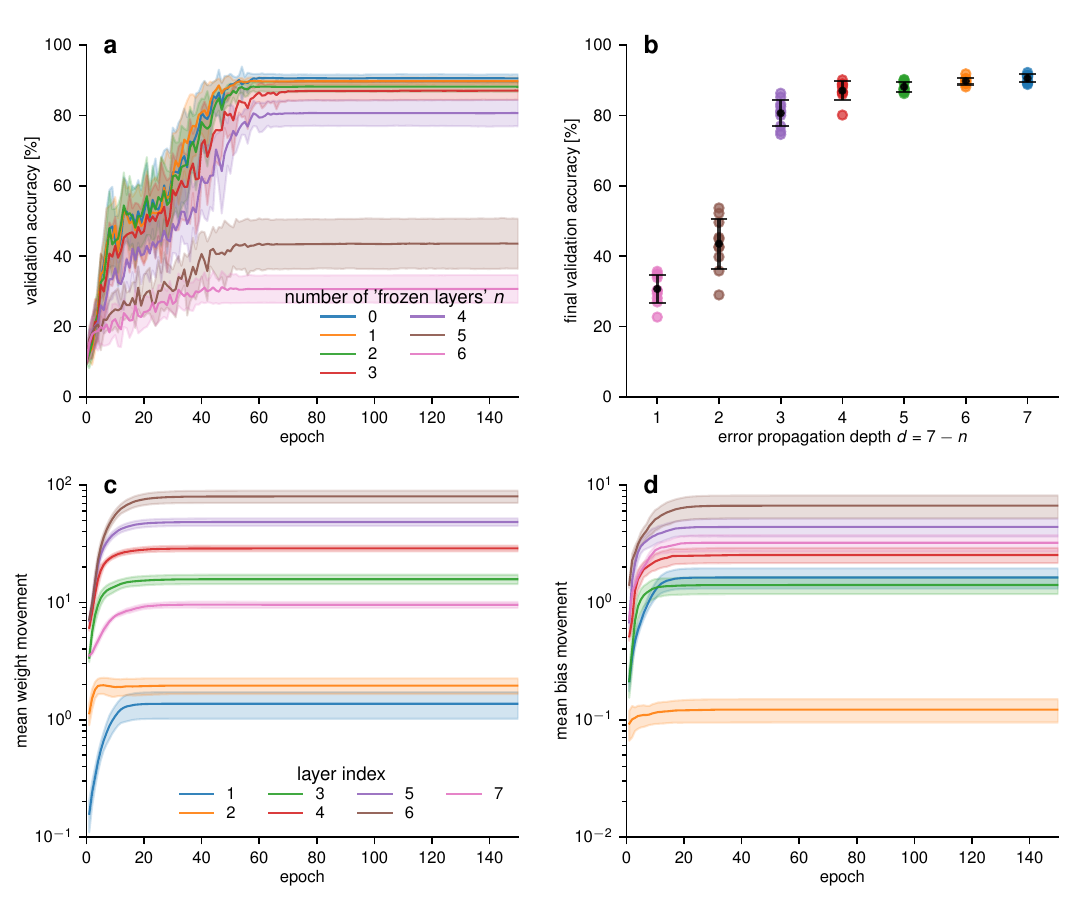}
\caption{
    \textbf{Deep error propagation is required for learning the MNIST-1D experiment.}
    \textbf{(a)} Validation accuracy (mean and standard deviation over $10$ runs) during training for a $7$-layer network with a varying number of ``frozen'' layers $n$, where layers $1, \dots, n$ are not trained.
    \textbf{(b)} Final validation accuracy as a function of the error propagation depth, defined as $d=7-n$.
    \textbf{(c)} Mean weight movement per layer, averaged over $10$ independent training runs.
    \textbf{(d)} Mean bias movement per layer, also averaged over $10$ runs.
}
\label{fig:frozen_mnist1d}
\end{figure}

To investigate the necessity of deep credit assignment, we trained $7$-layer networks on the \mnistId task (compare \cref{fig:largescale}d) while freezing the initial $n$ layers, with $n$ ranging from $0$ (all layers trained) to $6$ (only the final layer trained).
While ``shallow training'' involving only the last one or two layers ($d \leq 2$) leads to severely decreased performance (\cref{fig:frozen_mnist1d}a,b), achieving top performance percentiles requires exploiting learning across the full network depth.
This finding invalidates the hypothesis that lower layers can act as a static reservoir, demonstrating instead that deep error propagation is required for successful learning on this task.

To further investigate this conclusion, we quantify the movement for each parameter tensor (weight matrix $W_\ell$ or bias vector $b_\ell$ for layer $\ell\in\{1,\dots,7\}$) by computing the Frobenius norm of the difference from its initial state at $t=0$, i.e., $\Delta W_\ell(t) = \| W_\ell(t) - W_\ell(0) \|_F$ (here $t$ denotes training time in epochs) and equivalently for biases.
Significant weight and bias changes happen in all but the lowest two layers (\cref{fig:frozen_mnist1d}c,d), demonstrating that learning is distributed throughout most of the network's layers.
We note that the smaller movement in the final layer (7) may be partly due to its smaller number of synaptic connections, as we plot the total norm of the layer's parameter tensor.
Collectively, these experiments not only confirm that effective, deep credit assignment is a prerequisite for learning this task, but also demonstrate \gls{gle} ability to achieve it.

\section{Adjoint equations and parameter updates for the feedforward pathway in \gle networks}
\label{si:adjoint}

The general problem we aim to solve is to minimize some cost $\mathcal{C}(\bm u, \Udot, \bm \theta)$ through adaptation of the parameters $\bm \theta\in\{\bm W, \bm b, \bm \taum, \bm \taur\}$ of a dynamical system with presynaptic inputs $\bs\rin$, which is constrained by the dynamics of state variables $\bm u$.
For that, we employ the adjoint method (\gls{am}), as described, e.g., in \cite{bradleyPDEconstrainedOptimizationAdjoint2019}, in order to derive adjoint equations for the Lagrange multipliers $\bm \lambda$ and calculate the gradient of the cost functional $\mathcal{C}$ w.r.t.~the parameters $\bm \theta$.
The dynamics are described by a system of \glspl{ode} in implicit form
\begin{align}
    \bs h(\bs u, \Udot, \bs \theta, t) = \btaum \Udot +\bs  u - \bs  W \bs  \rin - \bs b= 0 \; .
    \label{eq:dynamic_constraint_SI}
\end{align}
Note that this is a special case of the \gle dynamics, where the error term $\bs e$ is zero and the solution can be written using the integral operator
\begin{equation}
    \bs u(t) = \Iminusm[\bs W \bs \rin + \bs b] \; .
\end{equation}
Recall that the \gls{gle} operators are defined as
\begin{align}
    \Iminus[\bs x(t)] &= \frac{1}{\tau} \int_{-\infty}^t \dd s \,\bs x(s) \, e^{-\frac{t-s}{\btau}}\, && \text{(leaky integrator)} \\
    \Dplus [\bs x(t)] &= \left(1 + \btau \frac{\dd}{\dd t}\right) \bs x(t)\;.  && \text{(lookahead)}
\end{align}
We define the cost functional \(\mathcal{C}\) as the integral of the instantaneous cost \(C\) over some time interval \([0, T]\)
\begin{equation}
    \mathcal{C}(\bs u, \Udot, \bs \theta) = \int_{0}^{T} C(t, \bs u, \Udot, \bs \theta)\,\dd t \; ,
\end{equation}
where the instantaneous cost can be defined as the mean-squared error between the target rate $\bs r^{\mathrm{trg}}$ and the actual rate $\bs r$ at time $t$ of some subset \(\mathcal{O}\) of all the neurons in the network $\mathcal{N}$,
$C(t, \bs u, \Udot, \bs \theta) = \frac{1}{2} \sum_{o\in\mathcal{O}} \| \bs r^{\mathrm{trg}}(t) - \bs \varphi(\bs u_{o} + \Taur_{o} \Udot_{o})\|^2$.
In practice, \(\mathcal{C}\)~could represent the integrated cost over a data sample or a window of a continuous input stream of length $T$.

In order to solve the constrained optimization problem with \gls{am}, we first introduce the Lagrangian functional $\mathcal{L}$
\begin{align}
    \mathcal{L} \equiv \int^T_{0} C\left(t,  \bs u(t), \Udot(t), \bs \theta\right)  +  \bs{\lambda}^{\TT}(t) \bs{h}(t,  \bs u(t), \Udot(t), \bs \theta) \,\dd t \;,
\end{align}
which incorporates the constraints of the dynamical system \cref{eq:dynamic_constraint_SI} into the cost functional \(\mathcal{C}\) via Lagrange multipliers $\bs \lambda(t)$.
The gradient of the Lagrangian $\mathcal{L}$ w.r.t.~the parameters $\bs{\theta}$ can be calculated using the chain rule (suppressing the explicit time-dependence from here on)
\begin{align}
    \frac{\dd\mathcal{L}}{\dd \bs \theta}  = \int^T_{0}
    \bigg\{
    \left[
    \frac{\partial C}{\partial \bs \theta}
    + \frac{\partial C}{\partial \bs u} \frac{\dd \bs u}{\dd \bs \theta}
    + \frac{\partial C}{\partial \Udot} \frac{\dd \Udot}{\dd \bs \theta}
    \right]
    + \underbrace{\frac{\dd \bs{\lambda}^{\TT}}{\dd \bs \theta} \bs  h}_{=0}
    + \bs{\lambda}^{\TT} \left(
    \frac{\partial \bs h}{\partial \bs \theta}
    + \frac{\partial \bs h}{\partial \bs u} \frac{\dd \bs u}{\dd \bs \theta}
    + \frac{\partial \bs h}{\partial \Udot} \frac{\dd \Udot}{\dd \bs \theta}
    \right)
    \bigg\}  \, \dd t \; .
\end{align}
The term proportional to $\bs h$ vanishes, as $\bs h=0$ is satisfied at all times $t$.
Most of the remaining terms can be simplified upon substituting one of the specific parameters $\bs \theta$.
In the following we derive the parameter updates for the synaptic weights $\bs W$.
We do so first for a single vector of output neurons (i.e., neurons receiving external targets; this corresponds to the output layer in a hierarchical network) and then extend the derivation to hidden layers.

\subsection{Output neurons}

We now calculate the gradient of the Lagrangian \(\mathcal{L}\) w.r.t.~the synaptic weights $\bs W$.
After integrating by parts the terms including $\dd\Udot/\dd \bs W$, that is, using $\int\dd t \,\bs x \frac{\dd \Udot}{\dd \bs W} = \int\dd t \,\frac{\dd}{\dd t} \left(\bs x \frac{\dd \bs u}{\dd \bs W}\right) - \int\dd t \,\frac{\dd \bs x}{\dd t} \frac{\dd \bs u}{\dd \bs W}$, we can rewrite the gradient as
\begin{align}
  \frac{\dd\mathcal{L}}{\dd \bs W} =
  \int^T_{0}
  \left\{
  \frac{\partial C}{\partial \bs W}
  + \bs \lambda^{\TT}\frac{\partial \bs h}{\partial \bs W}
  + \left[
  \frac{\partial C}{\partial \bs u}
  - \frac{\dd}{\dd t} \frac{\partial C}{\partial \Udot}
  +  \bs \lambda^{\TT}  \frac{\partial \bs h}{\partial \bs u}
  - {\dot{\bs{\lambda}}^{\TT}} \frac{\partial \bs h}{\partial \Udot}
  - {\bs{\lambda}^{\TT}} \frac{\dd}{\dd t}  \frac{\partial \bs h}{\partial \Udot}
  \right] \frac{\dd \bs u}{\dd \bs W} \right\} \dd t
  + {\left[ \left( \bs \lambda^{\TT} \frac{\partial \bs h}{\partial \Udot}  + \frac{\partial C}{\partial \Udot}\right)\frac{\partial \bs u}{\partial  \bs W} \right]}^{T}_{0} \;.
  \label{eq:adjoint_dyn_intermediate}
\end{align}
To set boundary terms of the dynamical system at $t=T$ to zero in \cref{eq:adjoint_dyn_intermediate}, we choose
\begin{align}
  \bs\lambda(T) = - \left.\left(\frac{\partial \bs h}{\partial \Udot}\right)^{-1}\frac{\partial C}{\partial \Udot}\right|_{t=T}\;
  \stackrel{\text{(\cref{eq:dynamic_constraint_SI})}}{=} \;\frac{\btaur}{ \btaum} \odot \bs\varphi'\left({\Dplusr}\{\bs u(T)\}\right) \odot \left(\bs r^\mathrm{trg}(T) - \bs\varphi\left({\Dplusr}\{\bs u(T)\}\right)\right)
  \label{eq:adjoint_dyn_final}
\end{align}
and obtain an initial condition for the adjoint system.
Similarly, one can account for the initial conditions $\left. \bs g(\bs u, \Udot, \bs W)\right|_{t=0}$ of the dynamical system at $t=0$ by adding another term to the Lagrangian, namely $\bs \mu^{\TT} \bs g(\bs u(0), \Udot(0), \bs W)$.
Both \(\bs \lambda\) and \(\bs \mu\) are vectors of Lagrangian multipliers that can be freely chosen because the constraints \(\bs h=0\) and \(\bs g=0\) are always satisfied by construction.
Inspecting \cref{eq:adjoint_dyn_intermediate}, we observe that we can avoid calculating the term $\frac{\dd \bs u}{\dd \bs W}$ if the terms in square brackets are zero for all times $t$, which defines the dynamics of the adjoint system:
\begin{align}
  \frac{\partial C}{\partial \bs u}
  - \frac{\dd}{\dd t} \frac{\partial C}{\partial \Udot}
  +  \bs \lambda^{\TT}  \frac{\partial \bs h}{\partial \bs u}
  - \dot{\bs{\lambda}}^{\TT} \frac{\partial \bs h}{\partial \Udot}
  - \bs{\lambda}^{\TT}\frac{\dd}{\dd t}  \frac{\partial \bs h}{\partial \Udot} = \bs 0 \;.
  \label{eq:adjoint_dynamics_general}
\end{align}
Substituting the dynamic constraint \(\bs h\) defined in \cref{eq:dynamic_constraint_SI} yields the corresponding adjoint dynamics
\begin{align}
  \btaum \frac{\dd}{\dd t} \bs \lambda (t) = \bs \lambda - \left(\bs 1 - \btaur \frac{\dd}{\dd t} \right) \left[\bs \varphi' \odot  (\bs r^\mathrm{trg} - \bs r) \right]\;.
\end{align}
This equation can be solved in a similar fashion to the forward dynamics (\cref{eq:dynamic_constraint_SI}).
First, we conveniently define the adjoint operators
\begin{align}
\qquad \bs{\mathcal{I}}^{+}_{\bs \tau} \left\{\bs x(t)\right\} &= \frac{\bs 1}{\bs \tau} \int^{\infty}_{t} \dd s \,\bs x(s) \, e^{\frac{t-s}{\bs \tau}}\, && \text{(discounted future)} \\
\qquad \bs{\mathcal{D}}^{-}_{\bs \tau} \left\{\bs x(t)\right\} &= \left(\bs 1 - \bs \tau \frac{\dd}{\dd t}\right) \bs x(t)\;.  && \text{(lookback)} \;.
\end{align}
Using these operators, we can write the dynamics of the adjoint variable \(\bs \lambda\) as
\begin{align}
  \bs \lambda(t) = \Iplusm\{\Dminusr\{\bs \varphi' \odot ( \bs r^\mathrm{trg} - \bs  r)\}\} \; .
  \label{eq:adjoint_top_sol}
\end{align}
together with a boundary condition at $t=T$ given by \cref{eq:adjoint_dyn_final}.
Note that $\Iplusm$ integrates the dynamics into the future; this is the crucial reason why an \gls{am} weight update cannot be calculated online.
Instead, we present a sample for the period $T$, and only afterwards integrate the adjoint dynamics \textit{backwards in time} to obtain the Lagrange multiplier $\bs \lambda$ at all times $t\in[0,T]$.
Finally, this allows us to calculate the gradient of the cost with respect to the synaptic weights
\begin{align}
  \nabla_{\bs W} \mathcal{L} =  \left(\frac{\dd \mathcal{L}}{\dd \bs W}\right)^{\TT} = \int^T_{0} \dd t\,\bs \lambda(t) \, \bs r^{\TT}_{\mathrm{in}}(t) \;.
  \label{eq:dLdW_sol}
\end{align}
\subsection{Hidden neurons}

We now extend the proof to hidden layers.
In order to do so, we introduce labels for all layers:
$\bs u_\ell$ denotes the somatic voltage of a given layer $\ell$.
Lookahead $\Dplusrell$ and low-pass filters $\Iminusmell$ operate with layer-specific time constants $\btaurell$ and $\btaumell$, respectively.
For each layer $\ell$, the Lagrange multipliers $\bs \lambda_\ell$ are required to enforce the dynamics
\begin{align}
  \bs h_\ell(\bs u_\ell,\Udot_\ell,\bs \theta, t) = \btaumell \Udot_\ell + \bs u_\ell - \bs W_{\ell} \bs \varphi\left(\Dplusrellmo[\bs u_{\ell-1}]\right)  -  \bs b_\ell = \bs 0 \; .
\end{align}
Since neurons in hidden layers are assumed to not be directly connected to the output layer, we can ignore the cost functional $\mathcal{C}$ in the derivation of the adjoint dynamics for the hidden layer $\ell$.
  Instead, neurons in hidden layers seek to minimize the layer-wise mismatch energy of the layer above, which we recall from \cref{eqn:GLEE} as
\begin{align*}
  E_{\ell}  = \frac{1}{2} \, \left\| \bs e_\ell \right\|^2 = \frac{1}{2} \, \left\|  \bs u_{\ell} + \btaur_{\ell}\Udot_{\ell} -\bs W_{\ell} \bs\varphi \left( \bs u_{\ell-1} + \btaur_{\ell-1}\Udot_{\ell-1}\right) - \bs b_\ell  \right\|^2 \; .
\end{align*}
Therefore, in order to derive a gradient descent update for the weights $\bs W_\ell$ of hidden layer $\ell$, we need to minimize the energy $E_{\ell+1}$ under the constraint of the dynamics of the layer $\ell$.
  This is done by introducing Lagrange multipliers $\bs \lambda_\ell$ for the hidden layer $\ell$ and calculating
\begin{align*}
  \frac{\dd\mathcal{L}}{\dd \bs W_{\ell}} =
  \frac{\dd}{\dd \bs W_{\ell}} \int_0^T \dd t \left[E_{\ell+1} + \bs \lambda_\ell^{\TT}   \bs h_\ell \right] \; .
\label{eq:layerwise-GD_general}
\end{align*}
Following the same reasoning as for the output layer, we integrate by parts the terms including $\dd\Udot_\ell/\dd \bs W_\ell$ and can avoid calculating $\dd \bs u_\ell/\dd \bs W_\ell$ if the Lagrange multipliers $\bs \lambda_\ell$ are determined by~(ref.~\cref{eq:adjoint_dynamics_general})
\begin{align}
  \frac{\partial E_{\ell+1}}{\partial \bs u_{\ell}}
  - \frac{\dd}{\dd t} \frac{\partial E_{\ell+1}}{\partial \Udot_{\ell}}
  + \bs \lambda_{\ell}^{\TT} \frac{\partial \bs h_{\ell}}{\partial \bs u_{\ell}}
  - \dot{\bs{\lambda}}_{\ell}^{\TT} \frac{\partial \bs h_{\ell}}{\partial \Udot_{\ell}}
  - \bs{\lambda}_{\ell}^{\TT} \frac{\dd}{\dd t}  \underbrace{\frac{\partial \bs h_{\ell}}{\partial \Udot_{\ell}}}_{= \btaumell} = \bs 0 \;.
\end{align}
Working out the derivatives
  \(\left( \frac{\partial E_{\ell+1}}{\partial \bs u_{\ell}} = \bs e^{\TT}_{\ell+1} \bs W_{\ell+1} \odot \bs \varphi'_\ell \right.\),
  \(\frac{\partial E_{\ell+1}}{\partial \Udot_{\ell}} = \bs e^{\TT}_{\ell+1} \bs W_{\ell+1}^\TT \odot \bs \varphi'_\ell \btaur_{\ell}\),
  \(\frac{\partial \bs h_{\ell}}{\partial \bs u_{\ell}} = \bs 1\) and
  \(\left. \frac{\partial \bs h_{\ell}}{\partial \Udot_{\ell}} = \btaum_\ell \right)\)
and assuming fixed time constants $\btaum_{\ell}$, that is, $\frac{\dd \btaum_{\ell}}{\dd t} = 0$, we find the adjoint dynamics of the hidden layer $\ell$ to be
\begin{align}
  - \left(\bs 1 - \btaurell \frac{\dd}{\dd t} \right) \left[\bs \varphi'_{\ell} \odot \bs e^{\TT}_{\ell+1} \bs W_{\ell+1} \right]
  + \bs \lambda^{\TT}_{\ell}
  - \btaumell \dot{\bs \lambda}^{\TT}_{\ell}
  = \bs 0 \;,
\end{align}
which after transposition can be written as
\begin{align}
  \btaumell \dot{\bs{\lambda}}_\ell =  \bs \lambda_\ell - \Dminusrell
  \left[\bs \varphi'_\ell  \odot  \bs W_{\ell+1}^\TT \bs e_{\ell+1}\right]\;,
\end{align}
and is solved by
\begin{align}
  \bs \lambda_\ell= \Iplusmell \Dminusrell
  \left[\bs \varphi'_\ell  \odot \bs W_{\ell+1}^\TT \bs e_{\ell+1}\right]\;.
  \label{eq:adjoint_hidden_sol}
\end{align}
Considering \cref{eq:adjoint_top_sol} and \cref{eq:adjoint_hidden_sol}, it now follows by induction that
\begin{align}
  \bs \lambda_\ell = \Iplusmell \Dminusrell
  \left[\bs \varphi'_\ell  \odot \bs W_{\ell+1}^\TT \bs \lambda_{\ell+1}\right]\;.
\end{align}
With that, we have derived \cref{eqn:lambda} from the main text for the adjoint dynamics of the hidden layer $\ell$.

    \section{Simulation details}

\subsection{A minimal GLE example}
\begin{figure}[bth]
    \centering
    \includegraphics[width=0.67\textwidth]{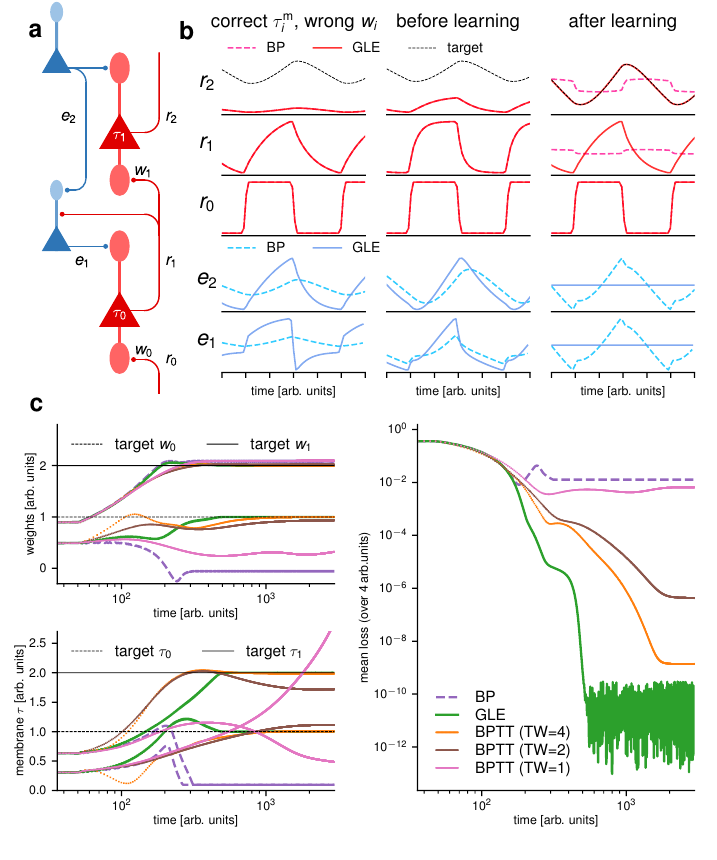}
    \caption{
        \tb{Learning with \gle in a simple chain.}
        \tb{(a)} Network setup.
        A chain of two retrospective representation neurons (red) learns to mimic the output of a teacher network (identical architecture, different parameters).
        In \gle, this chain is mirrored by a chain of corresponding error neurons (blue), following the microcircuit template in \cref{fig:mc}.
        We compare the effects of three learning algorithms: \gle (green), \bp with instantaneous errors (purple) and \bptt (point markers denote the discrete nature of the algorithm; pink, brown and orange denote different truncation windows (TW)).
        \tb{(b)} Output of representation neurons ($r_i$, red) and error neurons ($e_i$, blue) for \gle and instantaneous \bp.
        Left: correct membrane time constants but wrong weights.
        Middle: before learning (i.e., both weights and membrane time constants are wrong).
        Right: after learning.
        \tb{(c)} Evolution of weights, time constants and overall loss.
        Fluctuations at the scale of $10^{-10}$ are due to limits in the numerical precision of the simulation.
    }
    \label{fig:chain_SI}
\end{figure}
\Cref{fig:chain_SI} shows an extended version of \cref{fig:chain} from the main text.
The first column of panel (b), shows the output of the representation neurons (red) and the error neurons (blue) for \gle and instantaneous \bp in the case of correct membrane time constants but wrong weights.
This illustrates nicely the difference between instantaneous (dashed lines) and \gls{gle} prospective errors (solid lines).
Instantaneous \bp errors are in phase with the target errors as already described in \nameitref{sec:chain}.
In contrast, \gle errors are out-of-phase w.r.t. the target errors.
More concretely, since neurons in the feedforward pathway perform a low-pass filtering operation leading to a phase lag, error neurons in the feedback pathway must perform an inverse low-pass filtering operation leading to a phase advance.
The shape of both error signals $e_{i}$ over time matches nicely with the shape of the presynaptic signals $r_{i-1}$, which is important for plasticity to produce meaningful weight updates according to $\dot{w}_{i} \propto e_{i} r_{i-1}$.

\begin{figure}[bth]
    \centering
    \includegraphics[width=0.53\textwidth]{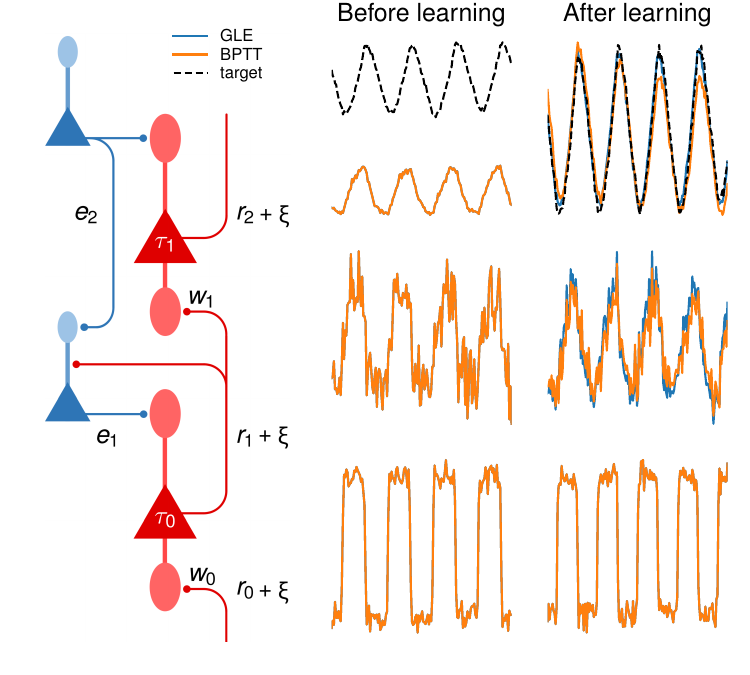}
    \caption{Additional lag-line experiment in the presence of noise on all feedforward rates $r_{i}$.
    }
    \label{fig:noise_chain_SI}
\end{figure}

In an additional experiment using the simple lag line setup, we added noise onto all neuronal outputs \( r_i \), as shown in \cref{fig:noise_chain_SI} to assess how fluctuations in internal activity levels influenced performance.
Despite the high level of noise of $10\%$ w.r.t. the amplitude of the input signal, \gls{gle} networks were still able to learn weights and time constants such that the output $r_{2}$ matched the target (dashed line).

\subsection{Small GLE networks}
\label{sec:supp-smallnets}
The main goal of \nameitref{sec:smallnetworks} is to compare the dynamics and frequency response of the adjoint variables given by \gls{am} with the errors of \gls{gle} networks.
In contrast to \nameitref{sec:chain}, where the \gls{bptt} baselines are obtained by backpropagating through the discretized forward dynamics of the \gls{gle} network, here we discretize the adjoint equations and integrate them numerically.
The pseudocode comprising the forward and adjoint dynamics, together with synaptic plasticity, is given in \cref{alg:AM}.

\begin{algorithm}
  \small
  \caption{Simulation of the Adjoint Method baselines in \nameitref{sec:smallnetworks}}
  \begin{algorithmic}
    \State initialize network parameters $\theta = \{W_{ij}, b_i, \taum_i, \taur_i\}$
    \State initialize network states at $t=0$: $u_i(0)$, $r_{i}(0)$
    \State initialize adjoint variable $\lambda$ states at $t=T$: $\lambda_i(T)$
    \For{time step $t \in [0, T]$ with step size $\dd t$} %
      \For{layer $\ell$ from $1$ to $L$}
        \State sum input currents:
        $\quad \bs I_\ell(t) = \bs{W}_{\ell}(t) \bs{r}_{\ell-1}(t) + \bs{b}_{\ell}(t)$
        \State integrate input currents into membrane:
        $\quad\Delta \bs{u}_{\ell}(t)  \gets {(\btaumell)}^{-1} \odot \left(-\bs{u}_{\ell}(t) + \bs I_{\ell}(t)\right)$
        \State update membrane voltages:
        $\quad\bs{u}_{\ell}(t + \dd t) \gets \bs{u}_{\ell}(t) + \dd t \Delta \bs{u}_{\ell}(t)$
        \State calculate prospective outputs:
        $\quad\bs{r}_{\ell}(t +\dd t) \gets \bs{\varphi}\left(\bs{u}_{\ell}(t) + \btaurell \odot \Delta \bs{u}_{\ell}(t)\right)$
        \State store variables and update time index:
        $\quad t \gets t + \dd t$
      \EndFor{}
    \EndFor{}
    \For{layer $\ell$ from $L$ to $1$}
      \For{time step $t \in [T, \dd t]$ with step size $\dd t$}
        \State calculate lookback \(\Dminusrell\):
          \If{$\ell = L$}
          \State $\DminusrL \gets \left(\bs 1 - \frac{\btaurell}{\dd t}\right) \odot \bs \varphi'_{L}(t) \odot \left( \bs r^\mathrm{trg}(t) - \bs  r_{L}(t) \right) + \frac{\btaurell}{\dd t} \odot \bs \varphi_{L}'(t-\dd t) \odot \left( \bs r^\mathrm{trg}(t -\dd t) - \bs  r_{L}(t -\dd t) \right)$
          \Else
          \State $\Dminusrell \gets \left(\bs 1 - \frac{\btaurell}{\dd t} \right) \odot \bs \varphi'_{\ell}(t) \odot \left( \bs W_{\ell +1}^{\TT} \bs \lambda_{\ell +1}(t)\right) + \frac{\btaurell}{\dd t} \odot \bs \varphi_{\ell}'(t-\dd t) \odot \left(\bs W_{\ell +1}^{\TT} \bs \lambda_{\ell +1}(t - \dd t)\right)$
          \EndIf
          \State calculate adjoint variable:
          $\quad\bs \lambda_{\ell}(t -  \dd t) \gets  \bs \lambda_{\ell}(t)  \odot e^{-\dd t/ \btaumell} + \Dminusrell \odot \dd t /\btaumell $
          \State update time index:
          $\quad t \gets t - \dd t $
      \EndFor
    \EndFor
    \State Update synaptic weights:
    $\quad\bs{W}_{\ell} \gets \bs{W}_{\ell} - \eta_{W}  \sum_{t=\dd t}^{T} \bs \lambda_{\ell}(t) \bs{r}^{\TT}_{\ell-1}(t)$
  \end{algorithmic}
  \label{alg:AM}
\end{algorithm}

\subsection{\mnistId reference baselines}
\label{sec:supp-mnist1d}
All baselines were trained offline with BP(TT) (using PyTorch \texttt{autograd}) on the classification loss $\mathcal{L}_{C} = \mathrm{CE}(\hat{y}, y)$, with the one-hot target $\hat{y}$ and the prediction $y = \text{Softmax}(\sum_{t=0}^{T} y_{t})$, where $T$ is the length of a sample.
In contrast, \gle only uses the instantaneous cost $C(t) = \mathrm{CE}(\hat{y}, y_{t})$.
For the \gls{tcn} and \gls{mlp} baselines this corresponds to classical, spatial \gls{bp} as they map time to (input) space and see the whole sequence at once.
Note that the \gls{gru} implementation maintains a record of all hidden states for each sample and subsequently trains a readout layer on top of them.
Therefore, despite the sequential processing, the \gls{gru} can only produce a prediction after the full sequence has been processed.

In particular, we use the following supersampling and model architectures:
\begin{itemize}
    \item \gls{mlp}: MNIST-1D sequences supersampled to 360 steps are fed into two consecutive hidden layers with 100 neurons each and ReLU activations in between, followed by a readout with 10 output neurons.
                     The learning rate is $0.005$.
    \item \gls{tcn}: MNIST-1D sequences supersampled to 360 steps are fed into three layers of dilated convolutions with 25 channels each, kernel sizes $[5,3,3]$, stride $2$, a dilation of $360/40 = 9$ followed by a linear readout with 125 input channels and 10 output neurons.
                     The learning rate is $0.01$.
    \item \gls{gru}: MNIST-1D sequences supersampled to 72 steps are streamed into single input unit that feeds into a single, bidirectional recurrent layer with 6 neurons followed by a linear layer on top of the hidden activations at all time steps, resulting in a bidirectional readout with $2*6*72=864$ input and 10 output neurons.
                     The learning rate is $0.01$.
\end{itemize}
\Cref{tab:mnist1d-results} summarizes the main characteristics and performance of all evaluated models.
\begin{table}[htb]
    \small
    \centering
    \begin{tabular}{|c|c|c|c|c|c|c|c|c|c|}
        \hline
        Model & GLE (42k) & GLE (15k) &  MLP$_{0}$ & TCN$_{0}$ & GRU$_{0}$ & MLP & TCN & GRU \\
        \hline
        number of parameters & 42040 & 14956 & 15210 & 5210 & 5134 & 47210 & 11960 & 8974 \\
        \hline
        online learning & yes & yes & no & no & no & no & no & no \\
        \hline
        streamed input & yes & yes & no & no & yes & no & no & yes \\
        \hline
        mean valid acc (last) / $\%$ & $93.5\pm0.9$ & $91.7\pm0.8$ & $65.0\pm1.2$ & $93.7\pm1.0$ & $90.3\pm2.5$ & $65.5\pm1.0$ & $96.7\pm0.9$ & $94.0\pm1.1$ \\
        \hline
    \end{tabular}
    \caption{
      Summary of the evaluated network models and their performance on the MNIST-1D dataset.
    }
    \label{tab:mnist1d-results}
\end{table}

\subsection{\gsc reference baselines}
\label{sec:supp-gsc}

As in the \mnistId dataset, both MLP and TCN networks were trained with spatial BP on a spatial representation of the temporal signal.
The GRU network was trained offline with BPTT.
\Cref{tab:gsc-results} summarizes the hyperparameters and performance of all evaluated models.

\begin{table*}[h]
    \centering
    \begin{tabular}{|c|c|c|c|c|c|c|c|c|}
        \hline
        Model & GLE & MLP L & TCN L & GRU L & LSTM L \\
        \hline
        train samples & 36923 & 36923 & 36923 & 36923 & 36923 \\
        validation samples & 4445 & 4445 & 4445 & 4445 & 4445\\
        test samples & 4890 & 4890 & 4890 & 4890 & 4890  \\
        MFS/MFCC & MFS & MFCC & MFCC & MFCC & MFCC \\
        bins/coeffs & 32 & 40 & 10 & 40 & 40 \\
        Fourier Window length & 64 ms & 40 ms & 40 ms & 40ms & 20 ms\\
        Fourier Window Stride & 25 ms & 40 ms & 20 ms & 20ms & 20 ms\\
        Batch size & 256 & 100 & 100 & 100 & 100\\
        Epochs & 320 & 320 & 320 & 320 & 320 \\
        \hline
        \# params & 1160262 & 495744 & 476734 & 498012 & 492620   \\
        \hline
        acc / $\%$ & $91.44\pm 0.23$ & $88.00\pm 0.25$ & $92.32\pm 0.28$ & $94.93\pm 0.25$ & $94.00\pm 0.19$ \\
        \hline
    \end{tabular}
    \caption{
      Summary of the evaluated network models and their performance on the \gsc dataset.
      Additional details on the baseline model hyperparameters can be found in Appendix A of~\cite{zhanghello2018}.
    }
    \label{tab:gsc-results}
\end{table*}

\printbibliography[heading=subbibliography]
\end{refsection}

\end{document}